\documentclass[12pt,preprint]{aastex}
\usepackage{epsfig}
\usepackage{natbib}                
\usepackage{lscape}
\usepackage{verbatim}              
\usepackage{graphicx}              
\usepackage{amsmath,amsthm,amsfonts,amsopn,amssymb} 
\usepackage{longtable,pdflscape}
\usepackage{afterpage}
\usepackage{rotating}					

\shorttitle{Diameters and Temperatures of Main Sequence Stars}
\shortauthors{Boyajian et al.}
\begin{document}


\title{Stellar Diameters and Temperatures 	\\
		I. Main Sequence A, F, \& G Stars} 

\author{Tabetha S. Boyajian\altaffilmark{1,2}, Harold A. McAlister\altaffilmark{1}, Gerard van Belle\altaffilmark{3}, Douglas R. Gies\altaffilmark{1}, Theo A. ten Brummelaar\altaffilmark{4}, Kaspar von Braun\altaffilmark{5} Chris Farrington\altaffilmark{4}, P. J. Goldfinger\altaffilmark{4}, David O'Brien\altaffilmark{1}, J. Robert Parks\altaffilmark{1},  Noel D. Richardson\altaffilmark{1}, Stephen Ridgway\altaffilmark{6}, Gail Schaefer\altaffilmark{4}, Laszlo Sturmann\altaffilmark{4}, Judit Sturmann\altaffilmark{4}, Yamina Touhami\altaffilmark{1}, Nils H. Turner\altaffilmark{4}, Russel White\altaffilmark{1}}

\altaffiltext{1}{Center for High Angular Resolution Astronomy and Department of Physics and Astronomy, Georgia State University, P. O. Box 4106, Atlanta, GA 30302-4106} 
\altaffiltext{2}{Hubble Fellow} 
\altaffiltext{3}{European Southern Observatory, Karl-Schwarzschild-Str. 2, 85748 Garching, Germany}
\altaffiltext{4}{The CHARA Array, Mount Wilson Observatory, Mount Wilson, CA 91023}
\altaffiltext{5}{NASA Exoplanet Science Institute, California Institute of Technology, MC 100-22, Pasadena, CA 91125}
\altaffiltext{6}{National Optical Astronomy Observatory, P.O. Box 26732, Tucson, AZ 85726-6732}



\begin{abstract}

We have executed a survey of nearby, main sequence A, F, and G-type stars with the CHARA Array, successfully measuring the angular diameters of forty-four stars with an average precision of $\sim 1.5$\%.  We present new measures of the bolometric flux, which in turn leads to an empirical determination of the effective temperature for the stars observed.  In addition, these CHARA-determined temperatures, radii, and luminosities are fit to Yonsei-Yale model isochrones to constrain the masses and ages of the stars.  These results are compared to indirect estimates of these quantities obtained by collecting photometry of the stars and applying them to model atmospheres and evolutionary isochrones.  We find that for most cases, the models overestimate the effective temperature by $\sim 1.5-4$\%, when compared to our directly measured values.  The overestimated temperatures and underestimated radii in these works appear to cause an additional offset in the star's surface gravity measurements, which consequently yield higher masses and younger ages, in particular for stars with masses greater than $\sim$ 1.3 $M_{\odot}$.  Additionally, we compare our measurements to a large sample of eclipsing binary stars, and excellent agreement is seen within both data sets. Finally, we present temperature relations with respect to ($B-V$) and ($V-K$) color as well as spectral type showing that calibration of effective temperatures with errors $\sim 1$\% is now possible from interferometric angular diameters of stars. 

\end{abstract}



\section{Introduction}                              

The direct measurement of the stellar angular diameter is a valuable key in determining fundamental properties of a star, particularly the linear radius and effective temperature.  These properties of a star provide the link between the theory of stellar structure and evolution to model atmospheres.  For nearby, main-sequence stars, where we know their distances well,  the angular diameters are difficult to measure due to their tiny sizes compared to their evolved counterparts.  The high angular resolution obtained through long-baseline optical/infrared interferometry has enabled us to resolve the photospheric disks of such nearby stars. 

Several decades ago, a survey carried out at the Narrabri Stellar Intensity Interferometer \citep{bro74, cod76} was conducted to measure the angular diameters of 32 stars. This survey extended from O to F type stars, eleven of which were roughly on the main sequence (luminosity class V or IV)\footnote{The average precision of these angular diameter determinations depended primarily on the brightness of the object, and was $\approx$ 6.5\% for the 32 stars measured.}. For several decades, luminosity class I, II, and III stars were observed with interferometers such as the Mark~III and the Palomar Testbed Interferometer, but no main-sequence star earlier than A7 were measured \citep{dav97}.  

As an update, the CHARM2 Catalogue \citep{ric05} provides a compilation of stellar diameters by means of direct measurements by high angular resolution methods, as well as indirect estimates.  The CHARM2 Catalogue includes all results as of July 2004, a total of 8231 entries.  Filtering out the entries to include only unique sources where direct measurements exist with errors in the angular diameter measurements of less than 5\%, this number drops to 242, and only 24 of these reside on the main sequence (luminosity class V or IV).  

In a recent work by \citet{hol08}, they remark that measurements of the angular diameters of main sequence F and G stars need to be better than 2\%, yielding temperatures to 1\%, in order for offsets in the color-temperature calibrations to be minimal.  At that time, only nine stars met this criterion.  This precision limit reiterates the target accuracy proposed by \citet{bla79} for the limits to the Infrared Flux Method, that a good $T_{\rm EFF}$ determination goal should be 1\% to match the best atomic data available for abundance determinations and $\log g$ estimates \citep{dav85, boo97}. 

The long baselines of the CHARA Array are uniquely suited for observing diameters of main-sequence stars to great precision.  In this paper, we present the angular diameters of 44 main sequence A-, F-, and G-type stars measured with the CHARA Array, the most extensive interferometric survey of main-sequence stars to-date. Details on the observing strategy and observations are in Section~2, and the results are presented and discussed in Section~3.  In Section~3 we also discuss the consistency of our results compared to the recent work of \citet{van09}, who used the Palomar Testbed Interferometer to observe a few dozen main sequence stars, where 14 of these stars are common sources with this work.  Section~4 introduces the fundamental stellar properties of linear radii, luminosities and effective temperatures for the stars observed.  In the discussion (Section~5) we use our data to derive empirical temperature relations to ($B-V$) and ($V-K$) colors and Spectral Type.  We also present masses and ages for the sample obtained via isochrone fitting, and masses computed by combining our radii with published values of surface gravities. Our results are compared to three surveys in the literature that have a large percentage of stars in common with our survey, as well as a large sample of non-interacting eclipsing binaries. We summarize in Section~6.

\section{Observations with the CHARA Array}

\subsection{Instrument}

The CHARA Array is an optical/infrared interferometric array located at Mount Wilson Observatory in the San Gabriel mountains of southern California (a detailed description of the instrument can be found in \citealt{ten05}).  Briefly, the CHARA Array consists of six, 1-meter aperture telescopes in a Y-shaped configuration spread across the mountaintop of the Observatory.  With the six telescopes, there are fifteen available baseline combinations, ranging from 34 to 331 meters, at a variety of position angle orientations $\psi$.  The CHARA Array currently is the longest baseline operational optical/infrared interferometer in the world.     

There are several beam combiners available for the CHARA Array, and for this project, the observations were made using the CHARA Classic beam combiner in two-telescope mode. CHARA Classic is a pupil-plane beam combiner, which is used in $K^{\prime}$-band, empirically measured in \citet{bow10} to have a central wavelength of $\lambda_{K^{\prime}} = 2.141 \pm 0.003 \mu$m.  Fringes are detected and recorded on the Near Infrared Observer (NIRO) camera, which is based upon a HgCdTe PICNIC Array read out at high speed.  Nearly all (98.5\%) of the observing for this work was performed remotely from Georgia State University's Cleon Arrington Remote Operations Center (AROC) in Atlanta, GA during the 2007, 2008, and 2009 observing seasons. 

\subsection{Target Selection}

The main goal of this survey is to determine the angular diameters of a large number of stars to high precision.  We limited the stars observed by selecting only those for which the predicted precision of the measured angular diameter will be better than 4\%.  The precision of a measurement of the stellar angular diameter depends on how far down the visibility curve one is able to sample.  The expression of the uniform disk visibility function of a single star (Equation~\ref{eq:Vis_UD}) is dependent on $B$ the projected baseline, $\theta$ the angular diameter of the star, and $\lambda$ the wavelength of observation:

\begin{equation}
V  =  \frac{2 J_1(\rm x) }{\rm x},
\label{eq:Vis_UD}
\end{equation}
\noindent where 
\begin{equation}
\rm x = \pi \emph{B} \theta \lambda^{-1}.
\label{eq:Vis_x}
\end{equation}

\noindent Where $J_1$ is the first order Bessel function. By knowing the $\lambda$ and $B$ utilized in a given observation, we can estimate the optimum resolution range resulting from the precision with which we can measure the object visibility. To ensure that we reach the precision goal of $\sigma\theta < $4\% for our observations, we find the approximate limiting resolution is $\theta \sim 0.65$~mas for $K^{\prime}$.  In reality, this number will vary (for example, see \citealt{bai08}), but its use to establish a first order cut-off for a preliminary sample selection is appropriate.


Each star we chose to observe was then hand selected from a {\it Hipparcos} Catalogue query.  This process was initiated from assumptions of the nominal linear size of a main sequence star from \citet{cox00} based on $B-V$ colors, in order to determine a maximum distance to be sampled for each spectral type before reaching the $\theta = 0.65$~mas resolution limit.  Additionally, the luminosity class of the star was restricted by apparent $V$ magnitudes to only admit roughly main sequence stars \citep{cox00}.  The declination limit to this survey was restricted to targets above $-$10$^{\circ}$ declination.  All stars were easily within the magnitude limits for observing with the CHARA Array, with magnitudes ranging from $2.5 < V < 6.4$ and $1.6 < K < 4.4$\footnote{Very conservative limits for observing with the CHARA Classic beam combiner require $V < 10$~mags and $K < 7$ mags.}. Table \ref{tab:sample} lists the names, coordinates, spectral types, magnitudes, metallicity  [Fe/H], and {\it Hipparcos} distance of the stars observed.   

\placetable{tab:sample}	




\subsection{Data Calibration}
\label{sec:data_calibration}

We follow the standard routine for interferometric observing where a calibrator star with known size is observed before and after every observation of a science object\footnote{This sequence of calibrator - object - calibrator is called a \textquotedblleft bracket\textquotedblright}, which enables us to remove the instrumental response of the system as well as effects from local seeing conditions.  A description of the ideal calibrator and the propagation of errors to the true visibility measurements of the science star, can be found in \citet{van05}. The calibrator stars used in this work were initially selected from the getCal web interface\footnote{http://nexsciweb.ipac.caltech.edu/gcWeb/gcWeb.jsp} and were restricted by angular distance in the sky to the object, magnitude, and its estimated angular size. In most cases, the calibrator star was less than $6^{\circ}$ from the object, and never lying further than $\sim 10^{\circ}$.  The closeness of the object to the calibrator is crucial for quickly acquiring brackets, where under typical observing conditions, there is approximately 4$-$5 minutes between observations. The magnitudes of the calibrator stars were chosen from observability limits with our telescopes and instrument.  Finally, the estimated angular size of the calibrator was the attribute that is most weighted in the calibrator searching process.  

With CHARA's long baselines, unwanted biases may be introduced if the calibrator is resolved. If possible, calibrators were chosen to have a angular diameter less than 0.45~mas, following the example described in \citet{van05} for CHARA's maximum baseline of 331~m.  In order to estimate the calibrator's angular size, $\theta_{\rm SED}$, a Kurucz model spectral energy distribution was fit to Johnson $UBV$ \citep{mer97}, Str\"{o}mgren $uvby$ \citep{hau98}, and 2MASS $JHK$ \citep{cut03} flux calibrated photometry\footnote{Transformations from  $UBV$, $uvby$, and $JHK$ magnitude to flux were made using the relations described in \citet{col96, gra98, coh03}, respectively.}. The calibrators used in this work are listed in Table~\ref{tab:calibs}.  Table~\ref{tab:calibs} also shows the calibrator Johnson, Str\"{o}mgren, and 2MASS magnitudes, and $\theta_{\rm SED}$ fit for each star.  The last column lists the science object(s) observed with that calibrator.   

\placetable{tab:calibs}	

Two-thirds of the calibrators used meet the $\theta_{\rm SED} < 0.45$~mas criteria\footnote{all have $\theta_{\rm SED} < 0.66$~mas, with a mean angular diameter error of 4.4\%}.  In many cases, the science stars were observed with more than one calibrator, on more than one occasion, in order to reduce the likelihood of biases being introduced (for example, see HD~30652; Table~\ref{tab:observations}).  In some instances, we were not so fortunate to observe stars with multiple calibrators having the ideal angular size of $\theta_{\rm SED} < 0.45$~mas.  For instance, the stars HD~5015, HD~6582, and HD~10780 were observed with only one calibrator, HD~6210 ($\theta_{\rm SED} = 0.519$~mas).  However, we note that the observations of the star HD~4614 were also obtained with this calibrator as well as two additional calibrators, one of which meets the ideal criterion for being completely unresolved (HD~9407, $\theta_{\rm SED} = 0.430$~mas).  We find that all the calibrated data sets for HD~4614 agree flawlessly in the final diameter fits. Thus, using the slightly resolved calibrator star HD~6210 is not a cause of concern.  Only four stars, HD~97603, HD~185144, HD~126660, and HD~114710 have observations with only one calibrator with $0.45 < \theta_{\rm SED} < 0.57$~mas, but consistency in the calibration process seen with stars observed with more than one calibrator (such as HD~4614) provides assurance that this issue is not a problematic one. Additionally, the maximum CHARA baseline of 331~m was never reached for a significant amount of these observations, thus pushing the theorized ideal maximum angular size for the calibrator to larger sizes.  

The observing log is shown in Table~\ref{tab:observations} and lists the identifications of the 44 stars observed in this work (column~1), UT date (column~2), CHARA baseline (column~3), number of brackets (column~4), and the calibrator(s) used on that date (column~5).  The specifics of the CHARA baseline configurations are displayed in Table~\ref{tab:CHARAbaselines}.  We include the observations of HD~6582, HD~10780, and HD~185144 here for completeness, because they are an original part of this survey, but their results have been previously presented in \citet{boy08}.  

\placetable{tab:observations}	
\placetable{tab:CHARAbaselines}	


\section{Angular Diameters}        %
\label{sec:diameters}

Angular diameters for each star were determined by fitting our interferometric measurements to the visibility curve for a single star's uniform disk $\theta_{\rm UD}$ and limb-darkened $\theta_{\rm LD}$ \citep{han74} angular diameters from the calibrated visibilities by $\chi^2$ minimization \citep{mar09}.  To account for limb-darkening, we use the $K$-band limb-darkening coefficients computed in \citet{cla00}.  Limb darkening corrections are at the two percent level in the infrared, and for these types of stars the uncertainty of this correction is at most a tenth of a percent, well within our error budget.  

\begin{figure}
\centering
\begin{tabular}{cc}

\epsfig{file=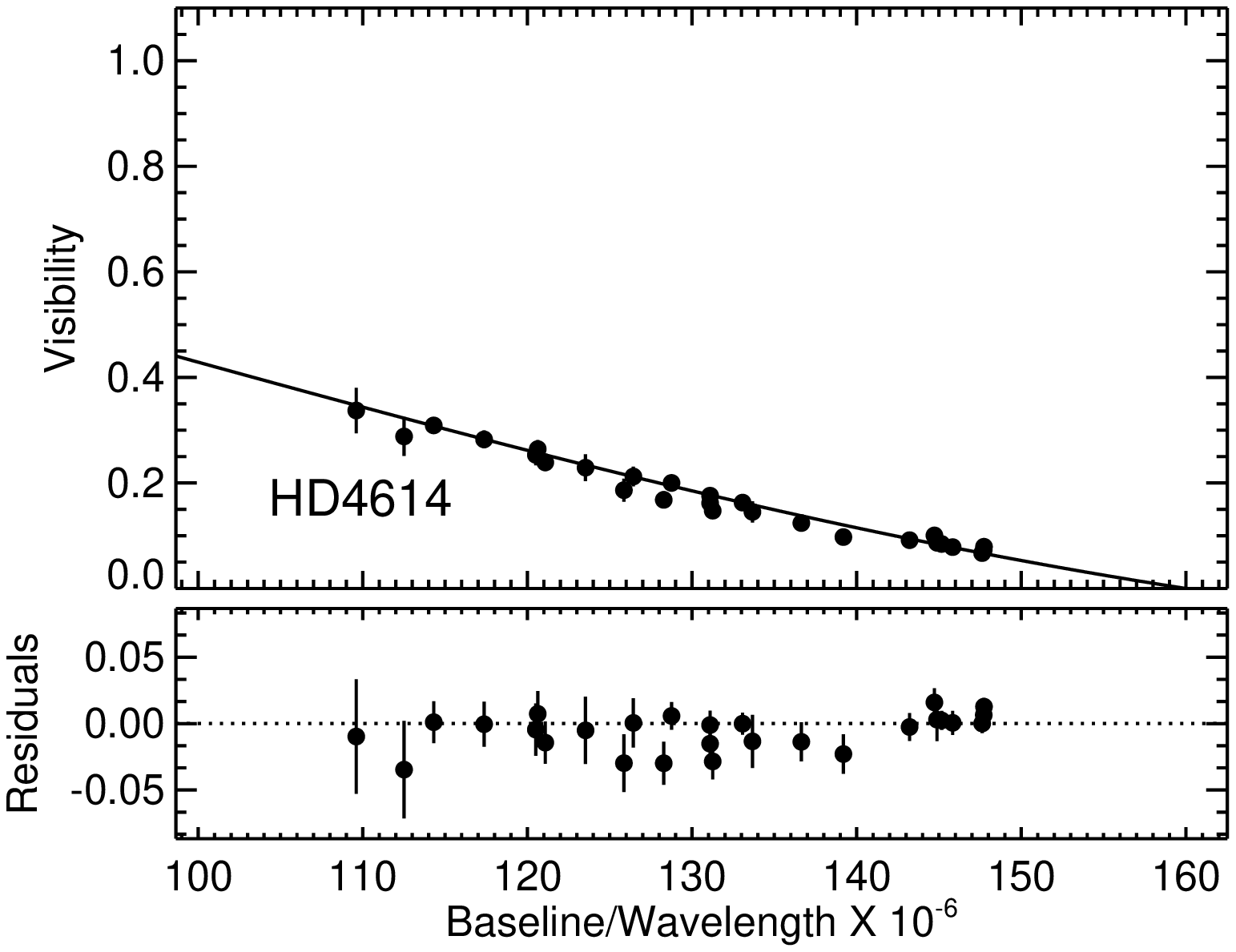,width=0.5\linewidth,clip=} &  
\epsfig{file=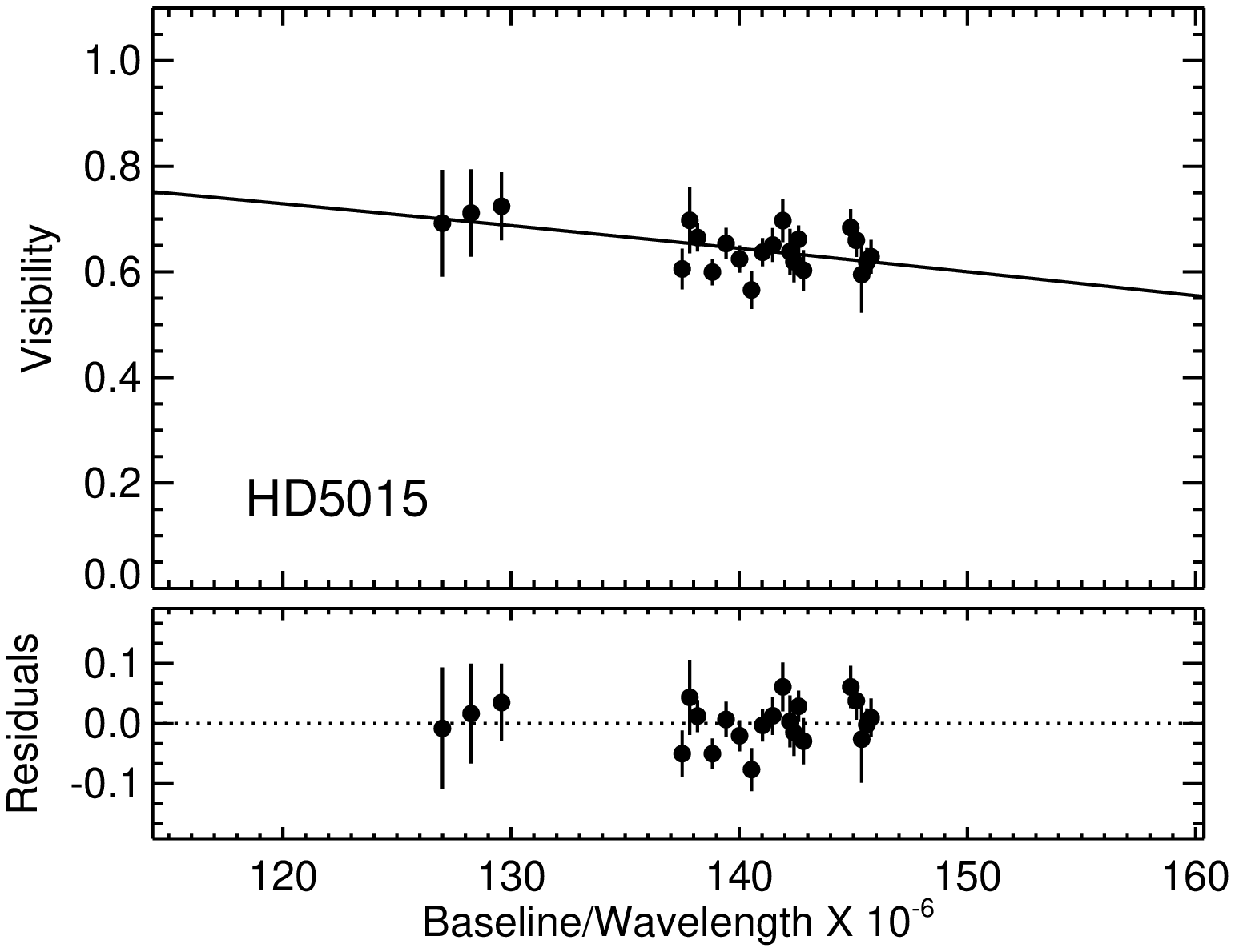,width=0.5\linewidth,clip=}   \\
\epsfig{file=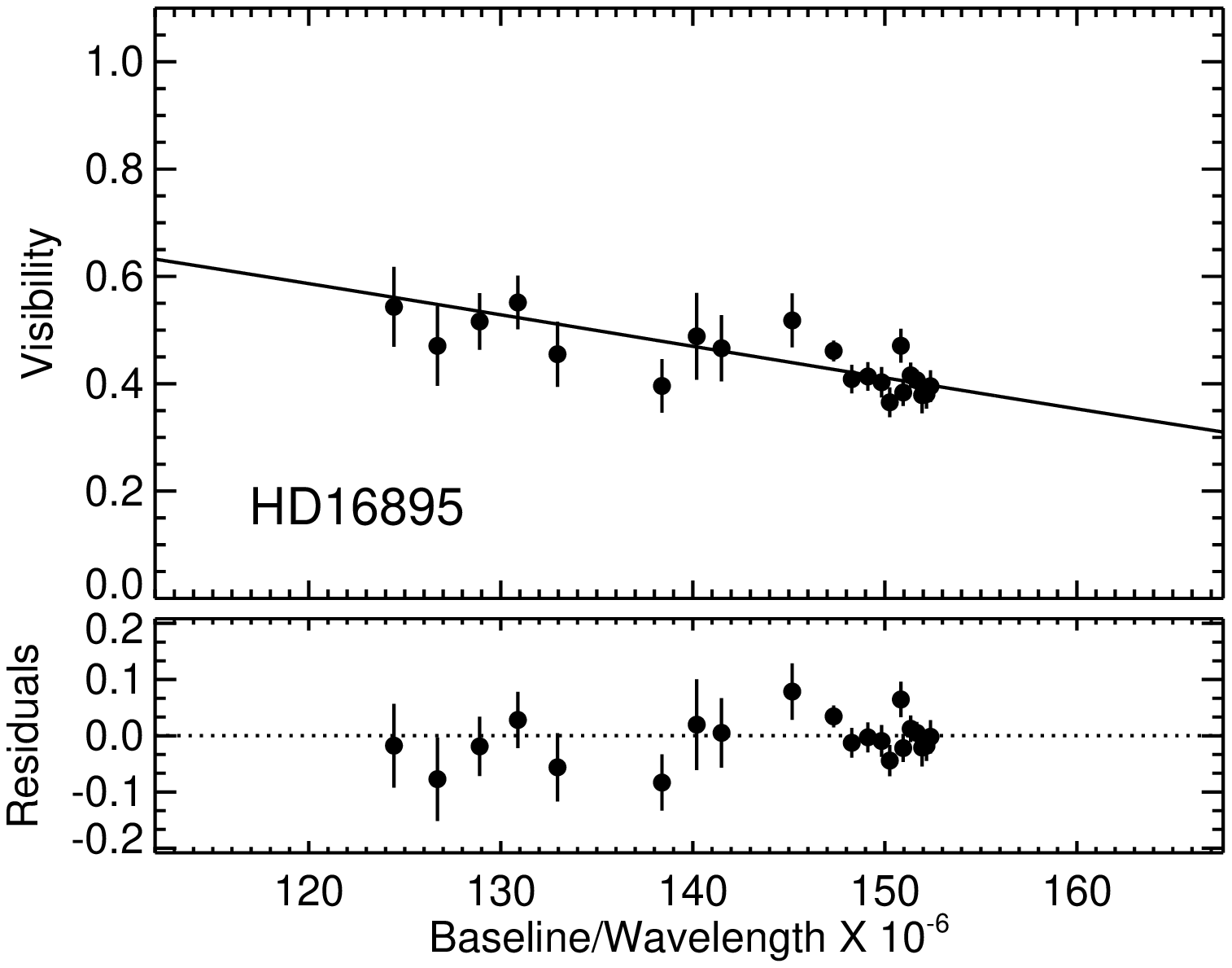,width=0.5\linewidth,clip=} &  
\epsfig{file=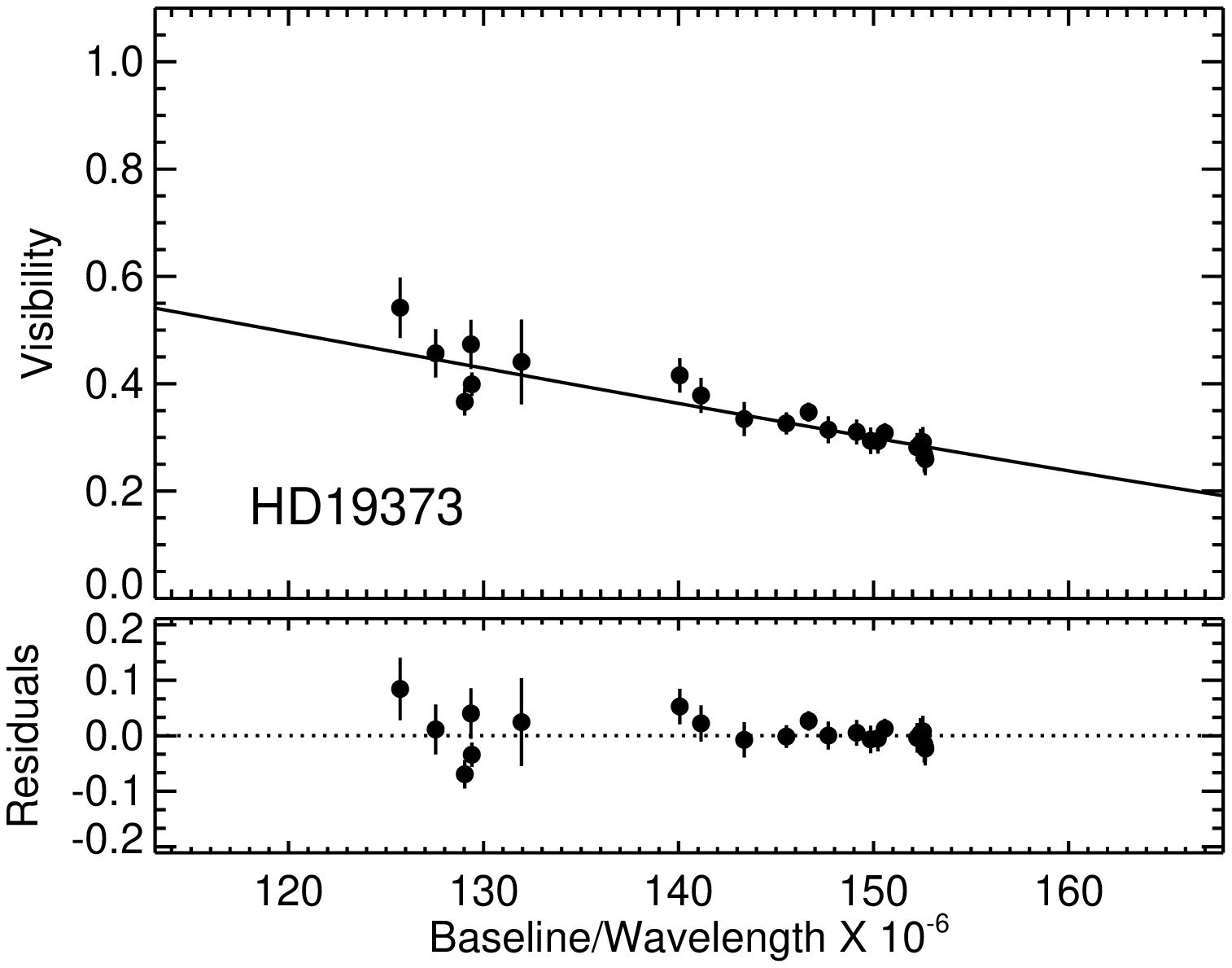,width=0.5\linewidth,clip=}   \\
\epsfig{file=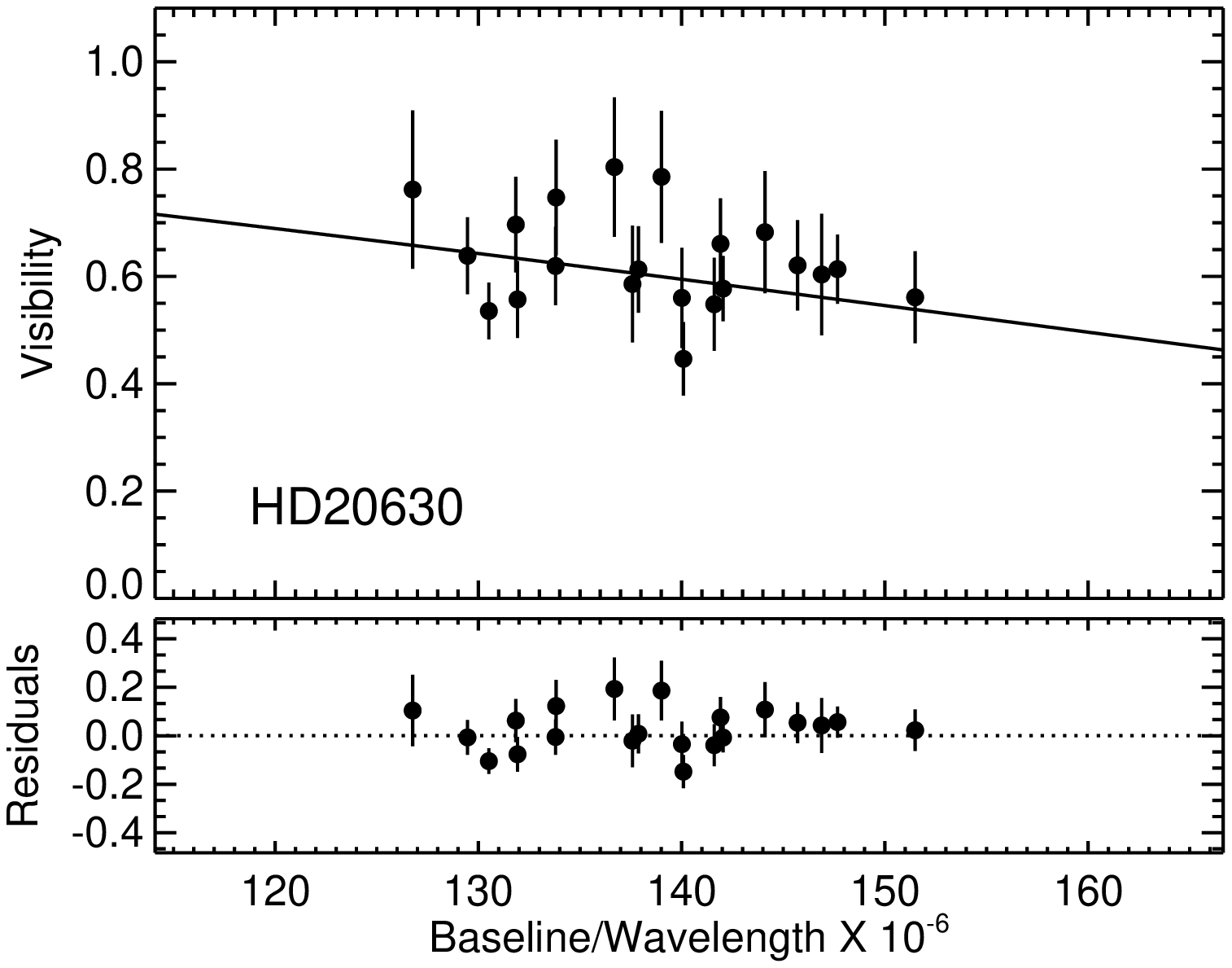,width=0.5\linewidth,clip=} &  
\epsfig{file=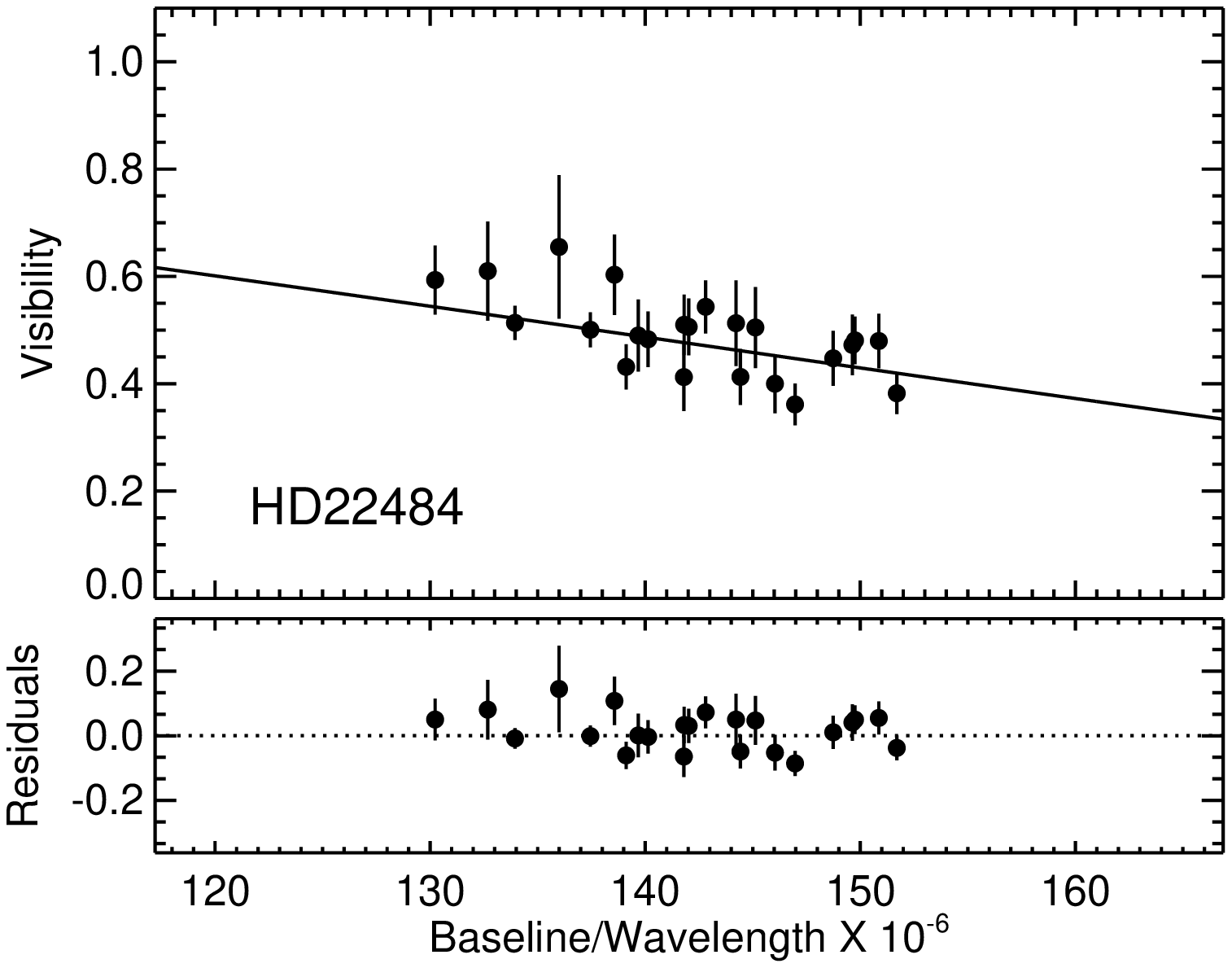,width=0.5\linewidth,clip=}
 \end{tabular}
\caption[Angular Diameters] {Calibrated observations plotted with the limb-darkened angular diameter fit for each star observed.  See Section~\ref{sec:diameters} and Table~\ref{tab:diameters} for details.}
\label{fig:diameters_1}
\end{figure}
\begin{figure}
\centering
\begin{tabular}{cc}
   
\epsfig{file=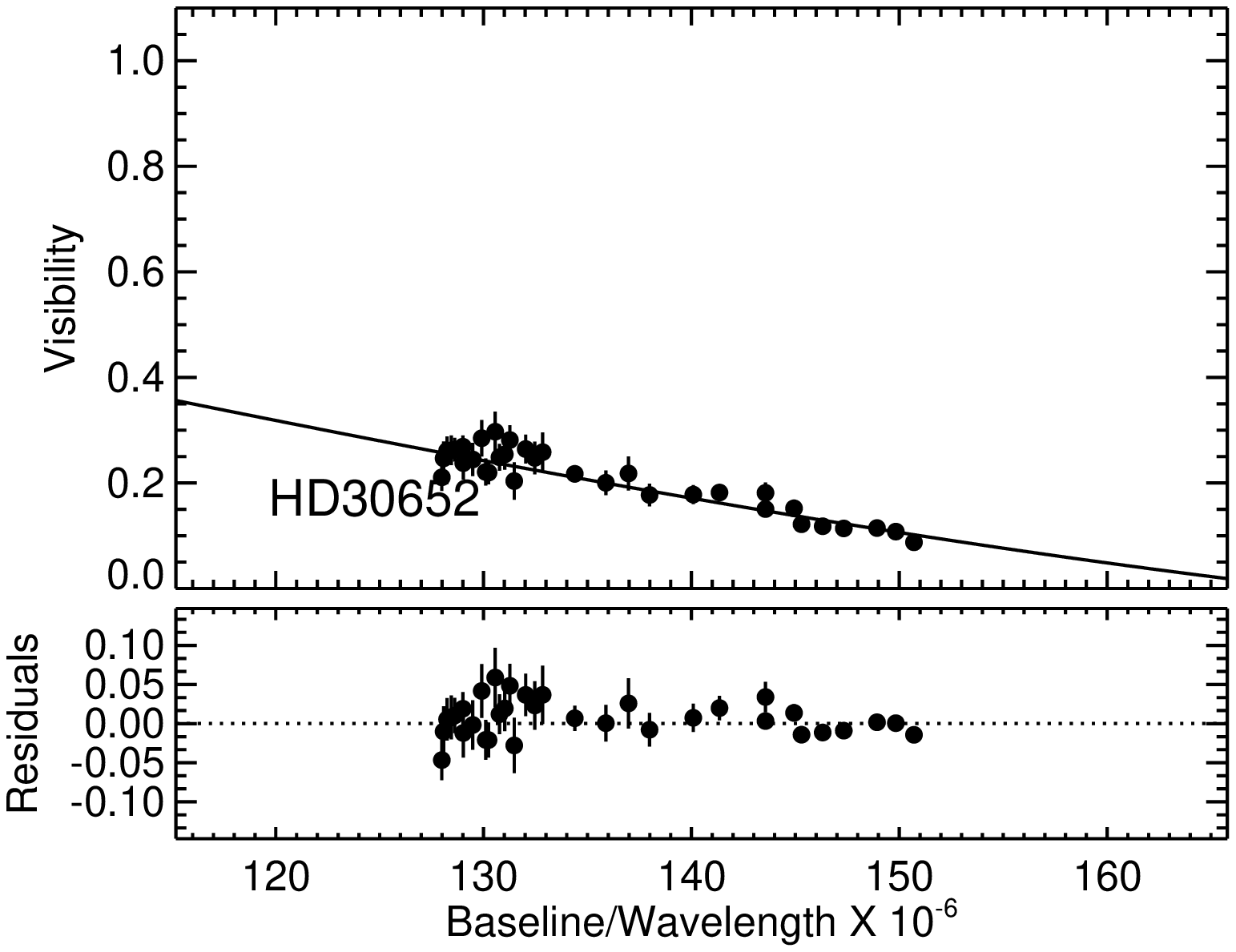,width=0.5\linewidth,clip=} &  
\epsfig{file=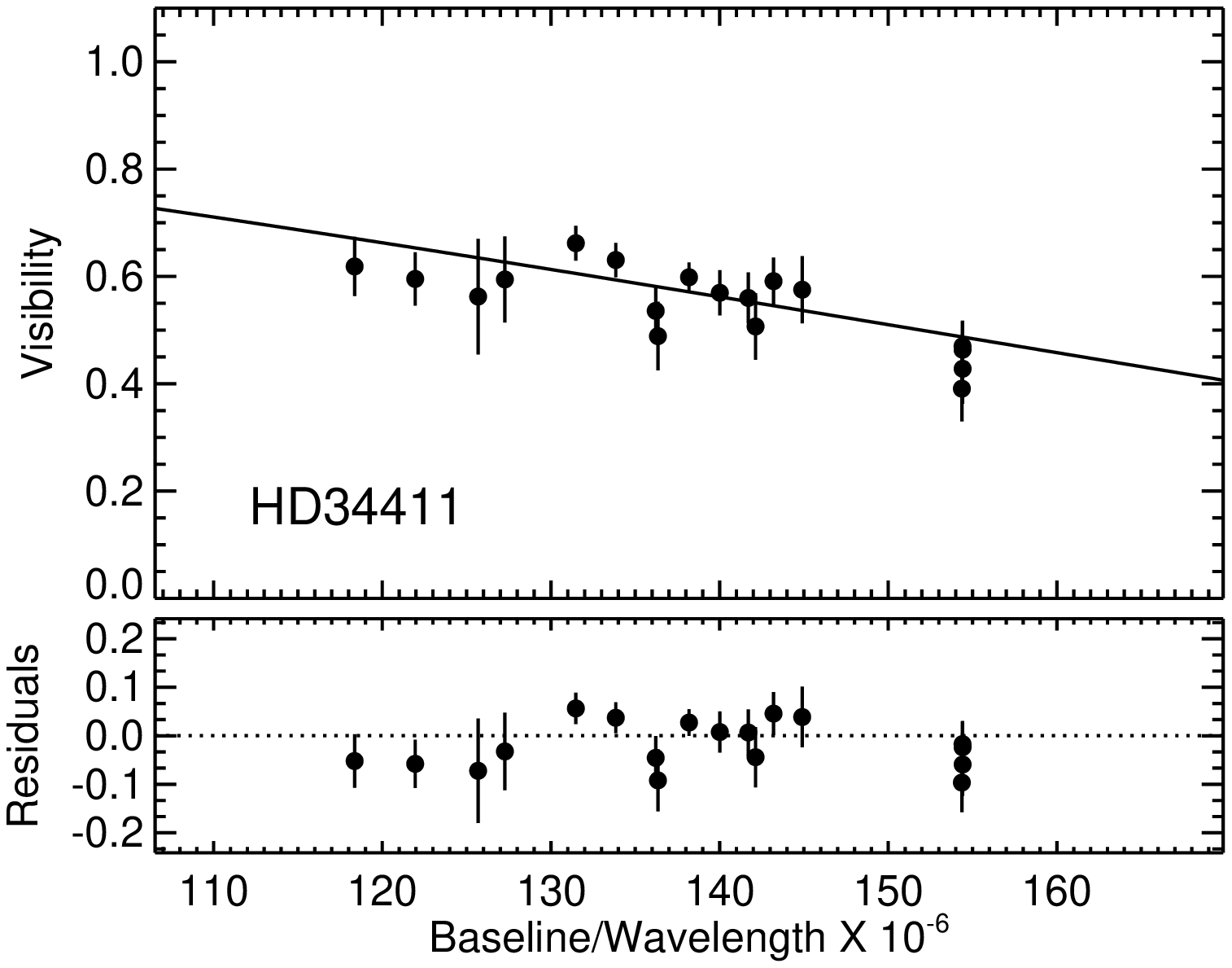,width=0.5\linewidth,clip=}   \\
\epsfig{file=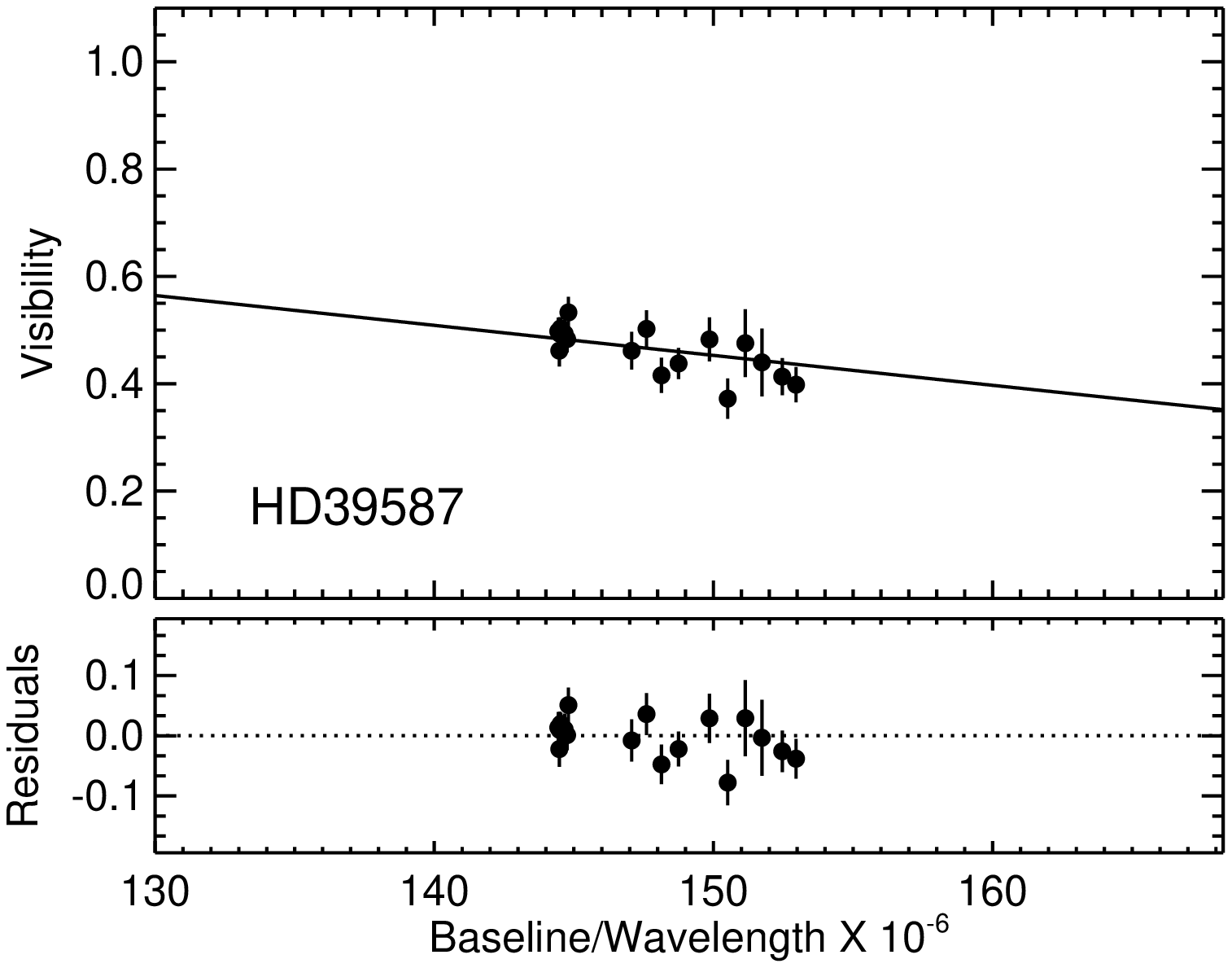,width=0.5\linewidth,clip=} &  
\epsfig{file=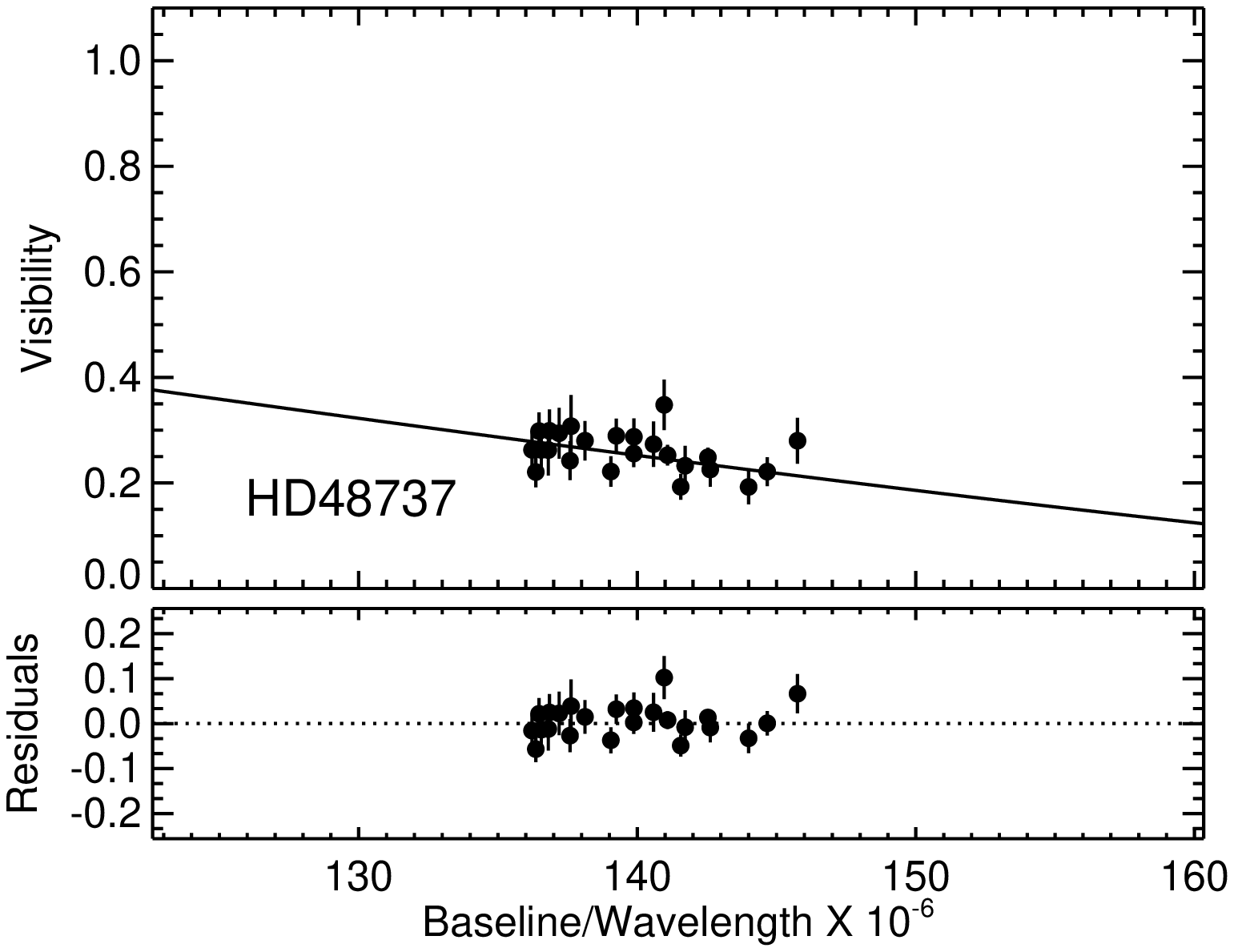,width=0.5\linewidth,clip=}   \\
\epsfig{file=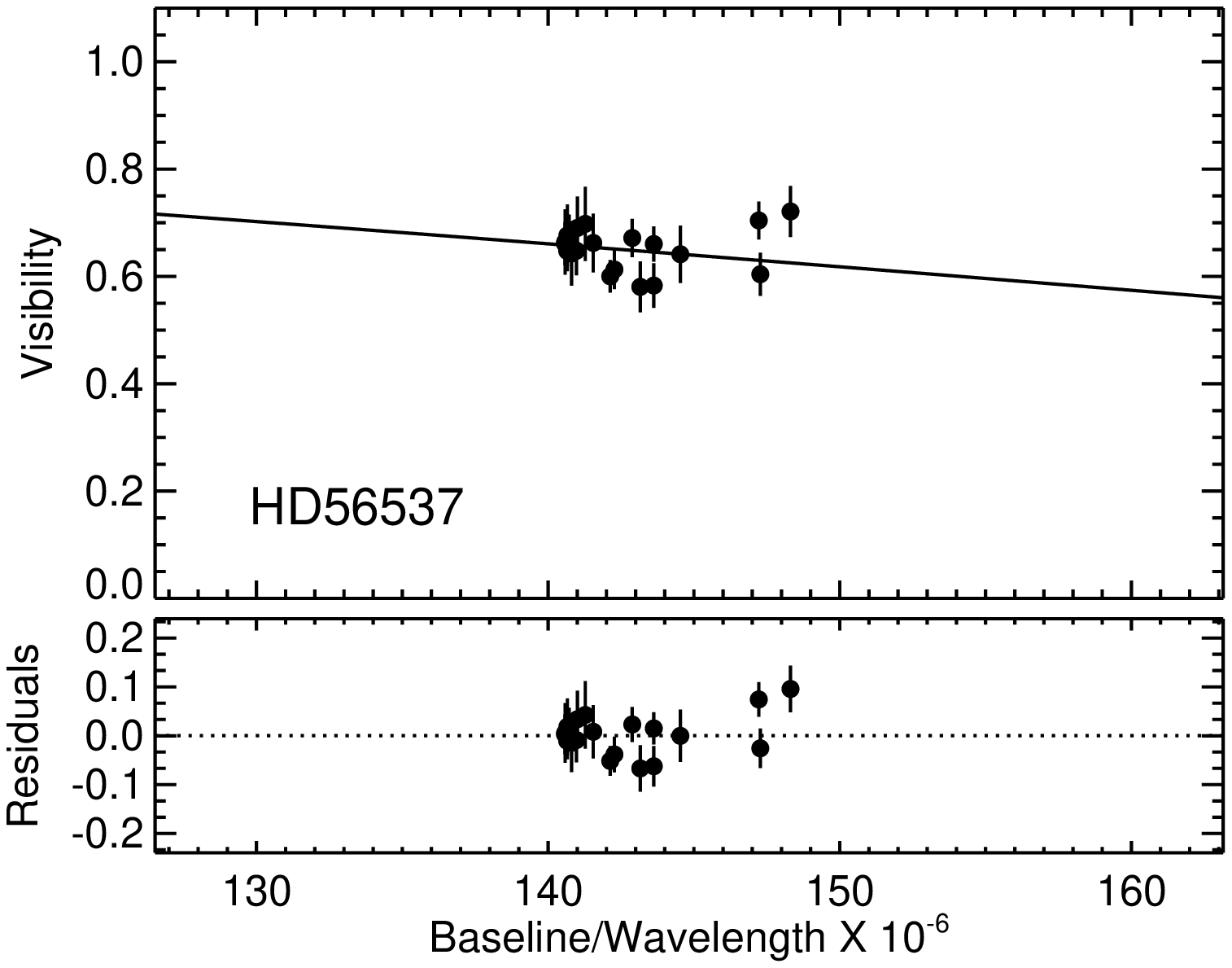,width=0.5\linewidth,clip=} &  
\epsfig{file=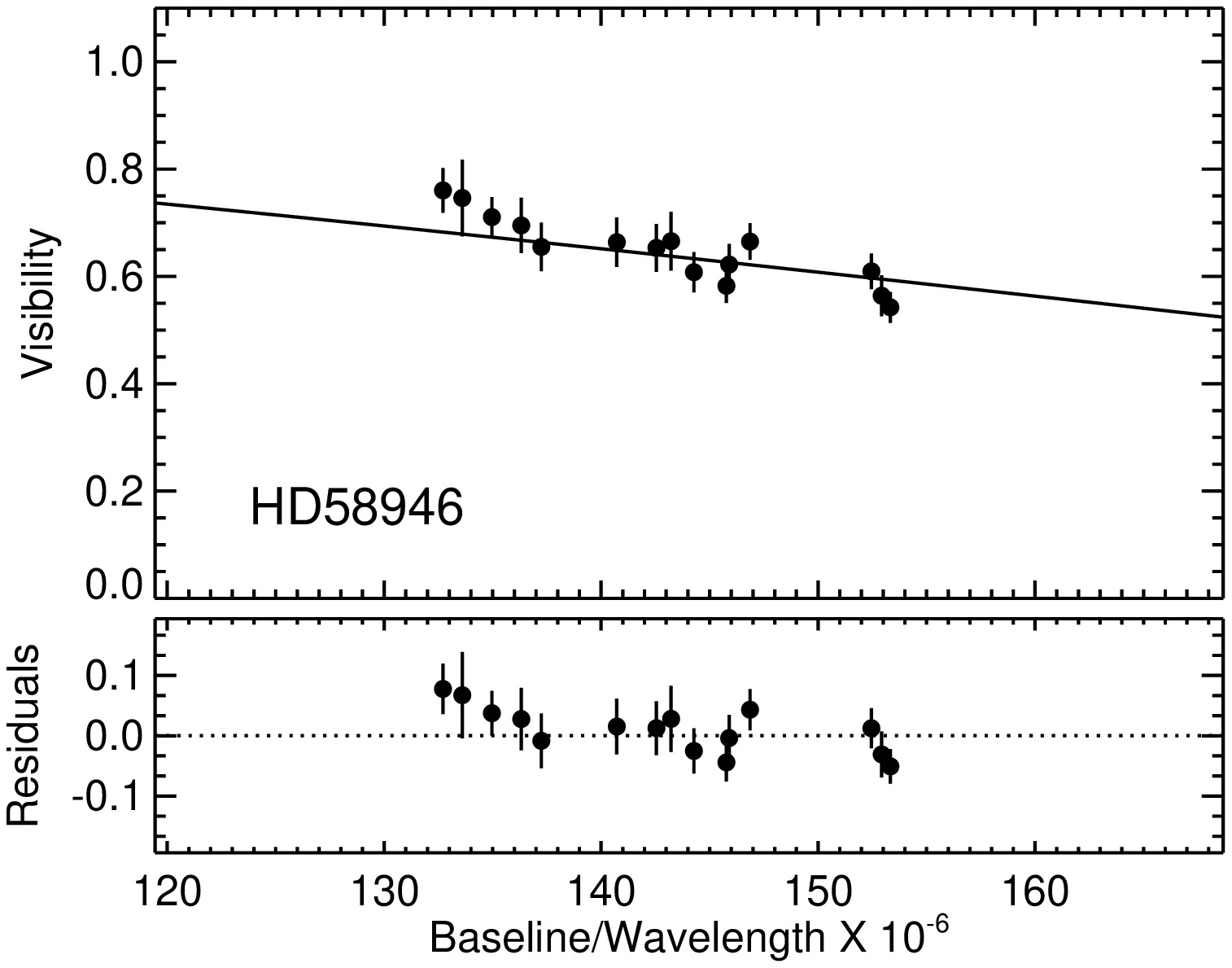,width=0.5\linewidth,clip=}
 \end{tabular}
\caption[Angular Diameters] {Calibrated observations plotted with the limb-darkened angular diameter fit for each star observed.  See Section~\ref{sec:diameters} and Table~\ref{tab:diameters} for details.}
\label{fig:diameters_2}
\end{figure}
\begin{figure}
\centering
\begin{tabular}{cc}
   
\epsfig{file=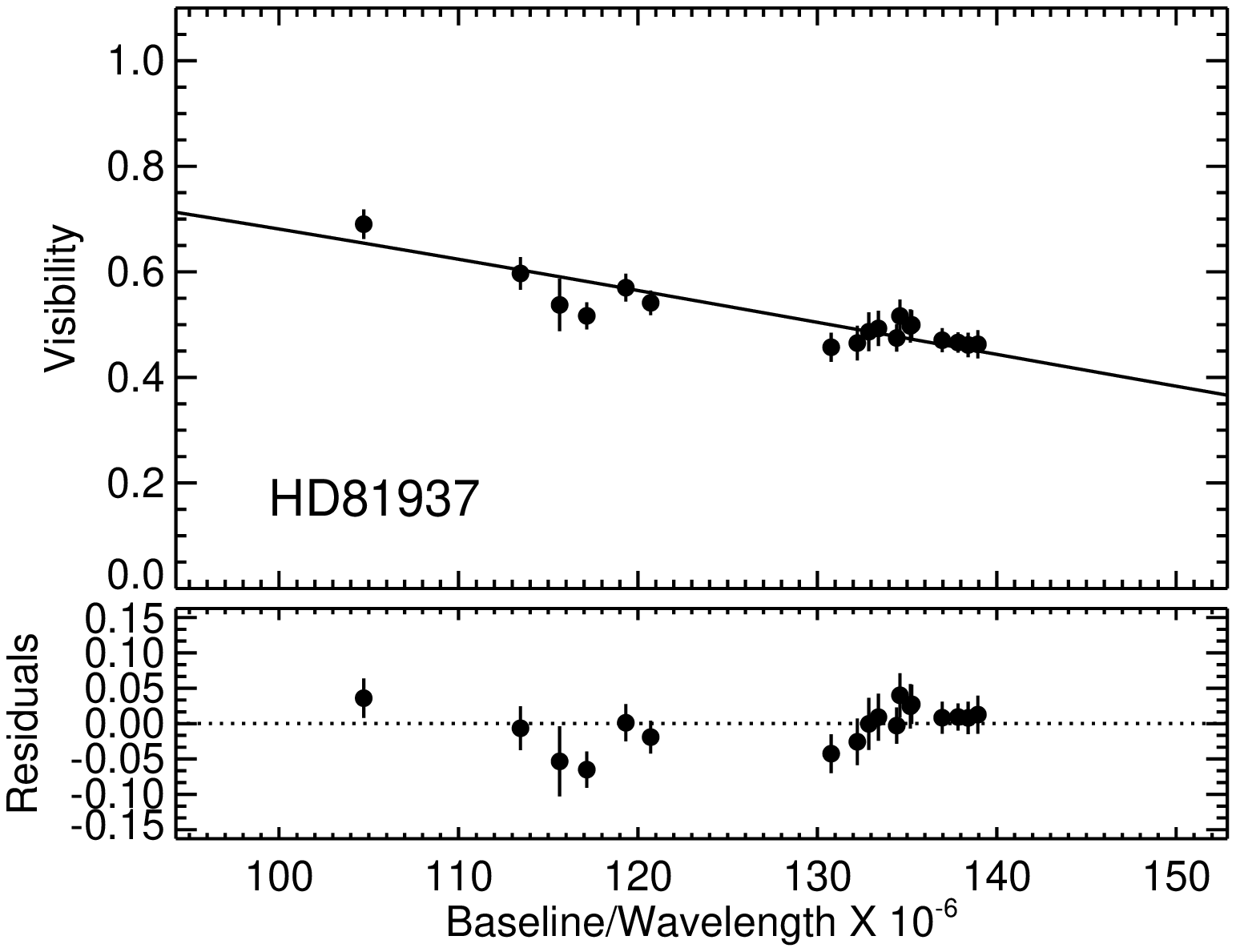,width=0.5\linewidth,clip=} &  
\epsfig{file=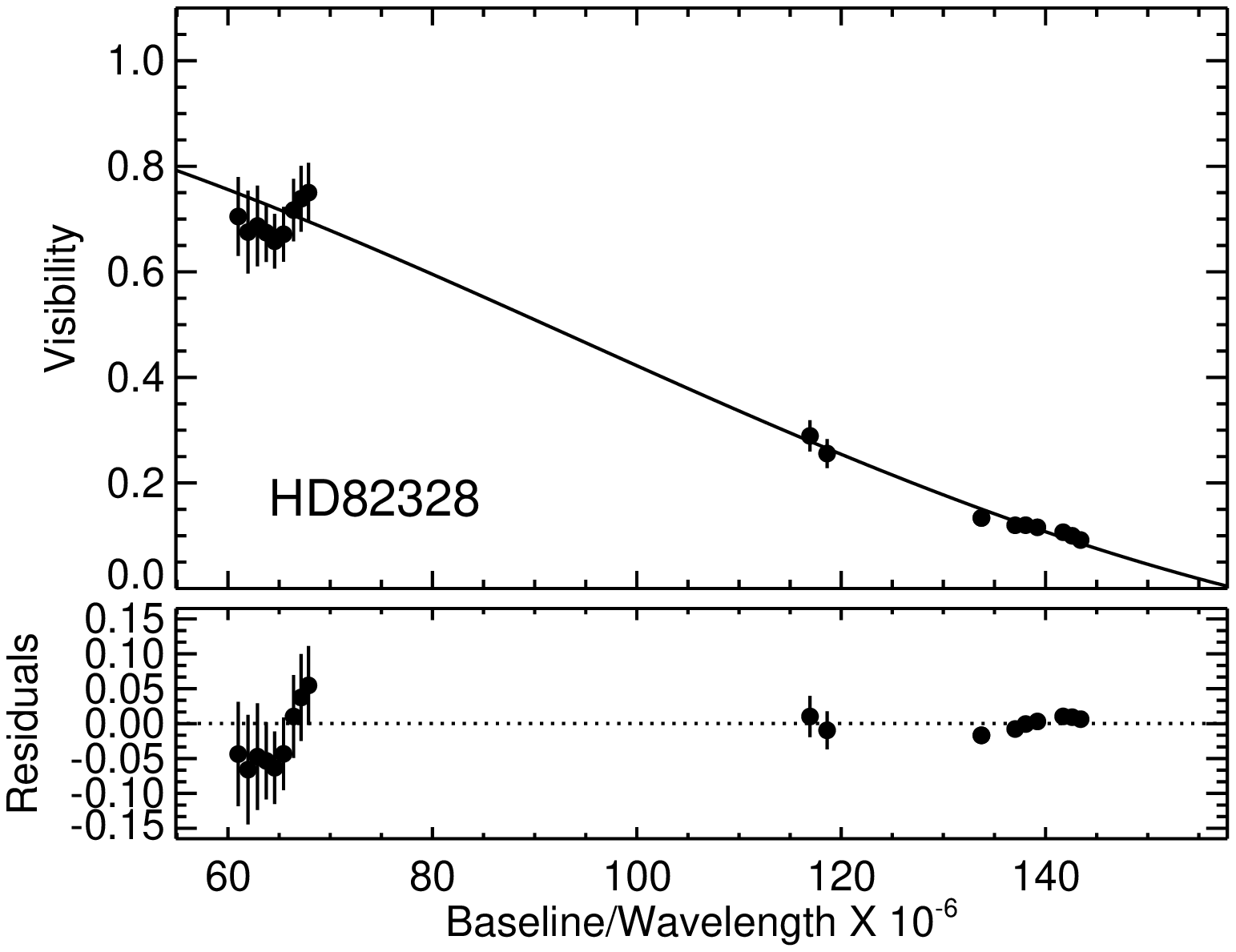,width=0.5\linewidth,clip=}   \\
\epsfig{file=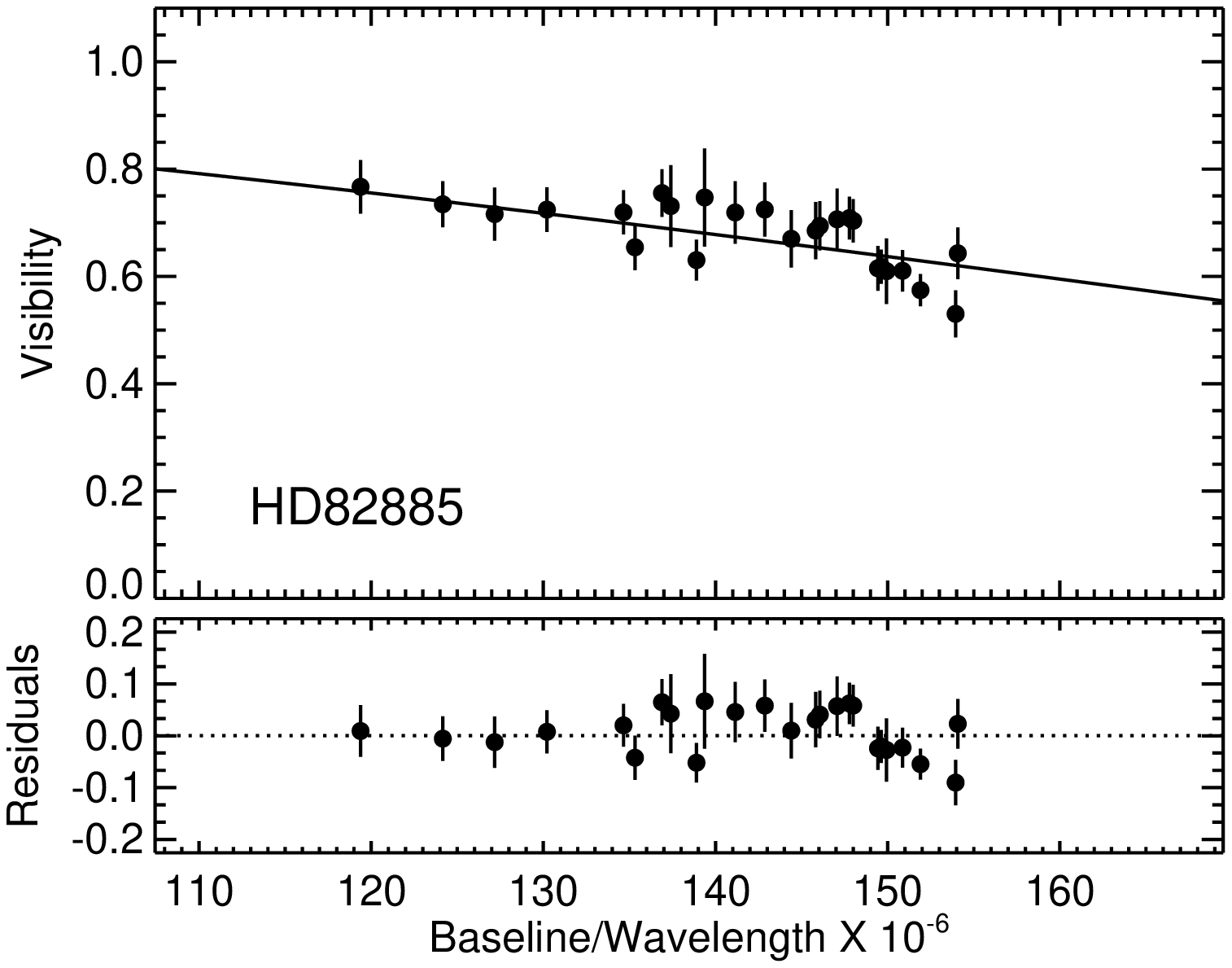,width=0.5\linewidth,clip=} &  
\epsfig{file=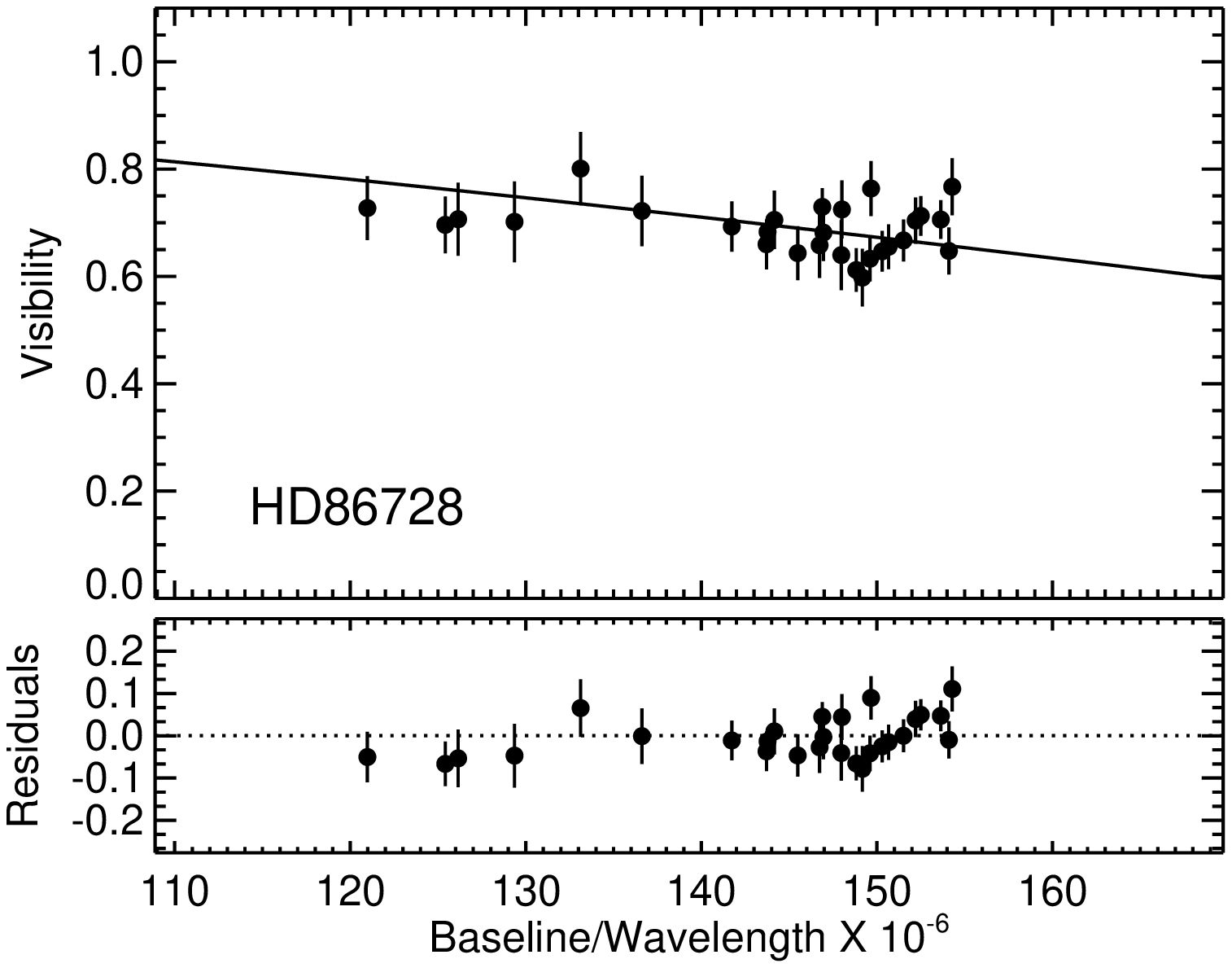,width=0.5\linewidth,clip=}   \\
\epsfig{file=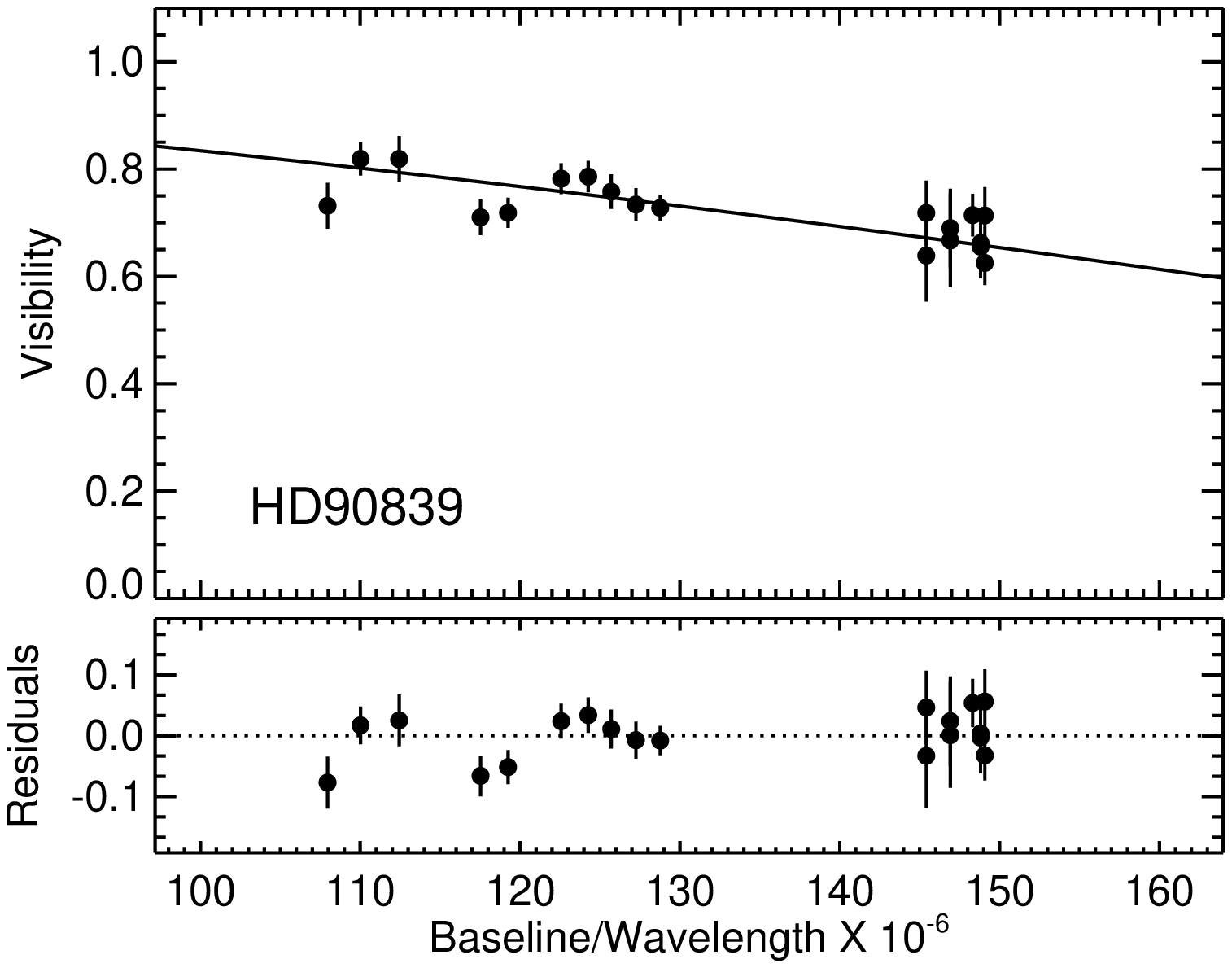,width=0.5\linewidth,clip=} &  
\epsfig{file=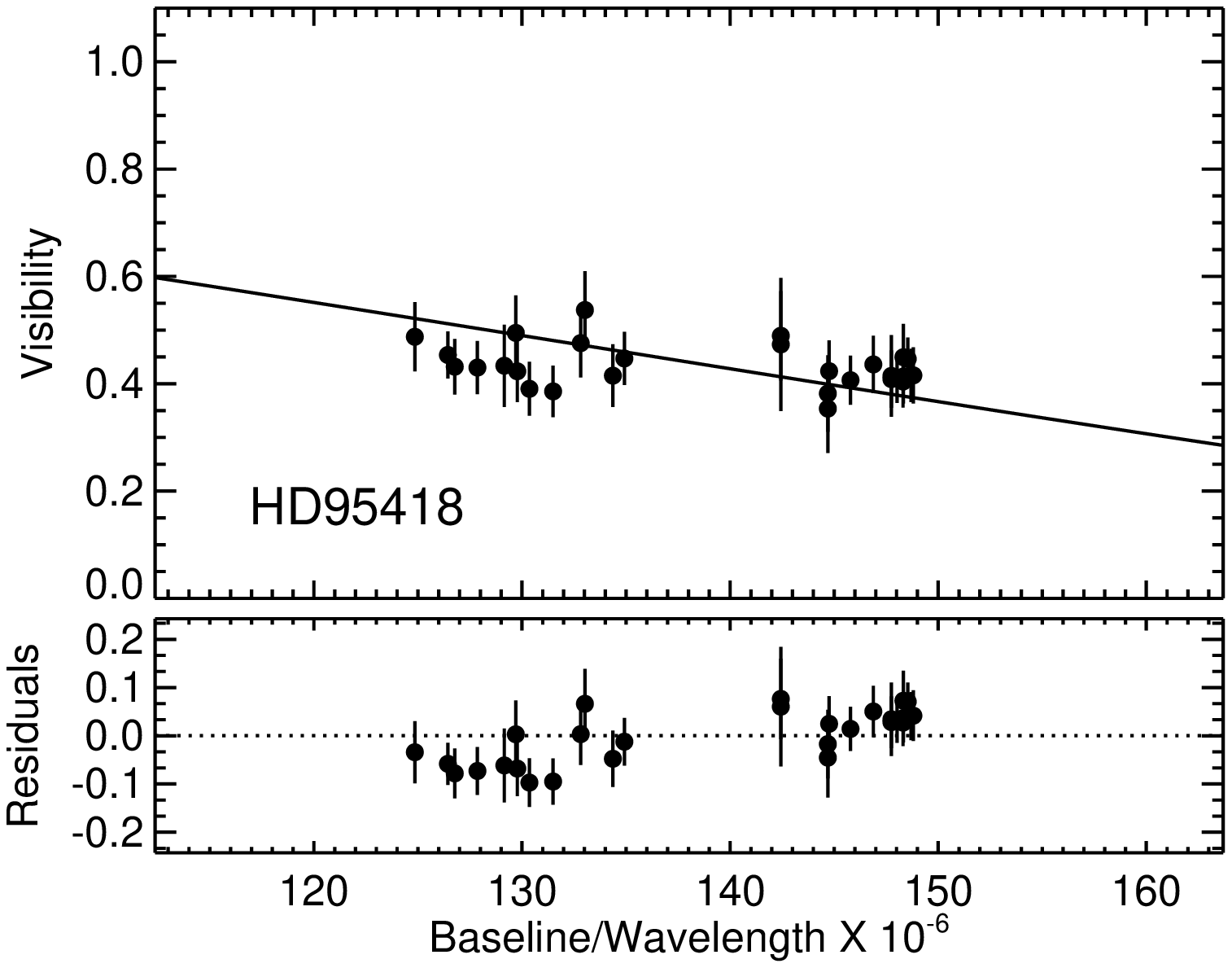,width=0.5\linewidth,clip=}
 \end{tabular}
\caption[Angular Diameters] {Calibrated observations plotted with the limb-darkened angular diameter fit for each star observed.  See Section~\ref{sec:diameters} and Table~\ref{tab:diameters} for details.}
\label{fig:diameters_3}
\end{figure}
\begin{figure}
\centering
\begin{tabular}{cc}
   
\epsfig{file=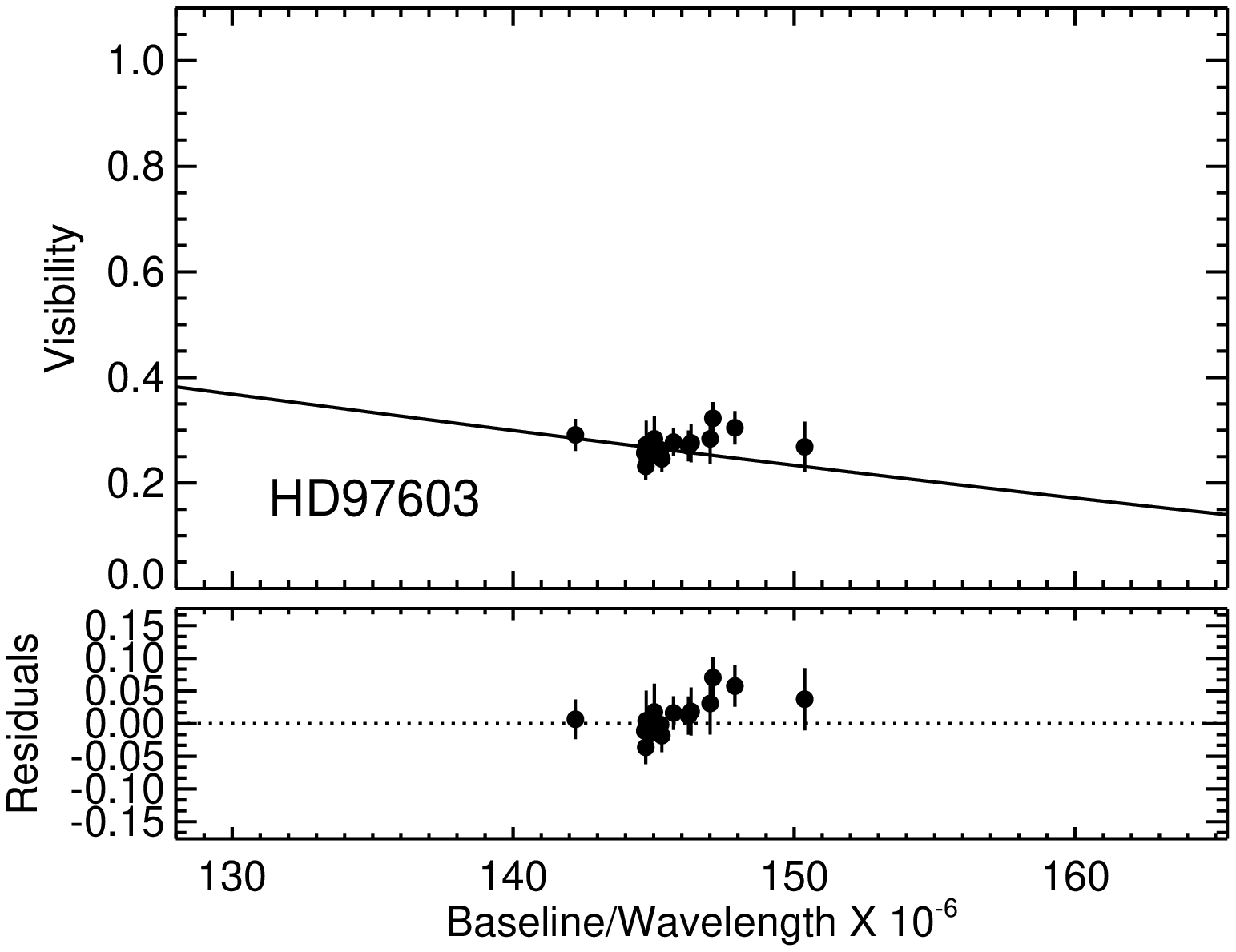,width=0.5\linewidth,clip=} &  
\epsfig{file=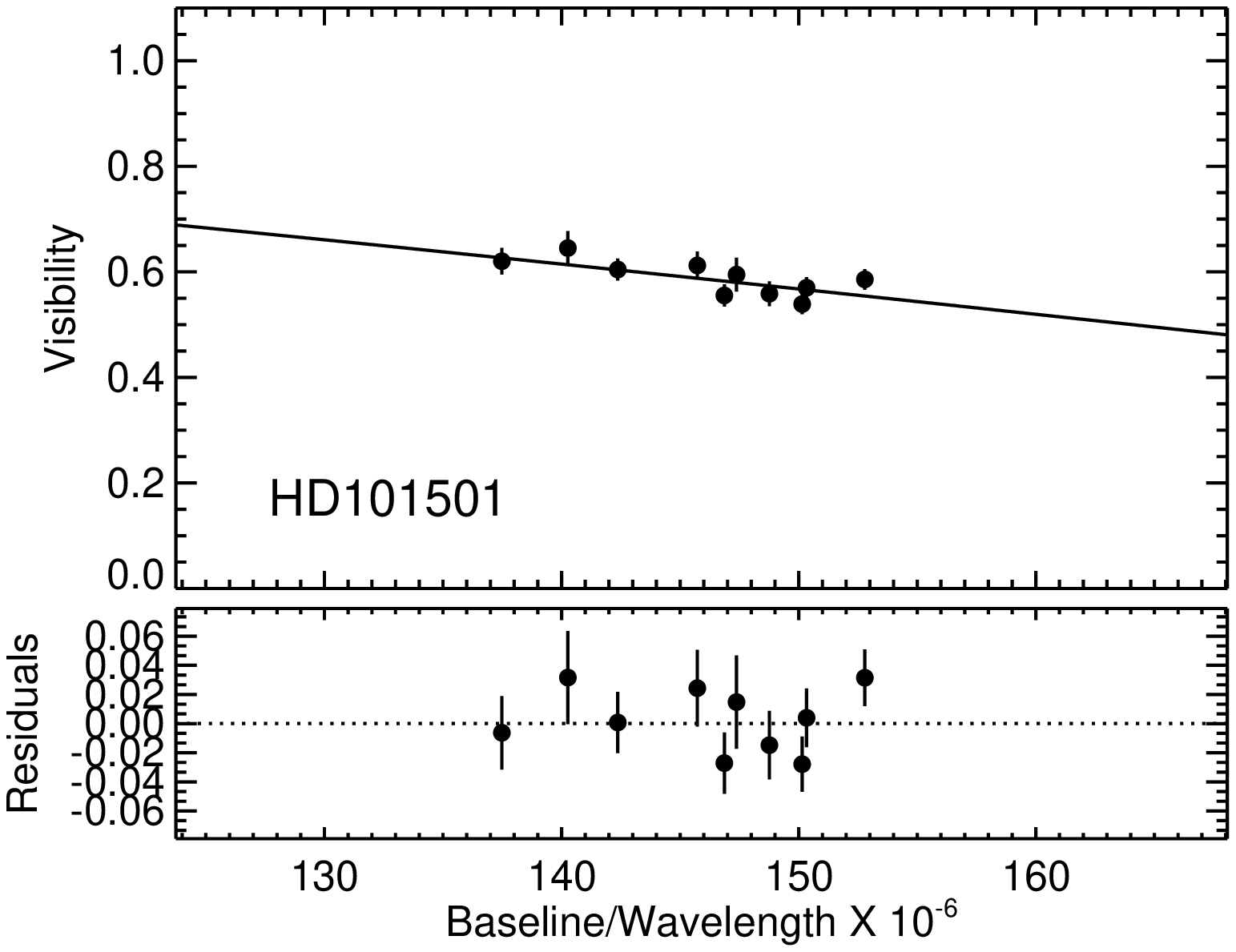,width=0.5\linewidth,clip=}   \\
\epsfig{file=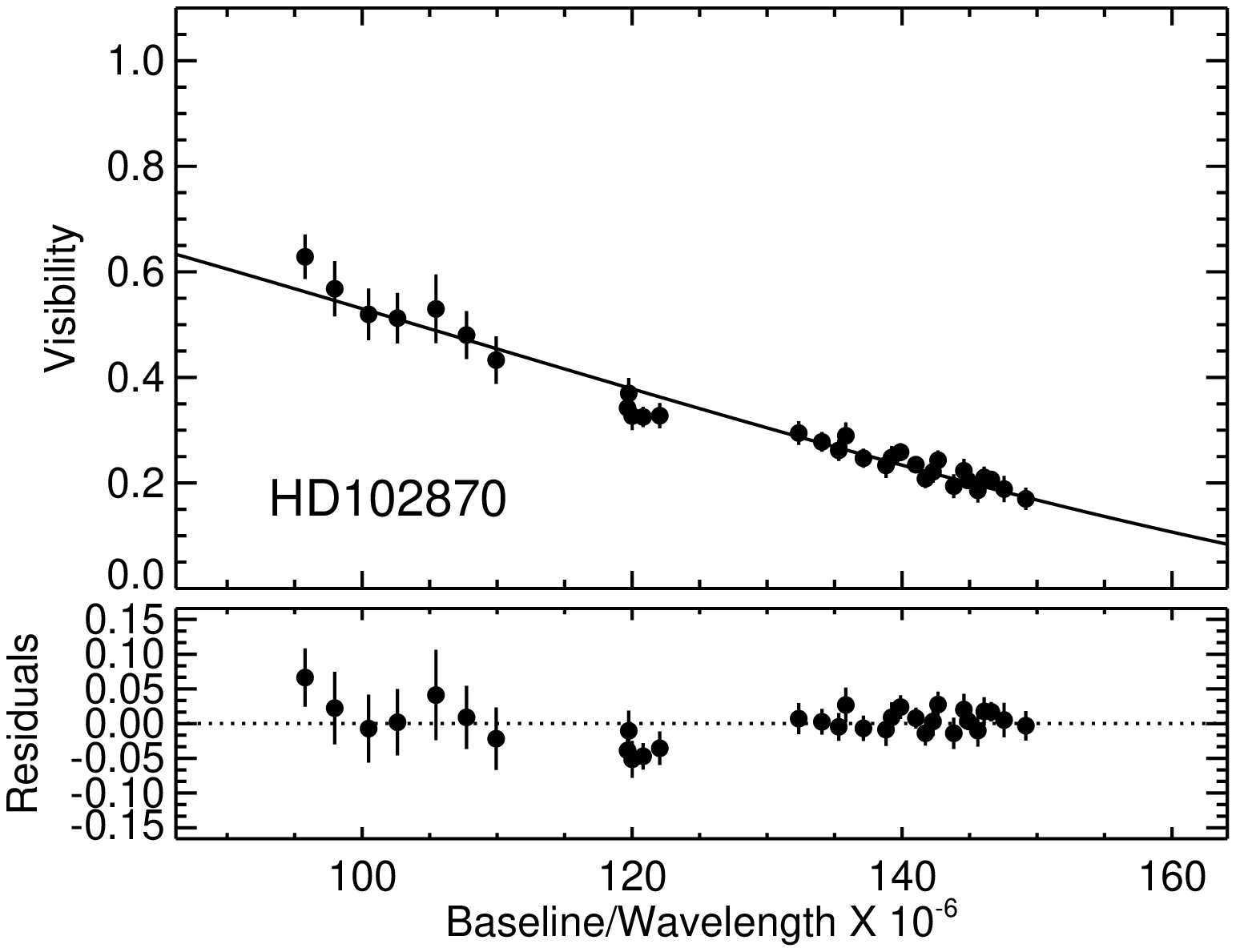,width=0.5\linewidth,clip=} &  
\epsfig{file=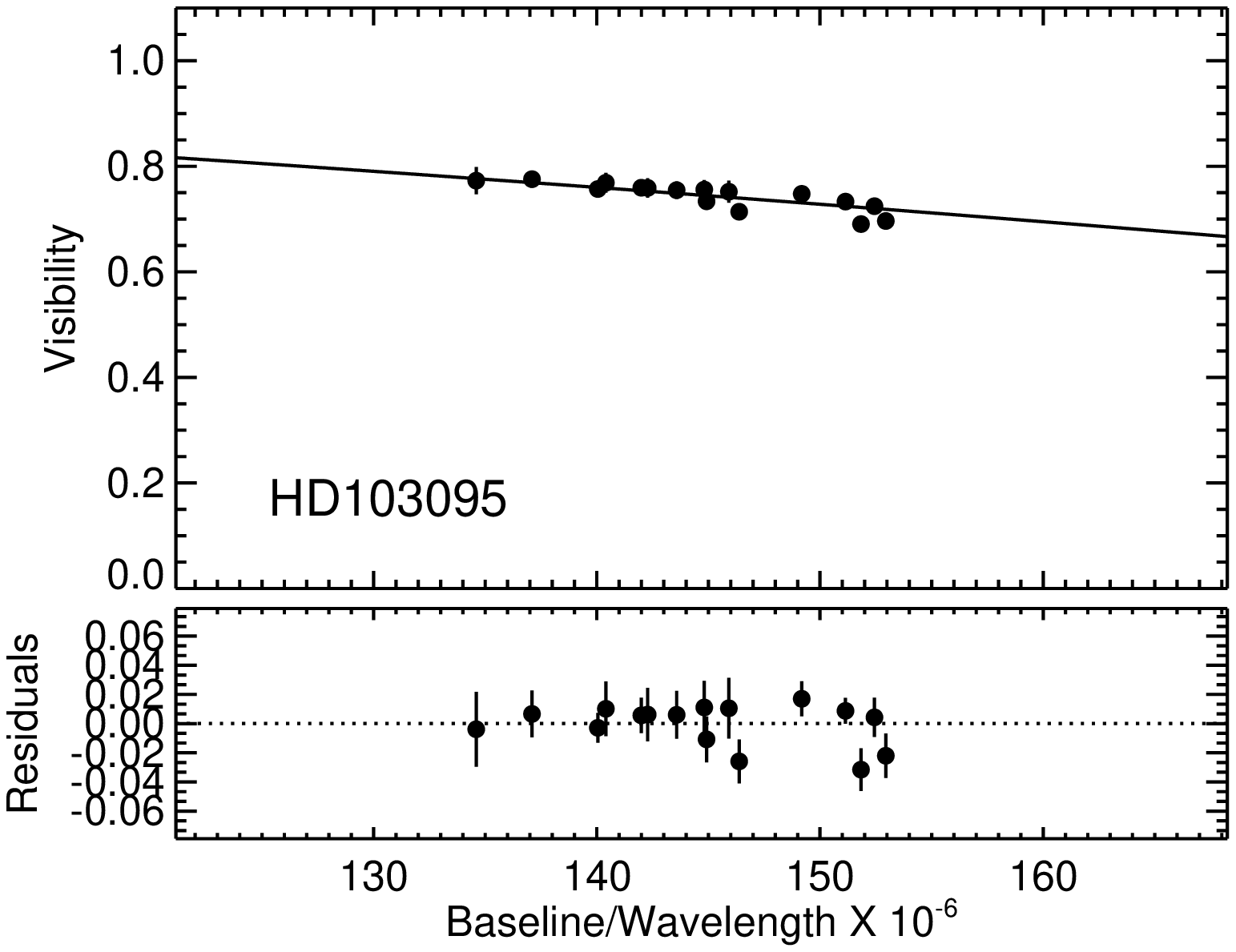,width=0.5\linewidth,clip=}   \\
\epsfig{file=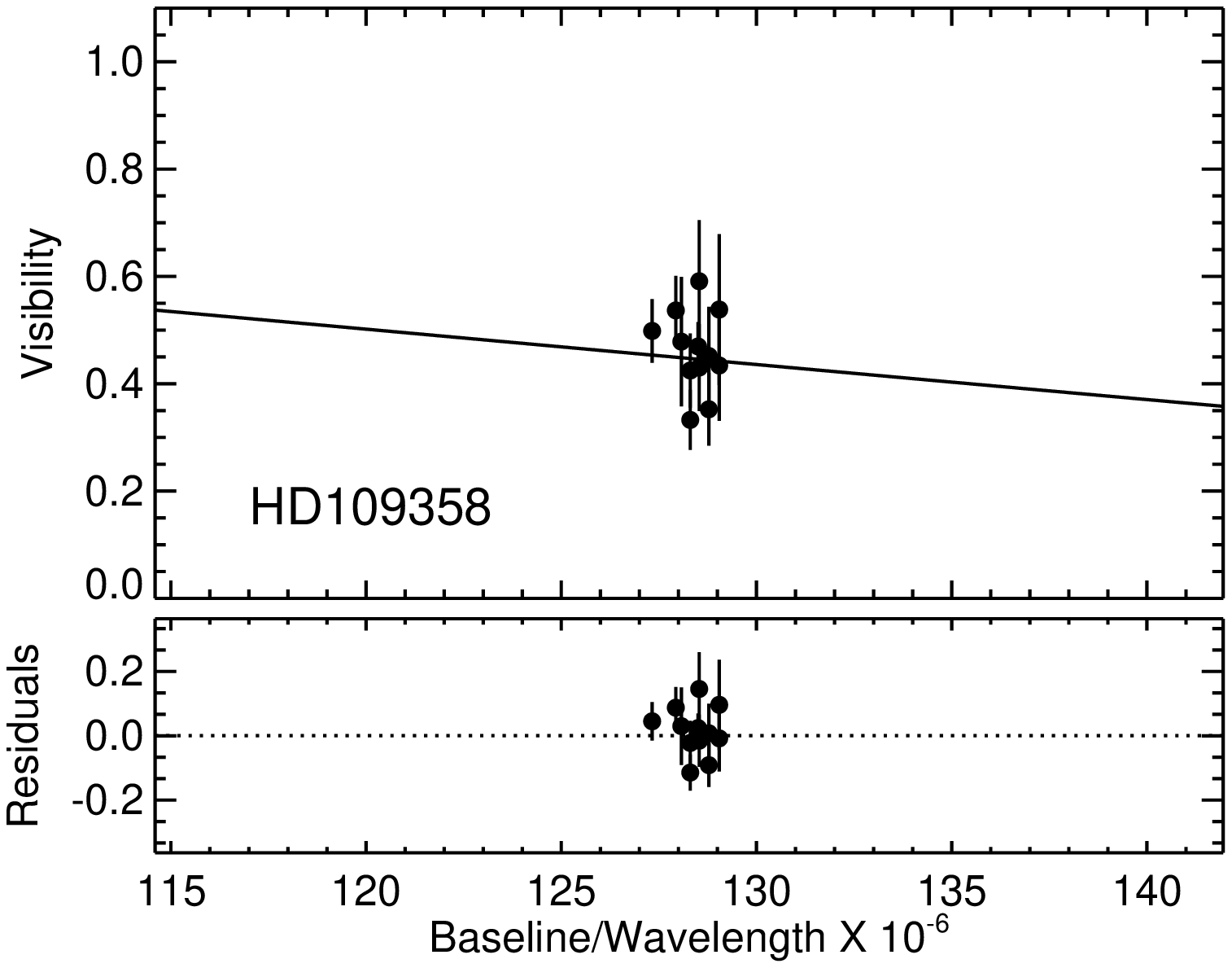,width=0.5\linewidth,clip=} &  
\epsfig{file=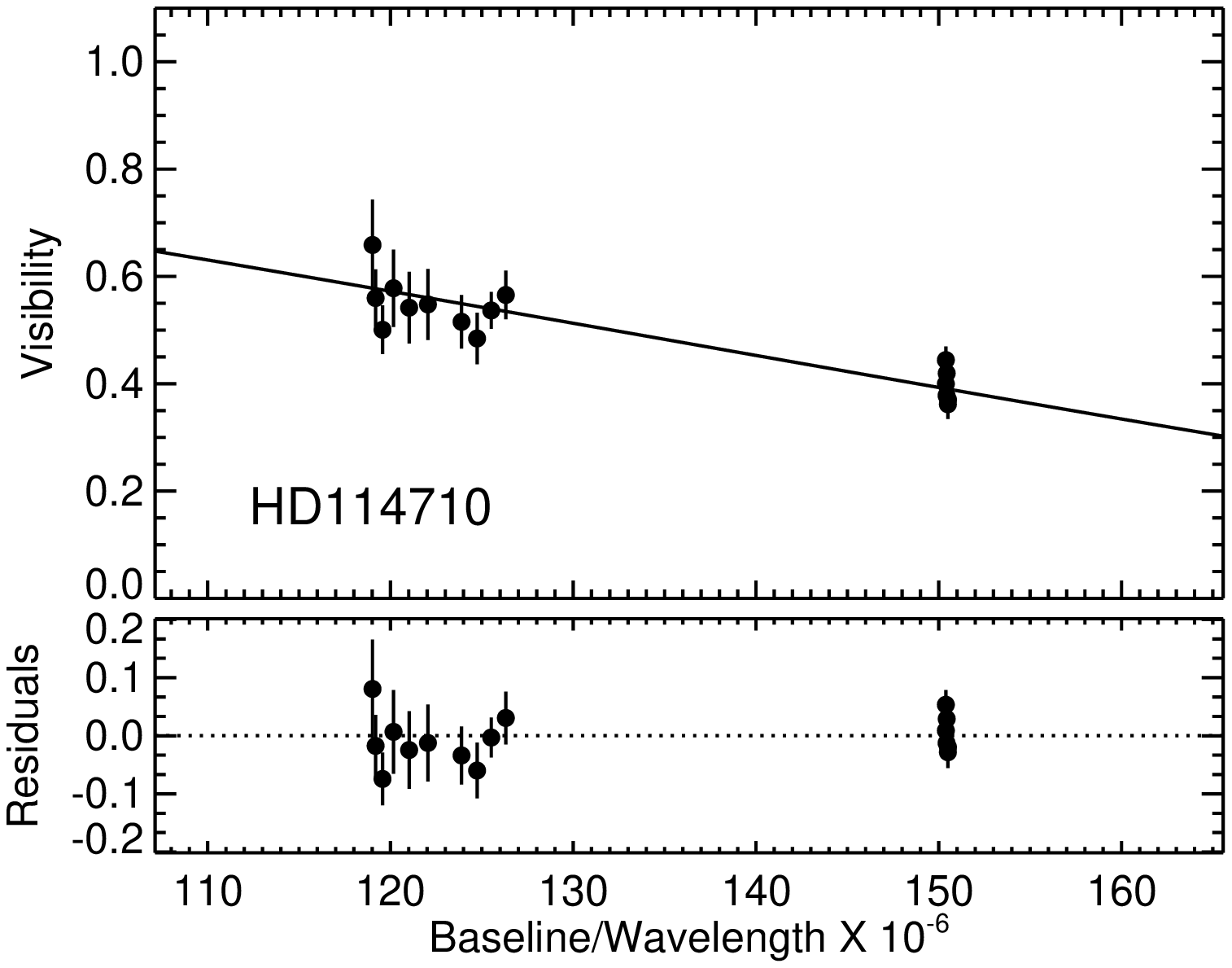,width=0.5\linewidth,clip=}
 \end{tabular}
\caption[Angular Diameters] {Calibrated observations plotted with the limb-darkened angular diameter fit for each star observed.  See Section~\ref{sec:diameters} and Table~\ref{tab:diameters} for details.}
\label{fig:diameters_4}
\end{figure}
\begin{figure}
\centering
\begin{tabular}{cc}
   
\epsfig{file=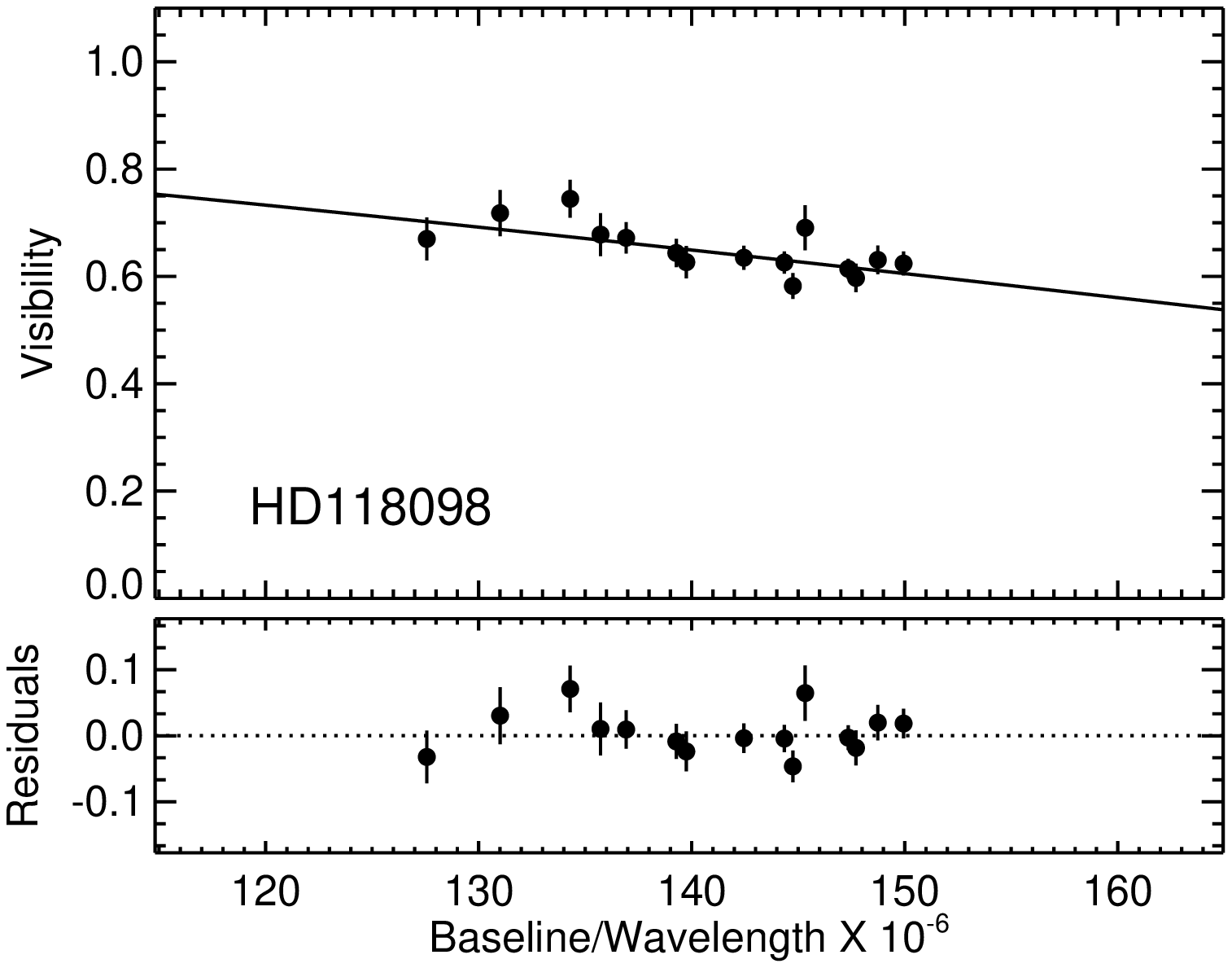,width=0.5\linewidth,clip=} &  
\epsfig{file=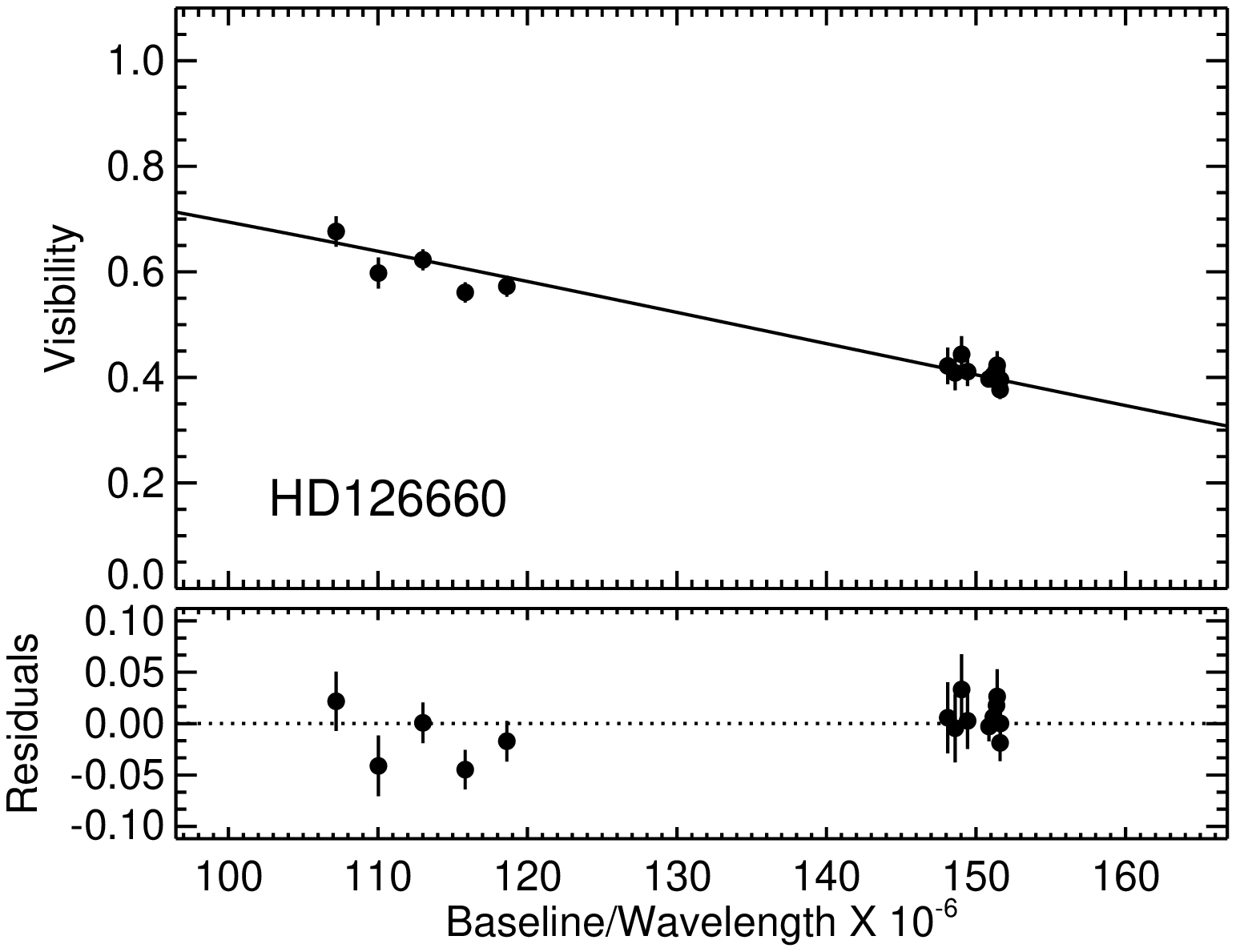,width=0.5\linewidth,clip=}   \\
\epsfig{file=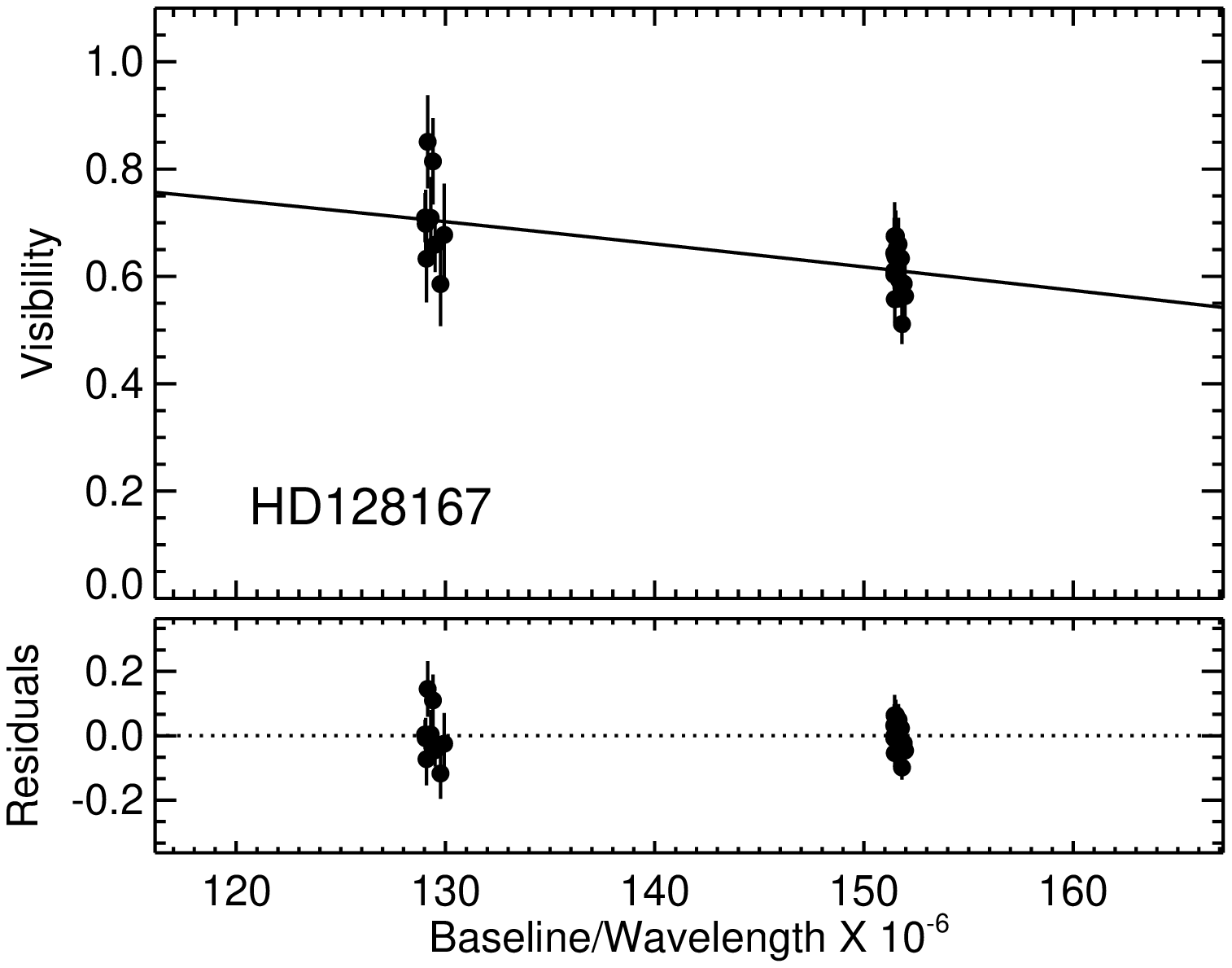,width=0.5\linewidth,clip=} &  
\epsfig{file=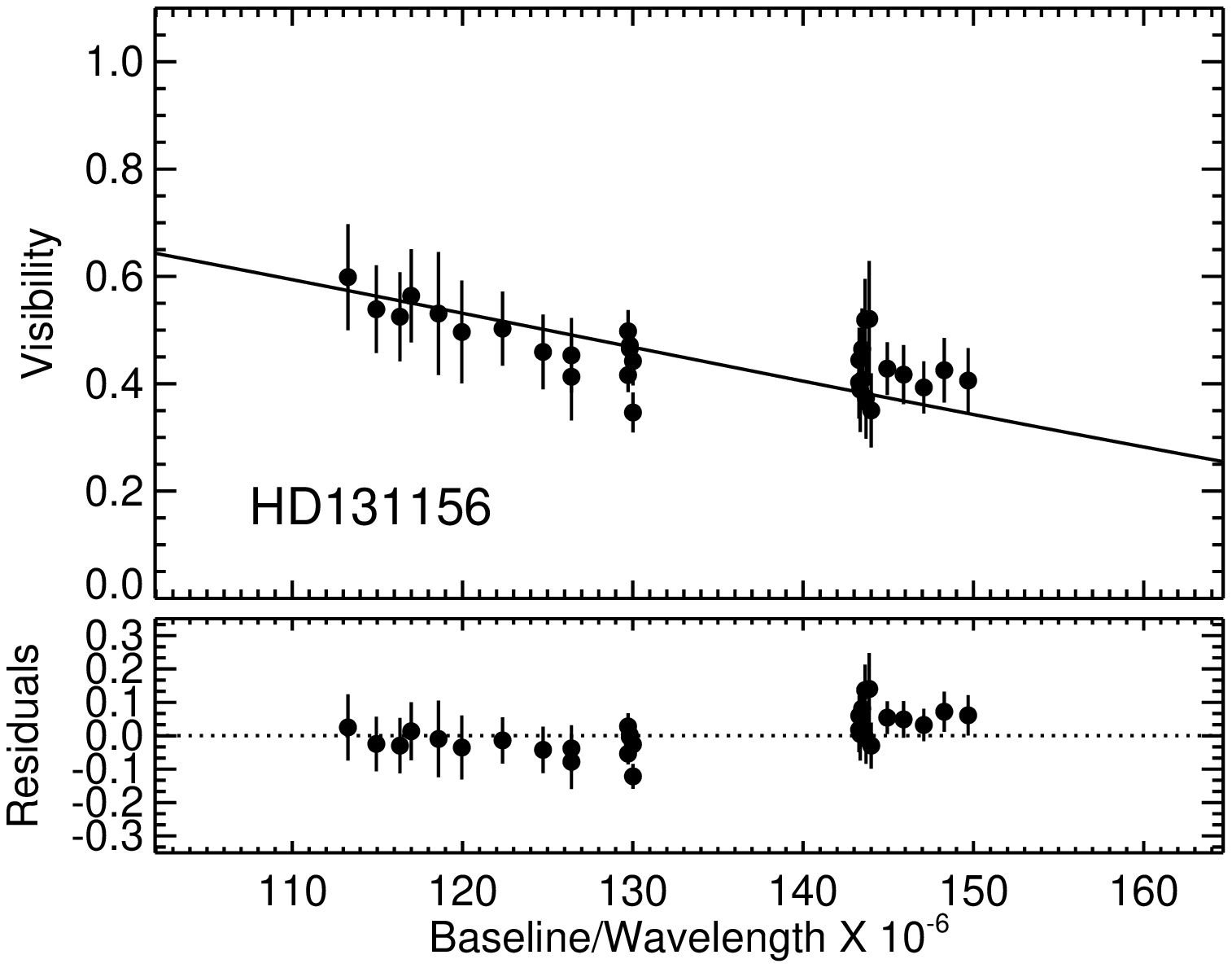,width=0.5\linewidth,clip=}   \\
\epsfig{file=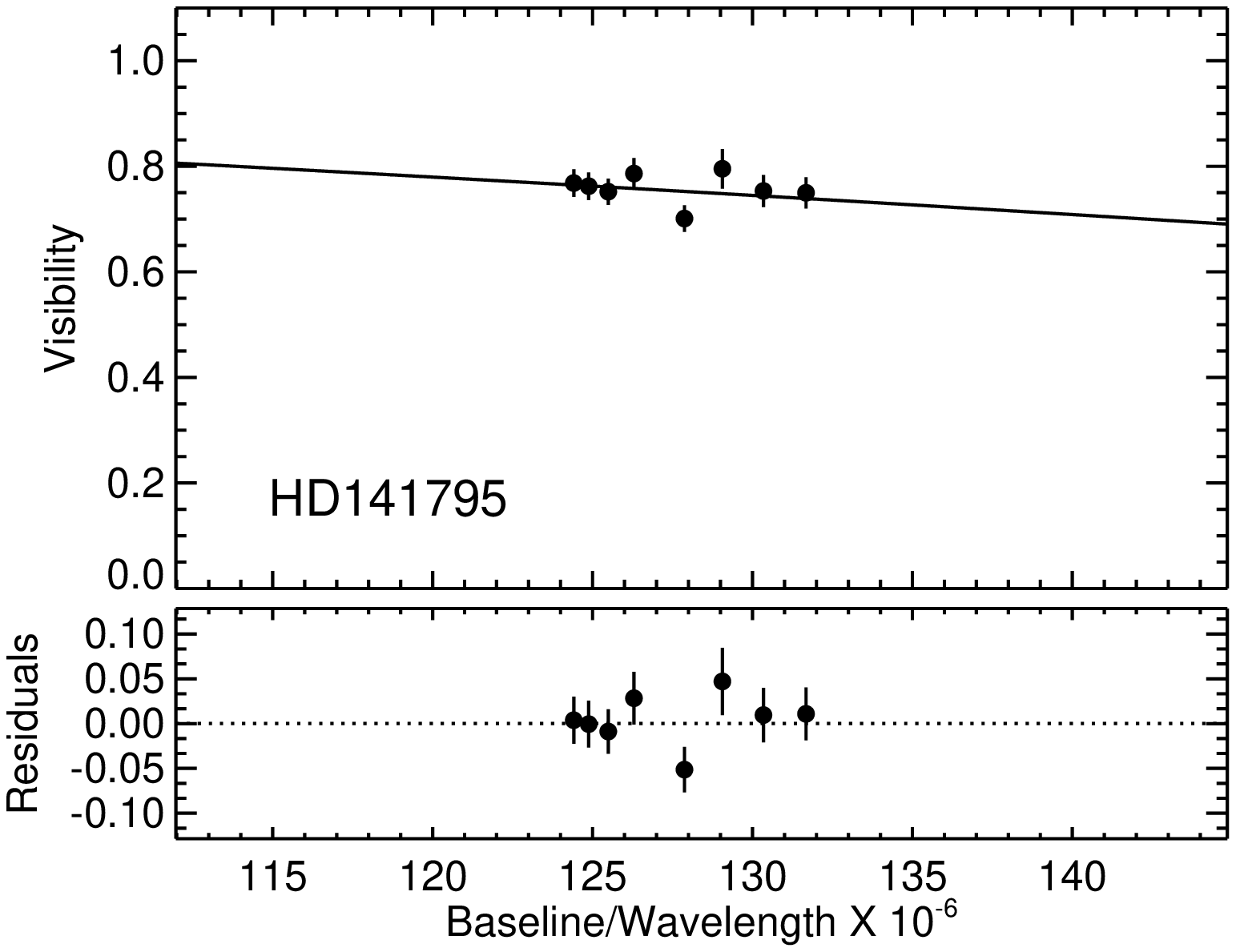,width=0.5\linewidth,clip=} &  
\epsfig{file=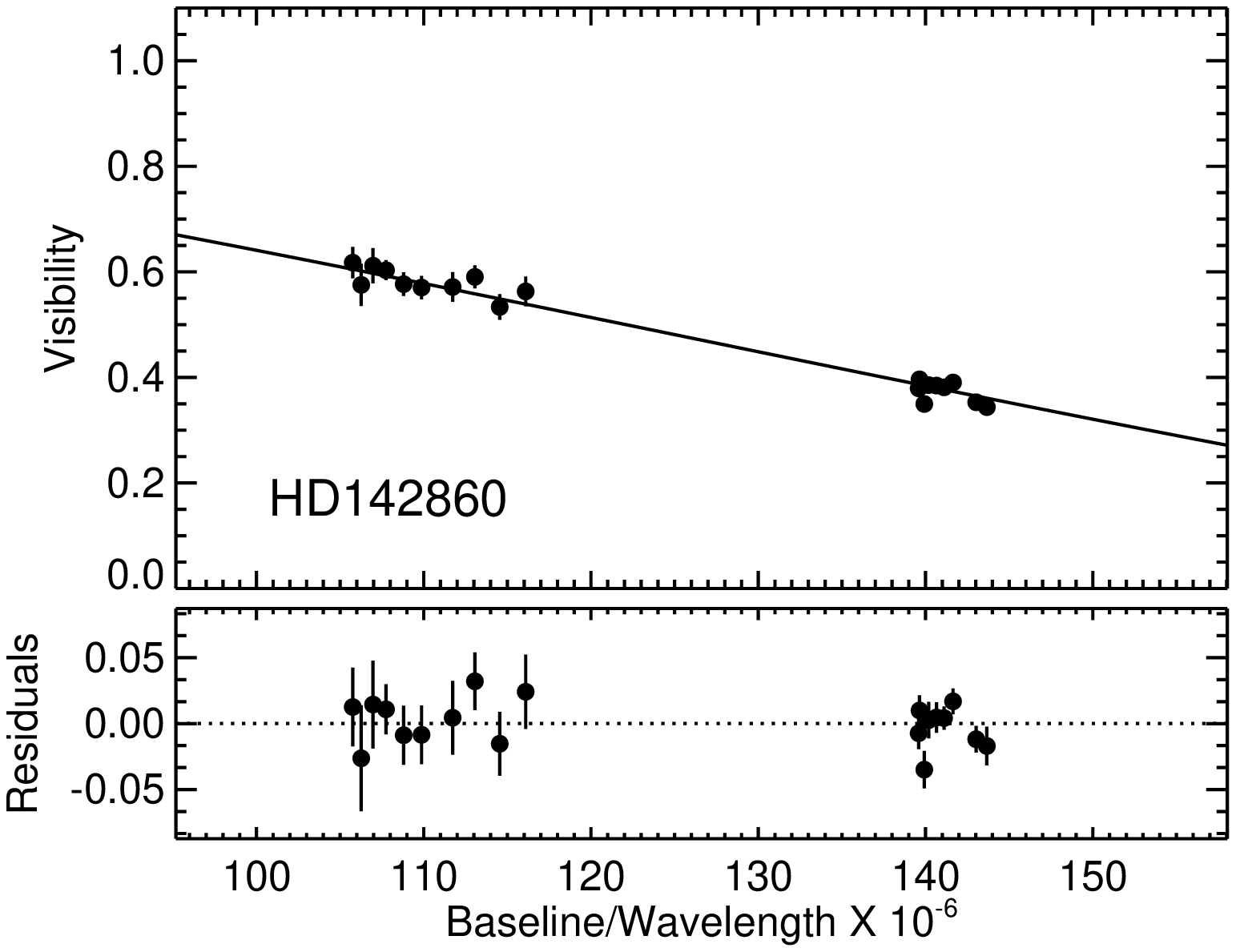,width=0.5\linewidth,clip=}
 \end{tabular}
\caption[Angular Diameters] {Calibrated observations plotted with the limb-darkened angular diameter fit for each star observed.  See Section~\ref{sec:diameters} and Table~\ref{tab:diameters} for details.}
\label{fig:diameters_5}
\end{figure}
\begin{figure}
\centering
\begin{tabular}{cc}
   
\epsfig{file=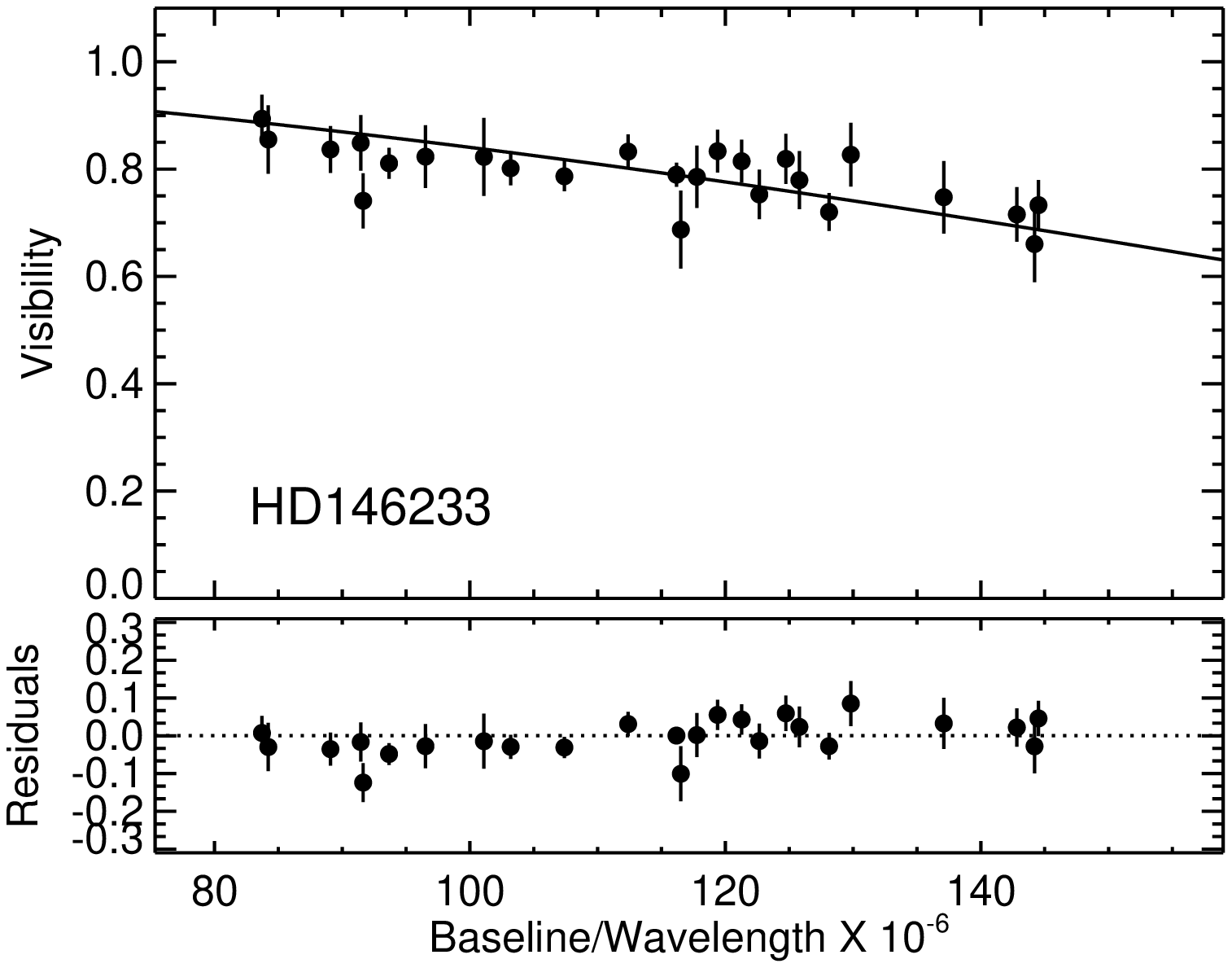,width=0.5\linewidth,clip=} &  
\epsfig{file=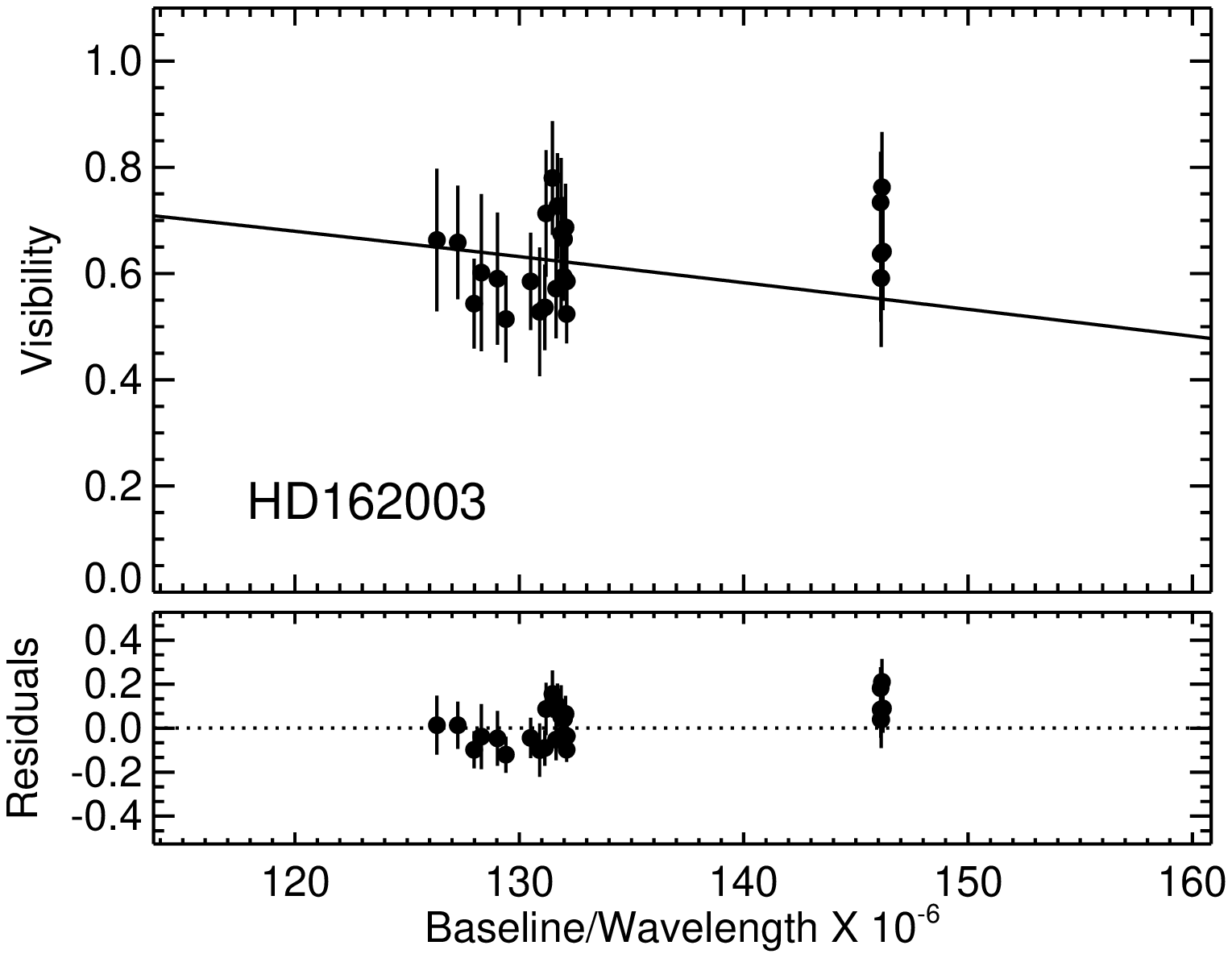,width=0.5\linewidth,clip=}   \\
\epsfig{file=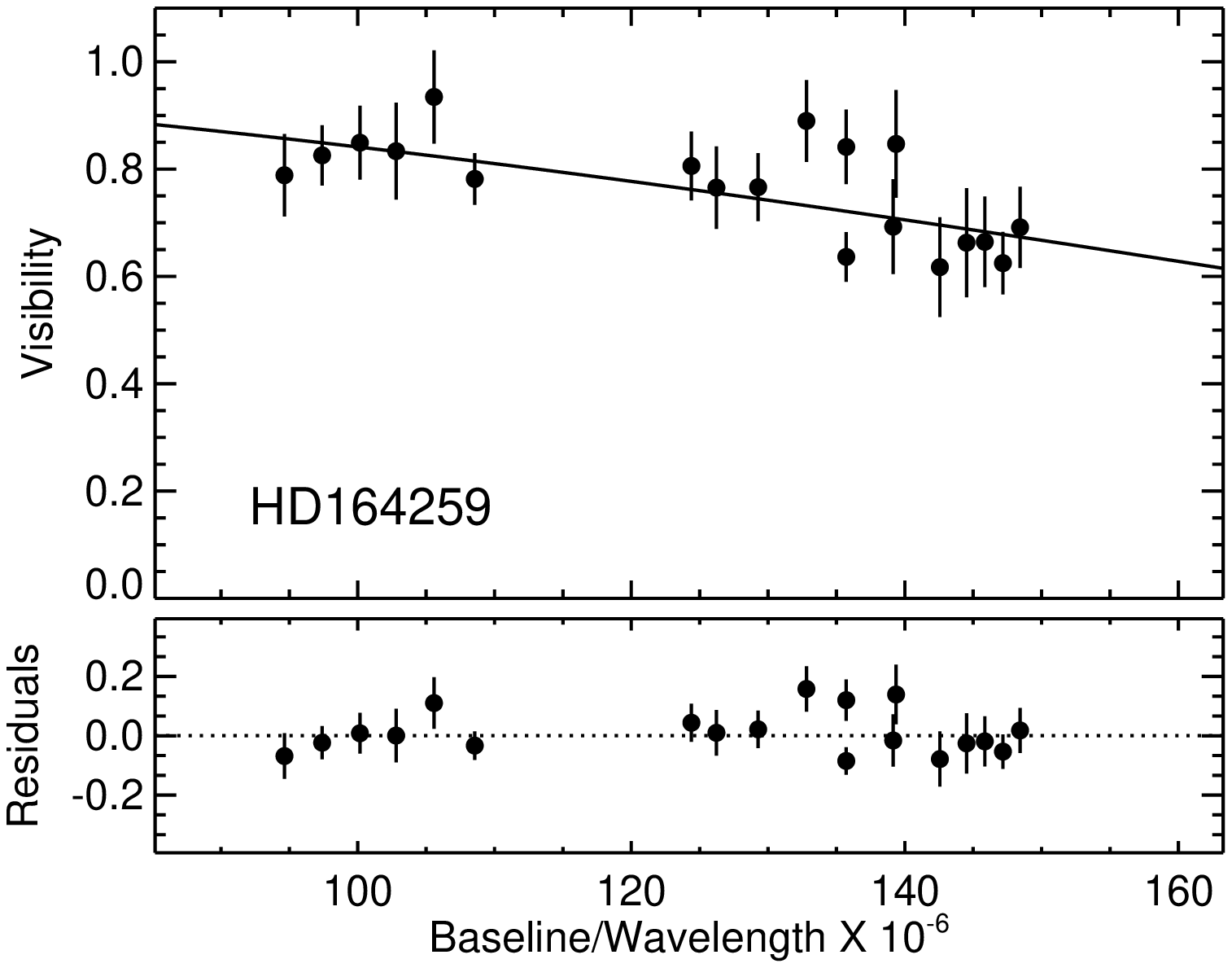,width=0.5\linewidth,clip=} &  
\epsfig{file=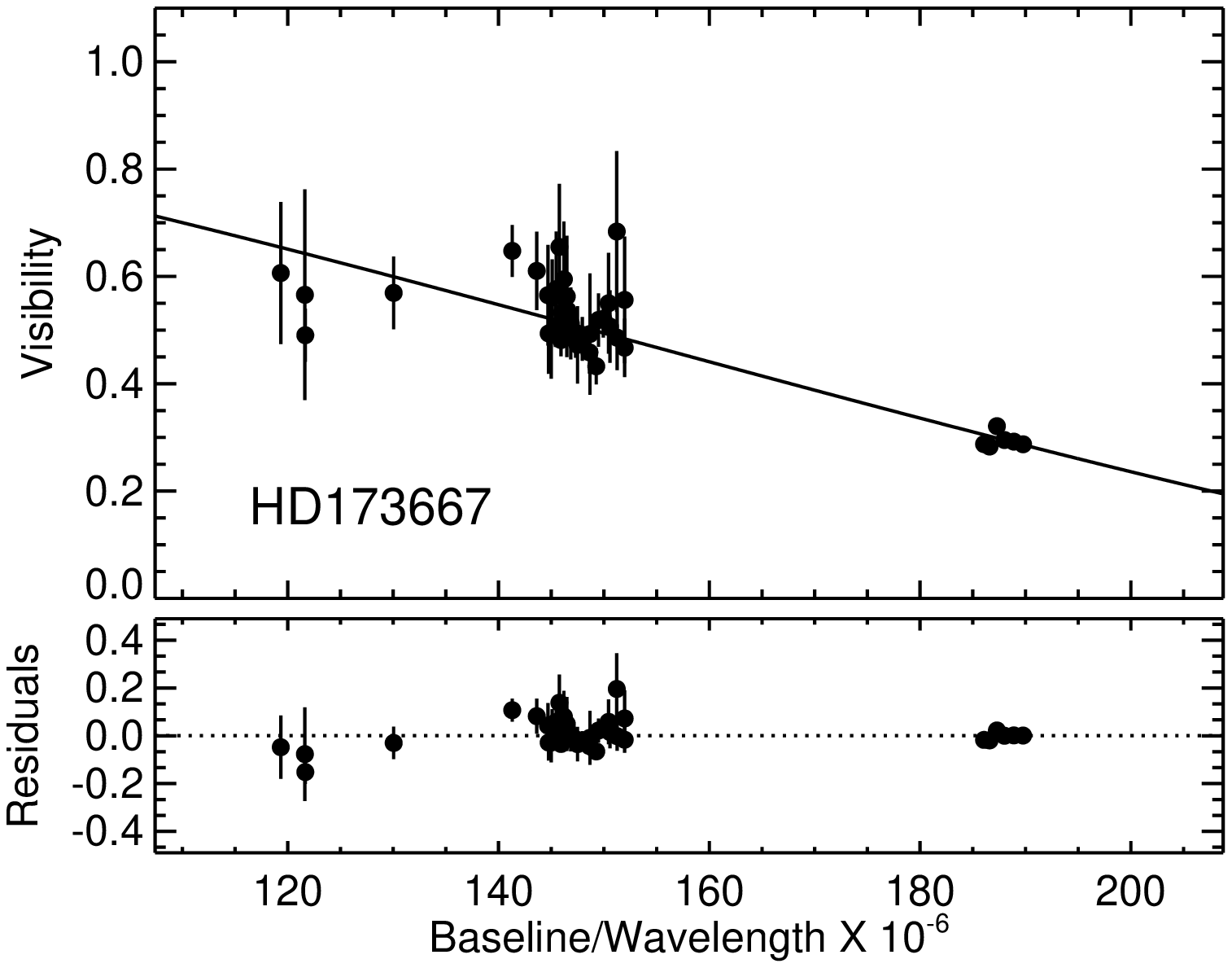,width=0.5\linewidth,clip=}   \\
\epsfig{file=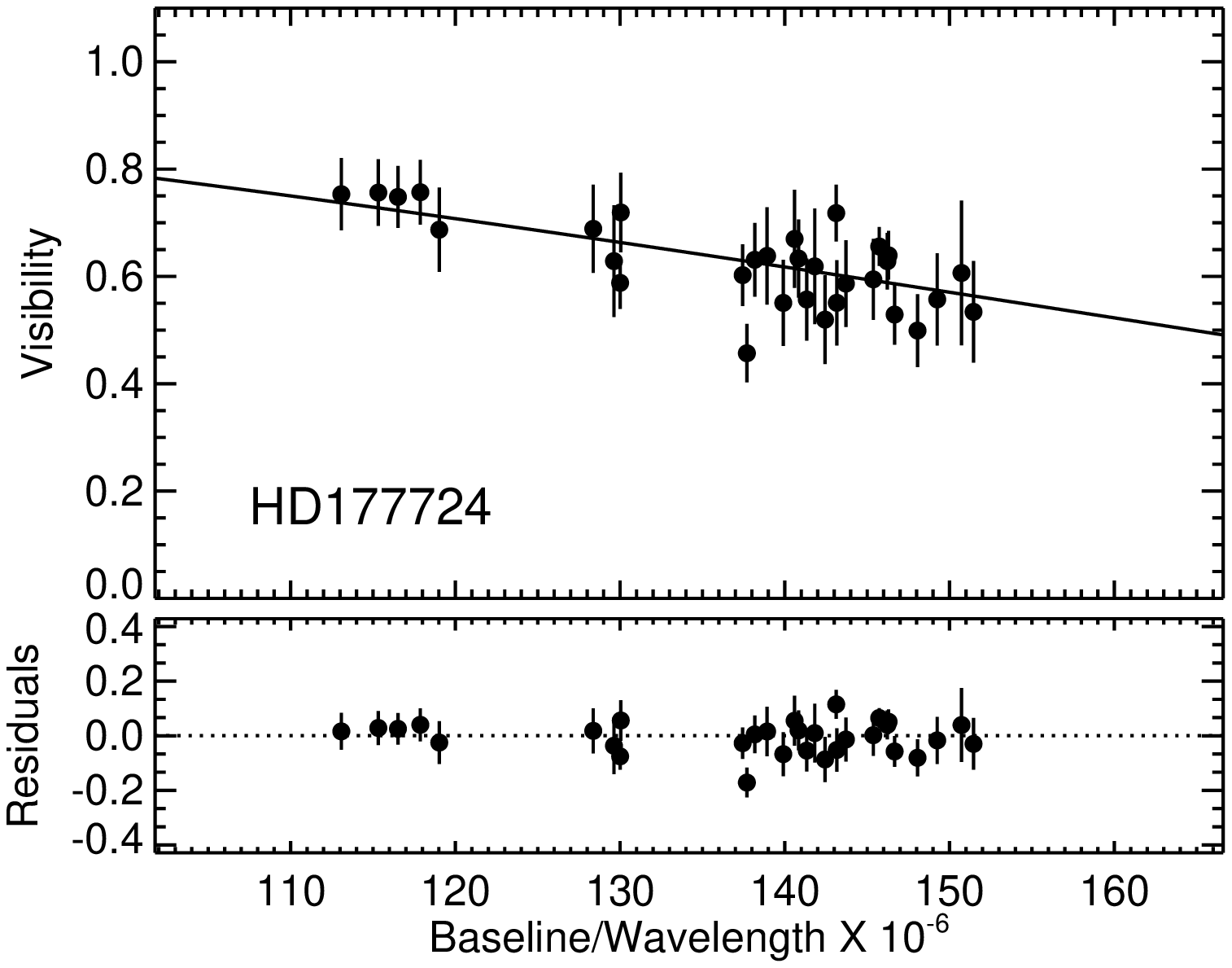,width=0.5\linewidth,clip=} &  
\epsfig{file=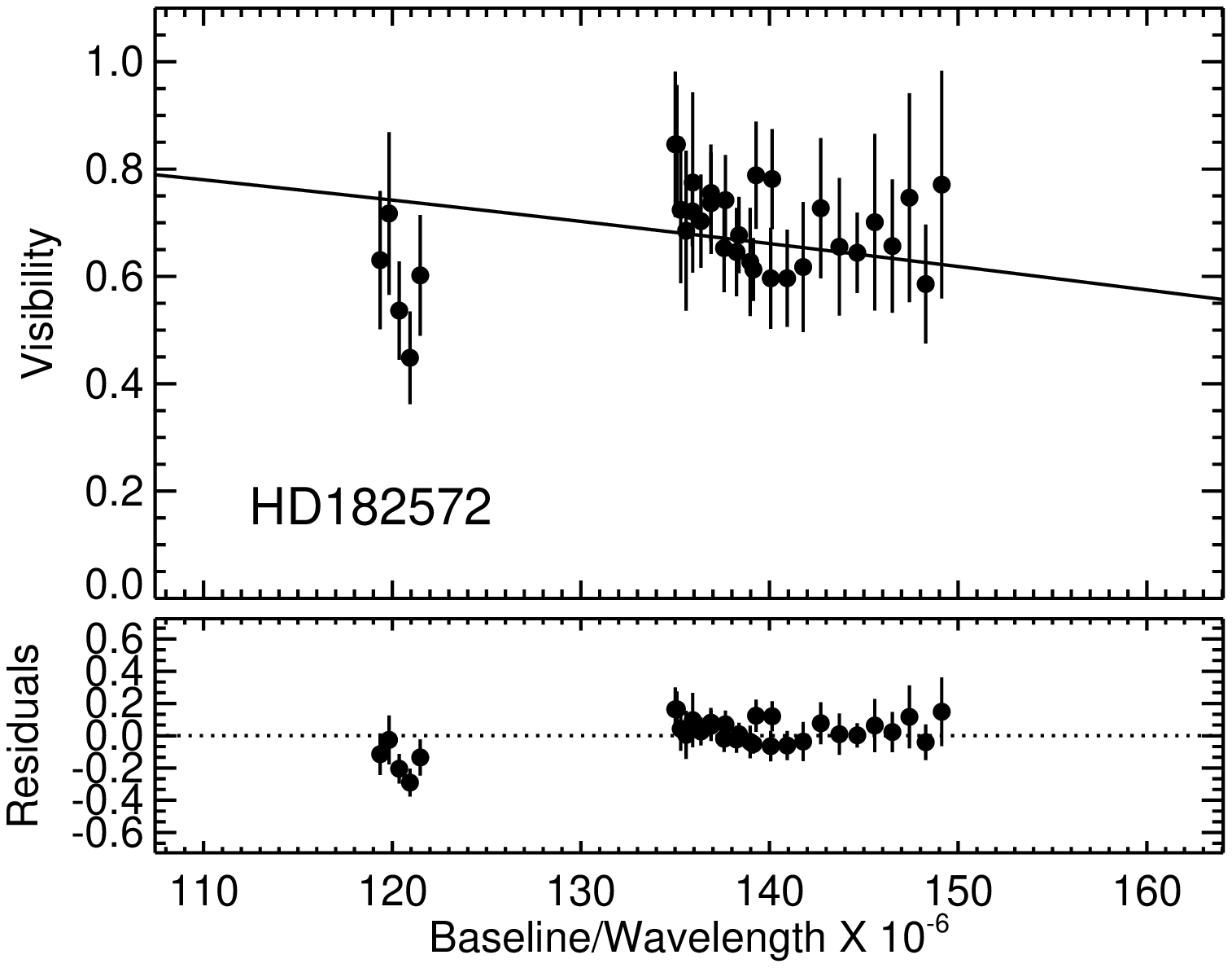,width=0.5\linewidth,clip=}
 \end{tabular}
\caption[Angular Diameters] {Calibrated observations plotted with the limb-darkened angular diameter fit for each star observed.  See Section~\ref{sec:diameters} and Table~\ref{tab:diameters} for details.}
\label{fig:diameters_6}
\end{figure}
\begin{figure}
\centering
\begin{tabular}{cc}

\epsfig{file=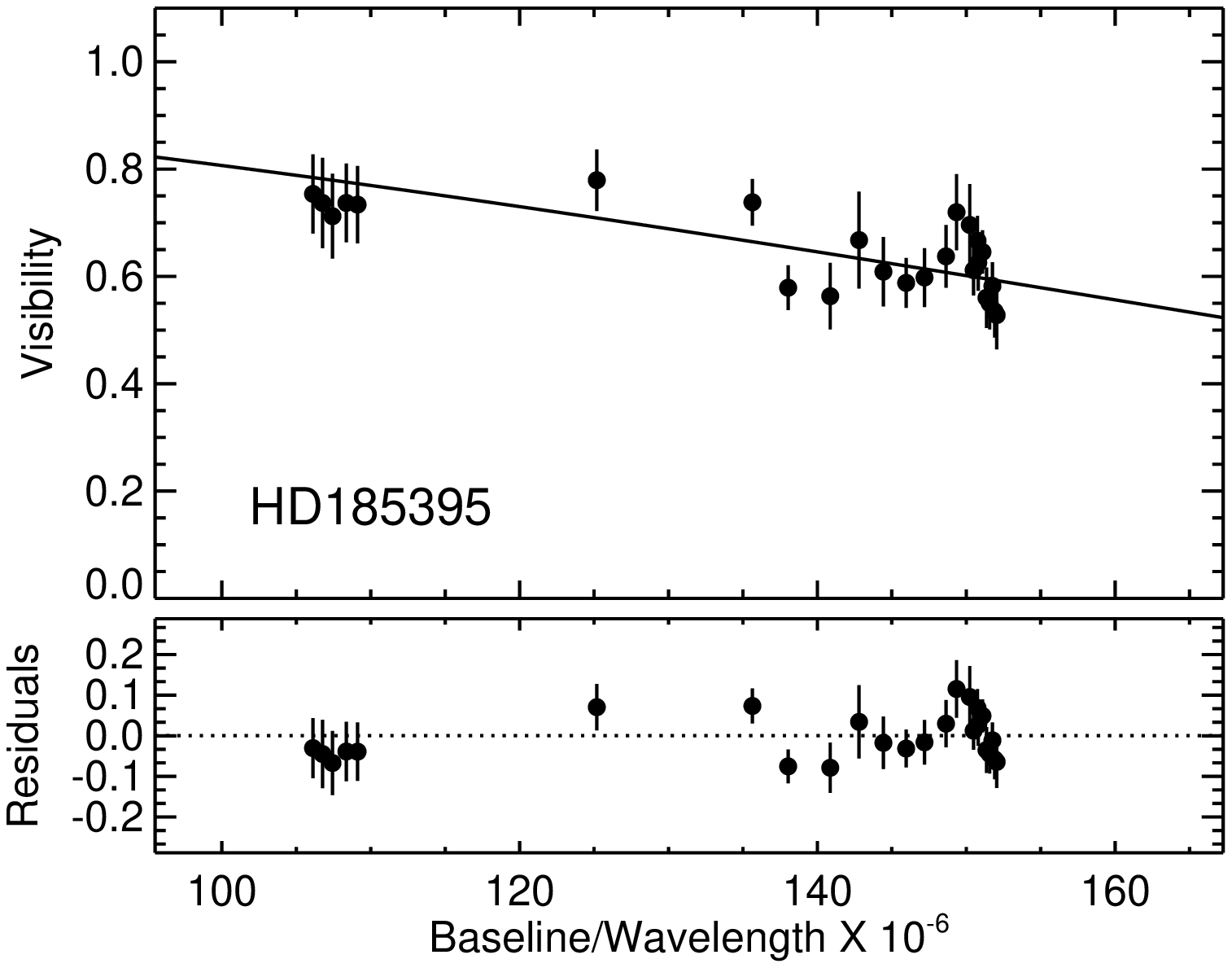,width=0.5\linewidth,clip=} &  
\epsfig{file=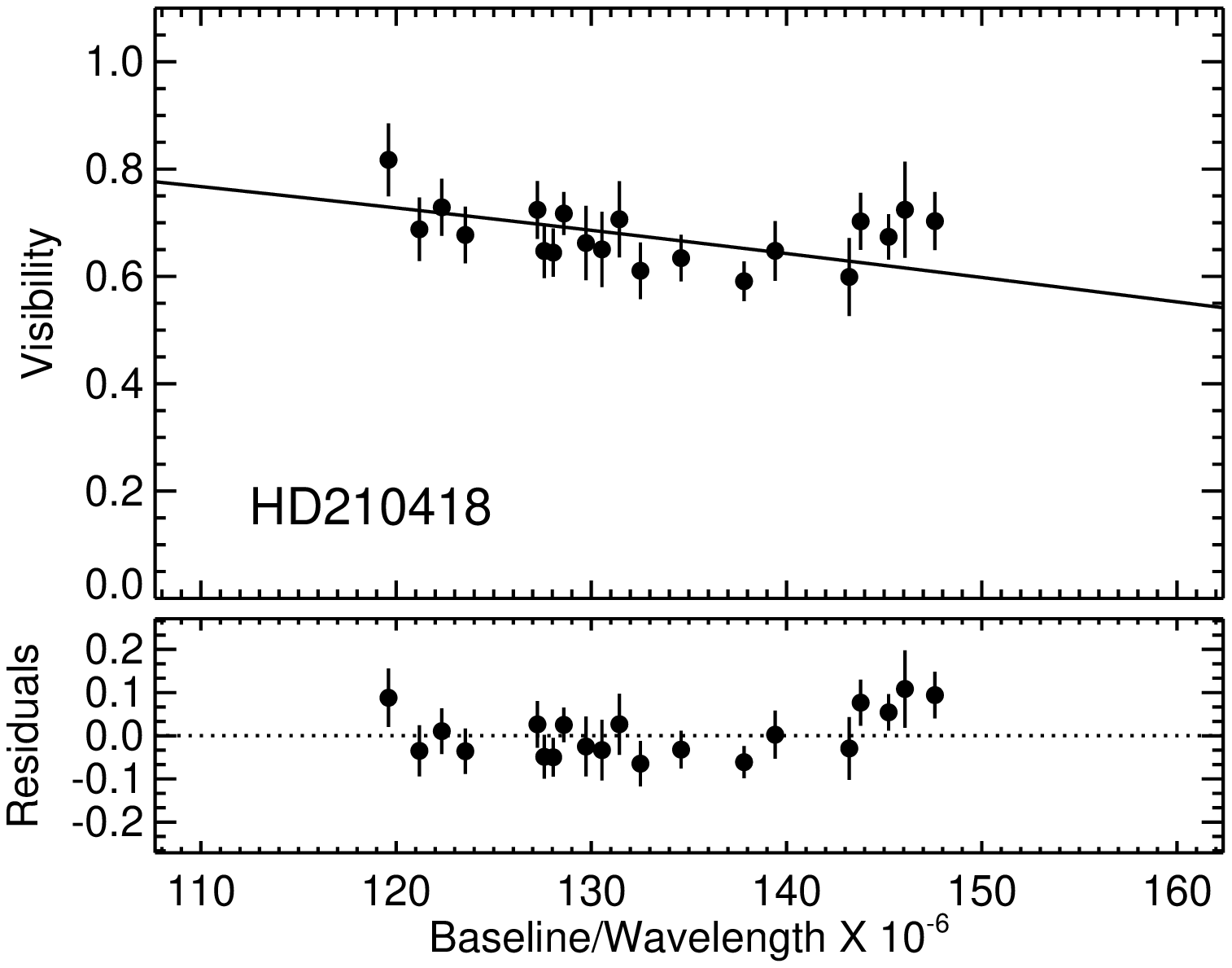,width=0.5\linewidth,clip=}   \\
\epsfig{file=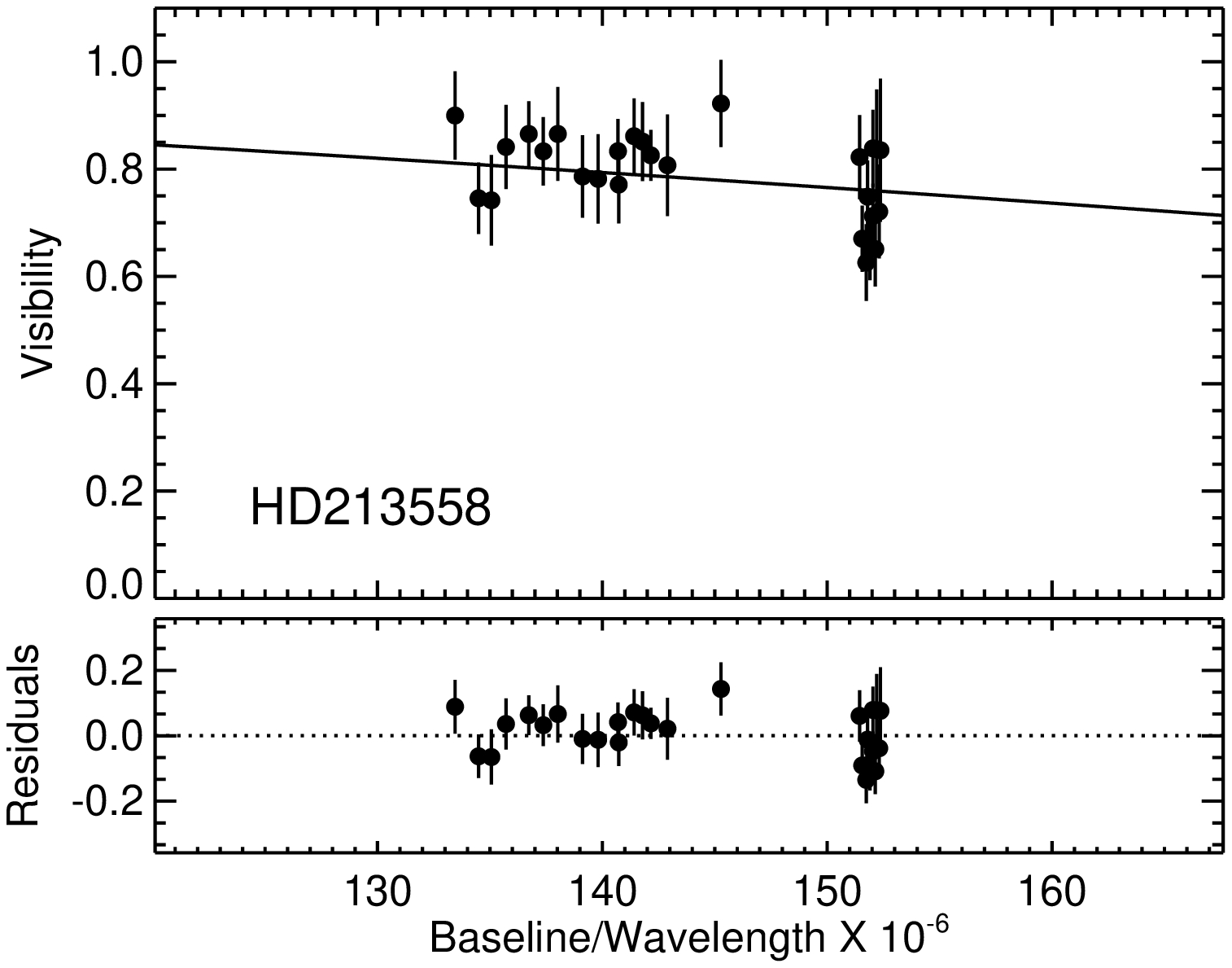,width=0.5\linewidth,clip=} &  
\epsfig{file=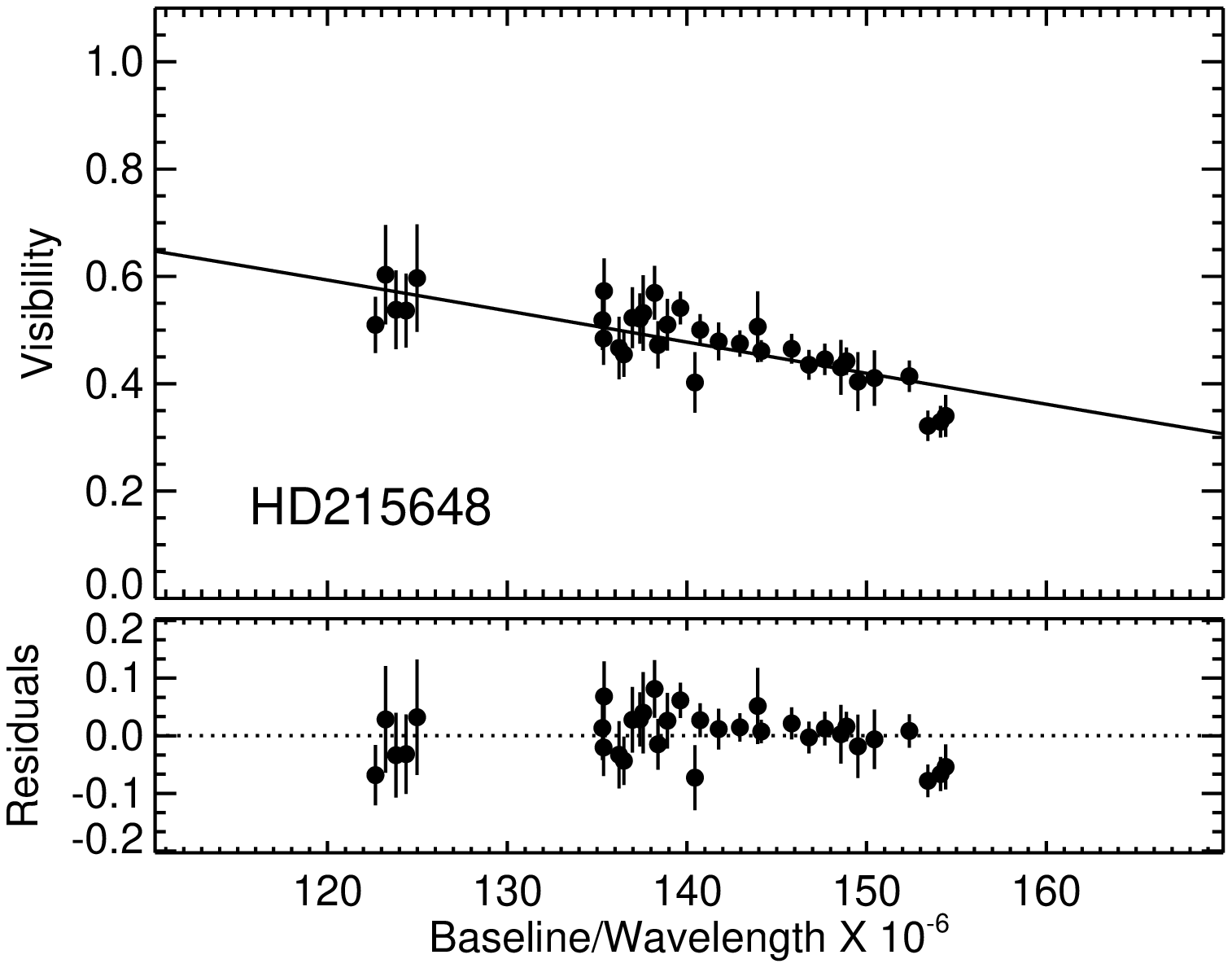,width=0.5\linewidth,clip=}  \\
\epsfig{file=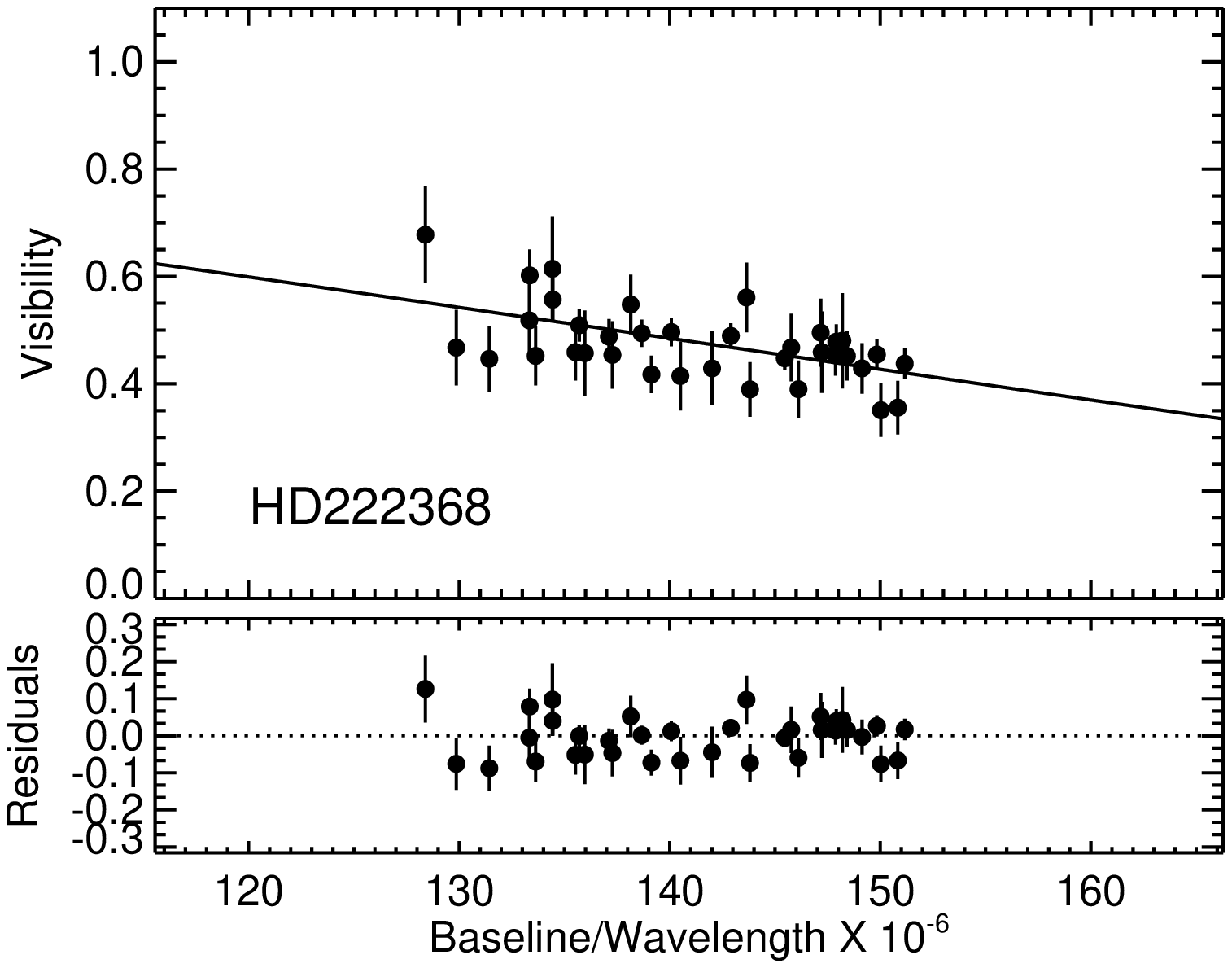,width=0.5\linewidth,clip=} 
 \end{tabular}
\caption[Angular Diameters] {Calibrated observations plotted with the limb-darkened angular diameter fit for each star observed.  See Section~\ref{sec:diameters} and Table~\ref{tab:diameters} for details.}
\label{fig:diameters_7}
\end{figure}

We find that in most cases that the value of the reduced $\chi^2$ is less than 1.0, meaning that we have overestimated the errors on the calculated visibilities for the star.  The results presented here adjust those error estimates to assume a reduced $\chi^2 = 1$ to compensate for the uncertainty in the visibility error estimates \citep{ber06}. Table~\ref{tab:diameters} presents the total number of observations, reduced $\chi^2$, uniform disk diameter $\theta_{\rm UD}$, limb-darkening coefficient $\mu_K$, and limb-darkened diameter $\theta_{\rm LD}$ for each star.  The resulting fit for each star are plotted in Figures~\ref{fig:diameters_1}-\ref{fig:diameters_7}.  Overall, we successfully measure the angular diameters of 8 A-stars, 20 F-stars, and 16 G-stars, with an average precision of $\sim 1.5$\%.  

\placetable{tab:diameters}	

\placefigure{fig:diameters_1}
\placefigure{fig:diameters_2}
\placefigure{fig:diameters_3}
\placefigure{fig:diameters_4}
\placefigure{fig:diameters_5}
\placefigure{fig:diameters_6}
\placefigure{fig:diameters_7}

\subsection{Notes on Individual Star Groups}

\subsubsection{A-Stars}

The results for the A-type stars (8 total) are special cases.  A-type stars are approaching the range at which stars begin to be seen with the highest rotational velocities (B-type stars), resulting in oblate shapes and apparent gravity darkening due to their rapid rotation (for example, see \citealt{zha09}).  This oblateness factor depends on the star's $\theta_{\rm LD}$, rotational velocity v~$\sin i$ and mass (see Equation~5 in \citealt{abs08}).  The most extreme case that we observed is HD~177724, which is among one of the fastest rotating A-stars, with a rotational velocity v~$\sin i$ =  317 km~s$^{-1}$ \citep{roy06}, which leads to a predicted apparent oblateness of 1.307 \citep{abs08}.  Using all measurements and assuming the object to be round, we present the mean diameter for HD~177724 of $\theta_{\rm LD} = 0.897 \pm 0.017$~mas (Table~\ref{tab:diameters}).  This value is in excellent agreement with the predicted mean angular diameter from \citet{abs08} of $\theta = 0.880 \pm 0.018$~mas.  The rotational velocity for all of the A-stars in this project (except for HD~141795; v~$\sin i$ = 47 km~s$^{-1}$) are fairly high (HD~56537 = 154 km~s$^{-1}$, HD~97603 = 180 km~s$^{-1}$, HD~118098 = 222 km~s$^{-1}$, HD~210418 = 144 km~s$^{-1}$, and HD~213558 = 128 km~s$^{-1}$, \citealt{roy06}).  Although their predicted oblateness is likely to be undetectable with the precision of our measurements, we should consider the angular diameter measured for these stars as the mean angular diameter.  The rotational velocities of the F and G-type stars observed are far below critical, therefore they are assumed to be round.

\subsubsection{Multiplicity}

Our targets were selected to only include single stars because incoherent light from a companion affects the visibility of the primary star and biases the diameter measurement.  We admit binary stars only with separations greater than 2~arcseconds or with large $\Delta K$~mags\footnote{The formal limits to detection of companions with our instrument is currently unknown, but is thought to be sensitive to binaries with $\Delta K\lesssim 2.5$ (D. Raghavan, private communication).}.  For example, the star HD~39587 (G0~V~$+$~M5~V; \citealt{han02}) is a single-lined spectroscopic binary.  Our data do not reveal the imprint of a companion (reduced $\chi^{2} = 0.33$), and the diameter is assumed to be unaffected by the presence of a secondary star with the assumption that the $\Delta$~magnitude is quite large ($\Delta K > 4$~mags).

The diameter fit for the star HD~162003 has the largest reduced $\chi^2$ of 2.23.  Very recent radial velocity survey work has identified HD~162003 to have a long term trend in radial velocity of $+$220~m/s/yr \citep{toy09}\footnote{\citep{toy09} comment that the $V_r$ trend is not attributed to the visual companion, HD~162004 $\sim 30^{\prime\prime}$ away.}. We suspect that the reason our diameter fit for this star shows the highest value of reduced $\chi^2$ is that we are actually detecting a mild signature of the secondary star in the visibilities.  Another target in this survey HD~173667 is identified by \citet{nid02} to have $\sigma_{rms} = 124$~m/s over a period of 5165~days. However, unlike the diameter fit for HD~162003, the reduced $\chi^2$ is typical of a fit from a single star (reduced $\chi^2 = 1.34$).  Lastly, we note that the stars HD~4614 and HD~109358\footnote{which coincidentally has been assigned the name {\it Chara} \citep{hof91}.} were once referred to as a spectroscopic binaries by \citet{abt76}. However, \citet{mor87} disputed this result, and HD~109358 is now used as a radial velocity standard (for example, see \citealt{beh09,kon05}).   



\subsection{CHARA Versus Palomar Testbed Interferometer Diameters}
\label{sec:CHARA_vs_PTI}

Angular diameters of a few dozen main sequence stars measured with the Palomar Testbed Interferometer (PTI) were presented by \citet{van09}.  Their work provides measurements of 14 stars in common with the CHARA stars measured here and is the only alternate source of direct angular diameter measurements of our program stars.  The longest baseline obtainable with PTI is 110~m, a factor of three shorter than those of the CHARA Array, and accurate measurements were quite difficult with this instrument due to the small angular sizes of these stars.  

Table~\ref{tab:pti_chara} lists the 14 stars in common with \citet{van09}, the limb-darkened angular diameters and errors, and how many $\sigma$ the two values differ from each other.  For these stars, the errors on the PTI angular diameters are anywhere from 2$-$12 times (with an average of 6.5 times) the errors on the CHARA angular diameters.  However, this comparison can still point to any systematic offsets in the results from each instrument.  Comparing the angular diameters from this work and \citet{van09}, we find that the weighted mean ratio of CHARA to PTI diameters is $\theta_{\rm CHARA} / \theta_{\rm PTI} = 1.052 \pm 0.062$.  \citet{van09} make this same comparison of their diameters compared to diameters from \citet{bai08}, who used the CHARA Array to measure the diameters of exoplanet host stars, and find that the ratio of the four stars they have in common is $\theta_{\rm CHARA} / \theta_{\rm PTI} = 1.06 \pm 0.06$, very similar to the results found here, indicating again that there is a slight preference for smaller PTI diameters, and larger CHARA diameters, although the displacement is at the $< 1$-$\sigma$ level.  

\placetable{tab:pti_chara}	

The preference to larger diameters compared to other interferometric measurements was also seen in \citet{boy09}, where the diameters of the four Hyades giants were measured with the CHARA Array.  In that work, two of the stars, $\epsilon$~Tau and $\delta^1$~Tau, were measured previously with other interferometers (Mark~III, NPOI, and PTI), all which lead to smaller diameters than those measured with CHARA.  However, \citet{boy09} find that models for the Hyades age and metallicity match flawlessly with the CHARA observations, and the smaller angular diameters from other works in turn lead to temperatures that are too hot for these stars.      
   
A main distinction that could lead to offsets in measured diameters is the estimated sizes of the calibrator stars.  For instance, \citet{van09} discuss the calibrator selection in their work compared to \citet{bai08}.  \citet{van09} set a limit to a sufficiently unresolved calibrator at CHARA to be $< 0.5$~mas in diameter, a criterion which all but a few calibrators in this work meet.  However, to investigate the possibility that the estimated size of the calibrators in this work are offset to the calibrators used in \citet{van09}, we compare the estimated sizes of the calibrators in the {\it Palomar Testbed Interferometer Calibrator Catalog} (PTICC, \citealt{van08}) to the ones derived here.  Twenty-nine of the 63 calibrators used in this work are included in the PTICC.  Overall, the ratio of the estimated diameter of the calibrator in this work to the PTICC is $0.97 \pm 0.06$, a less than 1-$\sigma$ difference.  

Twelve of the 14 stars in common with \citet{van09} were observed with calibrators whose diameters are also included in the PTICC. For each of these 12 calibrators, the estimated angular diameter $\theta_{\rm SED}$ is presented in Table~\ref{tab:pti_chara_calibs}, along with the ratio of the CHARA to PTI SED diameters.  The object employing the calibrator is also listed in Table~\ref{tab:pti_chara_calibs} along with the ratio of the CHARA to PTI measured limb-darkened diameters.  Here, there is no pattern in the calibrator SED diameter ratio and the object diameter ratio.   In fact, the effects of a slight offset in the calibrator's estimated diameter listed above (ratio $\theta_{\rm CHARA} / \theta_{\rm PTI} = 0.97 \pm 0.06$) would actually contribute counter-productively to the slight offset in the diameter measurements (ratio $\theta_{\rm CHARA} / \theta_{\rm PTI} = 1.05 \pm 0.06$).  For instance, in the case of our data, the size of the calibrator $\theta_{\rm SED}$ is typically smaller, thus the true visibility of the calibrator would be greater (i.e., it would be more unresolved).  If the true visibility of the calibrator is greater, it would in turn make the true visibility of the object larger in the calibration process.  Thus, the object would appear more unresolved (having larger calibrated visibilities) if we were using a SED diameter of the same calibrator but with a larger value.  Because we do not see the case of smaller CHARA diameters, this indicates that the calibrators are not the cause of any offset, if present, in each data set.  

\placetable{tab:pti_chara_calibs}	

\begin{figure}[!hrt]		
  \centering 
  \includegraphics[width=0.8\textwidth] 
  {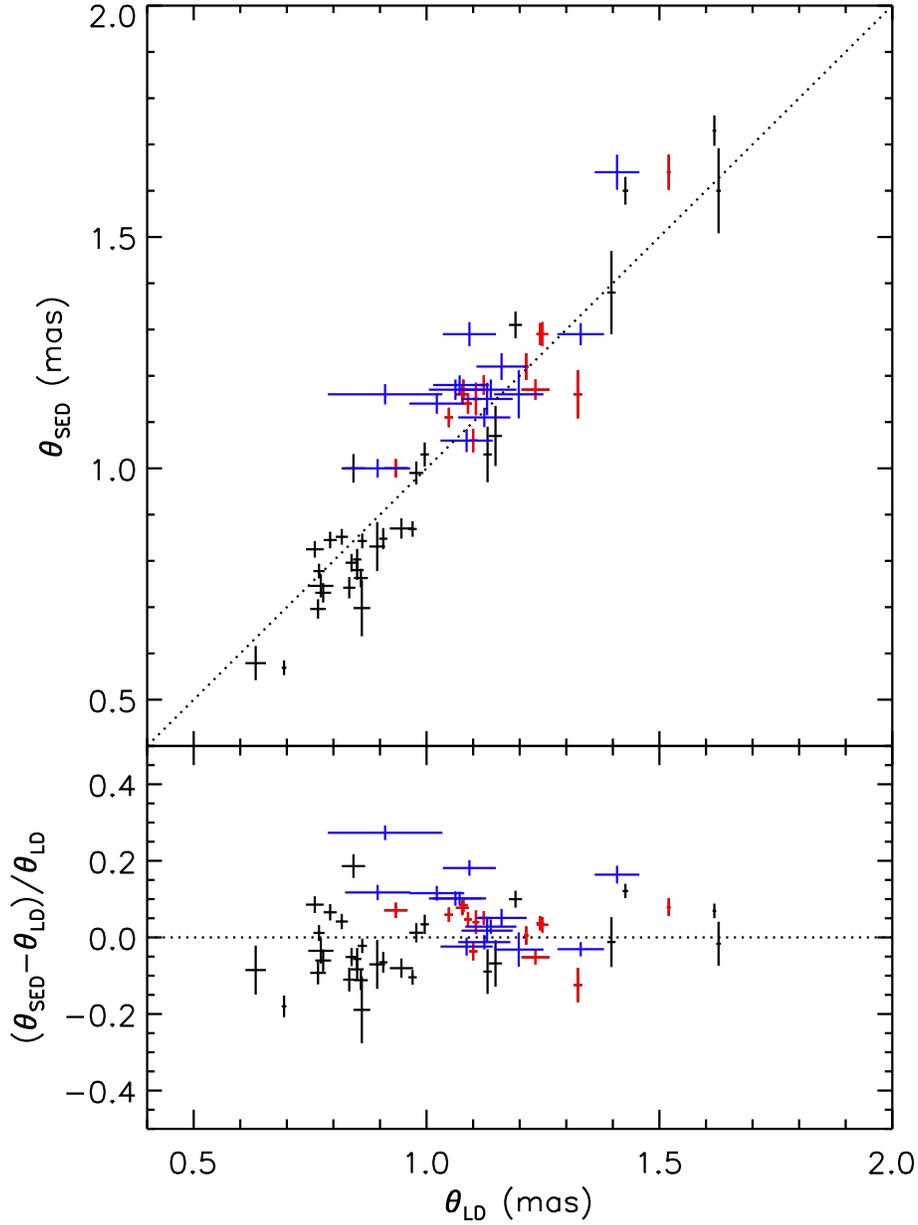}
  \caption{TOP: Plot showing the measured limb-darkened angular diameters versus the SED angular diameters and the 1-$\sigma$ errors ({\it black}).  Measurements for the 14 stars in common with this work are highlighted in color with {\it red} indicating a CHARA measurement presented in this paper and {\it blue} for a PTI measurement measured in \citet{van09}.  BOTTOM: Plot showing the fractional difference between the SED and limb-darkened angular diameters.  The dotted line shows an equal agreement of both measurements.  See Section~\ref{sec:CHARA_vs_PTI} and Tables~\ref{tab:diameters} and \ref{tab:pti_chara} for details.}
  \label{fig:theta_VS_color2}
\end{figure}

In Figure~\ref{fig:theta_VS_color2} we show a plot of the measured limb darkened angular diameters presented here compared to the computed SED diameters for the stars in this survey.  The 14 stars in common with the \citet{van09} work are highlighted in color; red indicating a CHARA measurement and blue for a PTI measurement.  It is apparent here that the majority of the PTI diameters are deceptively low compared to the SED estimates.  

\begin{figure}[!hrt]		 
  \centering 
  \includegraphics[width=0.9\textwidth] 
  {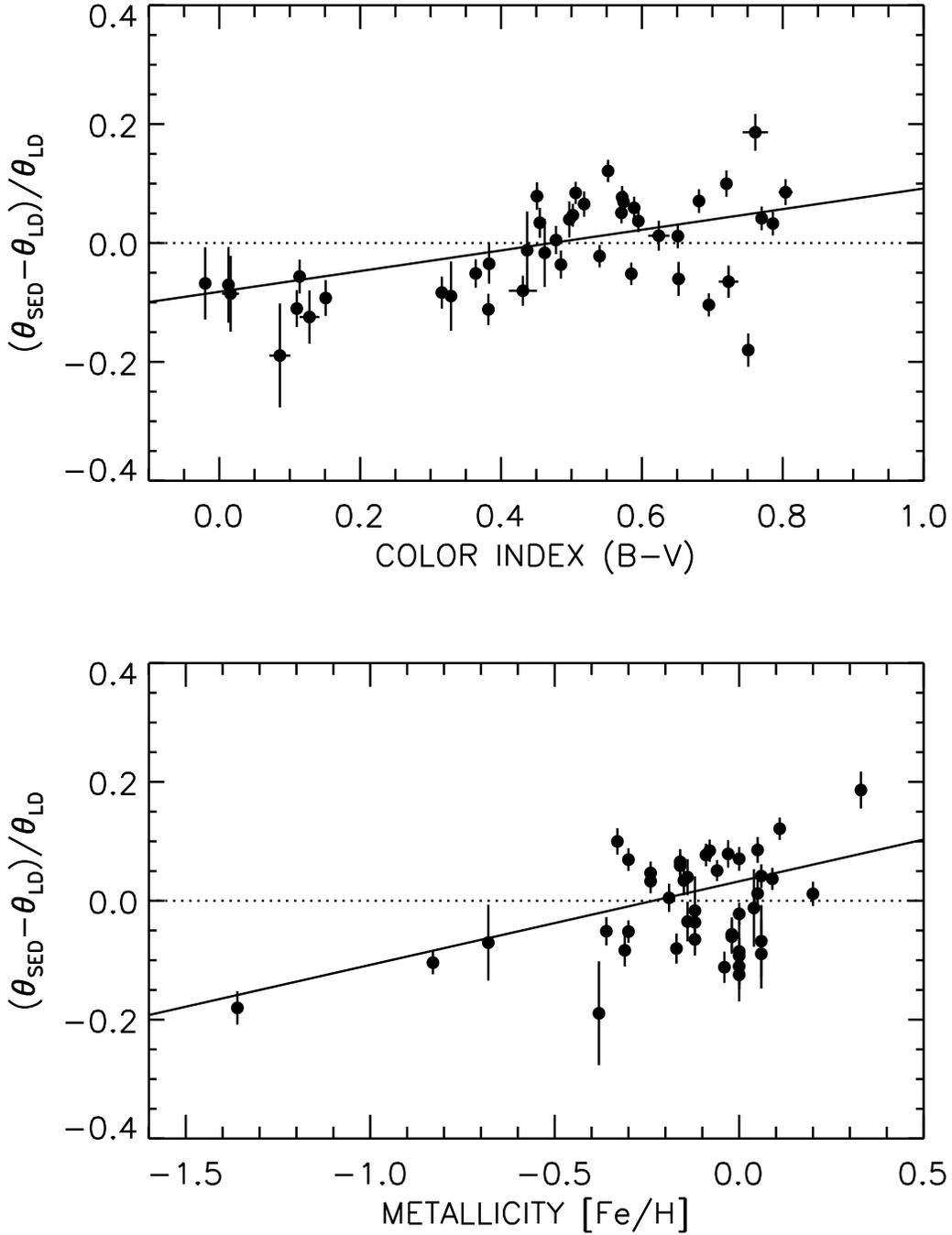}
  \caption{TOP: The difference between limb-darkened angular diameters and the SED angular diameters from this work versus ($B-V$) color index.  Bottom: Same as above with respect to metallicity. The dotted line shows an equal agreement of both measurements.  A provisional fit to the data is shown as a solid line, and is suspect to a cautious interpretation due to the sparse amount of data available. See Section~\ref{sec:diameters} for details.}
  \label{fig:theta_VS_color7}
\end{figure}

Figure~\ref{fig:theta_VS_color7} shows the fractional difference of the measured limb-darkened versus SED angular diameter as a function of ($B-V$) color index.  For stars with a ($B-V$)~$ < 0.4$, the measured $\theta_{\rm LD}$ is always larger than the estimated $\theta_{\rm SED}$.  There is a lot of scatter for stars redder than ($B-V$)~$\sim 0.4$, but no systematics are seen.  However, in this region the two most metal-poor stars, HD~6582 and HD~103095, are the most extreme outliers, where $\theta_{\rm LD} > \theta_{\rm SED}$ in both cases.  Additionally, HD~182572 ($B-V$ = 0.761; [Fe/H] = 0.33) is the most metal-rich star observed in this work, and shows the largest disagreement here where $\theta_{\rm LD} < \theta_{\rm SED}$.  Perhaps there is a lack of sufficient data to show if there is in fact a trend in either the color-index or metallicity of a star and the estimated $\theta_{\rm SED}$.

\placefigure{fig:theta_VS_color2}	
\placefigure{fig:theta_VS_color7}	


\section{Stellar Parameters}        %
\label{sec:stellar_params}

\subsection{Bolometric Fluxes}
\label{sec:fbols}

We present new measures of the bolometric flux $F_{\rm BOL}$ for each star in this work that follows the method used previously in support of interferometric observations \citep{van08, van09}.  The fit involves a collection of flux-calibrated photometry for each source available in the literature  (See Table~\ref{tab:Fbol_phot}) that is subsequently fit with a template spectra from the \citet{pic98} library.  Reddening is for the most part absent for the sample\footnote{distance of the furthest star is $\sim 30$ parsecs}, however we fit it here to the photometric data using reddening corrections based upon the empirical reddening determination described by \citet{car89}, which differs little from van de Hulst's theoretical reddening curve number 15 \citep{joh68, dyc96}.

\begin{figure}[!hrt]		 
  \centering 
  \includegraphics[width=0.9\textwidth] 
  {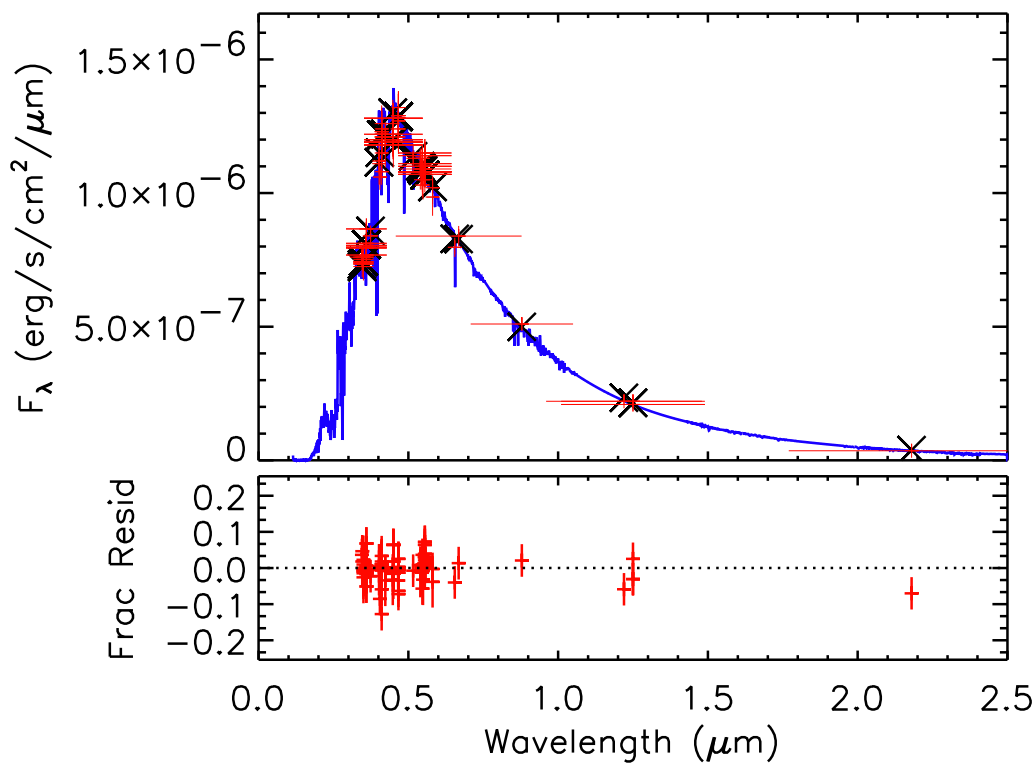}
  \caption{Example SED fit for HD~142860. The (red) plusses indicate flux calibrated photometry from the literature with corresponding errors (y-direction) and bandwidth of the filter (x-direction). The (black) crosses show the flux value of the spectral template integrated over the filter transmission for each point. The spectral template is plotted by a blue line. The lower panel shows the residuals. See Section~\ref{sec:stellar_params} for details.}
  \label{fig:HD142860_sedfit}
\end{figure}

Uncertainties in the $F_{\rm BOL}$ are for our data, on average 1.9\% and tend to be dominated by uncertainty in the reddening fit and poor photometry in the $0.6 - 1.2$~$\mu$m range.  Each science object is listed in Table~\ref{tab:sample_fbols} along with the fitting parameters: \citet{pic98} template spectral type, number of photometry points, and reduced $\chi^2$ of the fit, as well as the resulting integrated $F_{\rm BOL}$ and $A_V$.   In Figure~\ref{fig:HD142860_sedfit} we show an example SED fit. 

\placetable{tab:Fbol_phot}		
\placetable{tab:sample_fbols}	
\placefigure{fig:HD142860_sedfit}	
 

\subsection{Luminosities, Temperatures, and Radii}
\label{sec:l_t_r}

These $F_{\rm BOL}$ measurements are then simply used in combination with the {\it Hipparcos} distance $d$ to solve for the absolute luminosity $L$: 

\begin{equation}
L = 4 \pi d^2 F_{\rm BOL} .
\label{eq:luminosity}
\end{equation}

\noindent Additionally, we can express the effective temperature of a star as defined through the Stephan-Boltzmann law

\begin{equation}
F = \sigma T_{\rm EFF}^{4}
\label{eq:sb}
\end{equation}

\noindent where $F$ is the total emergent flux of the star and $\sigma$ is the Stefan-Boltzmann constant.  Transforming this equation to observables at Earth, we arrive at the expression:

\begin{equation}
F_{\rm BOL} = \frac{1}{4} \theta^{2} \sigma T_{\rm EFF}^{4}.
\label{eq:sb_Earth}
\end{equation}

\noindent The effective temperature $T_{\rm EFF}$ is found by solving Equation~\ref{eq:sb_Earth} in terms of $F_{\rm BOL}$ and  $\theta_{LD}$, where in Equation~\ref{eq:temperature}, $F_{\rm BOL}$ is in units of $10^{-8}$ erg~s$^{-1}$~cm$^{-2}$, and $\theta_{LD}$ is in units of mas.  This yields the relation 

\begin{equation}
T_{\rm EFF} = 2341 (F_{\rm BOL}/\theta^2)^{0.25} .
\label{eq:temperature}
\end{equation}

\begin{figure}[!hrt]		
  \centering 
  \includegraphics[width=0.9\textwidth] 
  {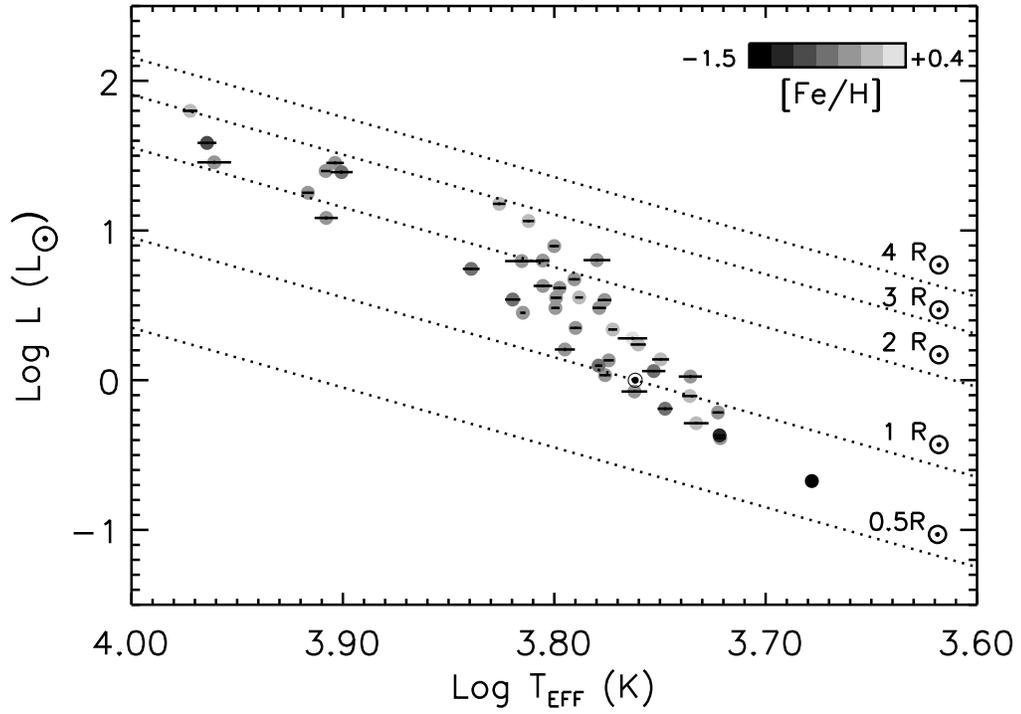}
  \caption[ ] {The luminosities and temperatures of the stars in the survey are plotted with their 1-$\sigma$ errors. Lines of constant radii are plotted as dotted lines.  The shading of the symbol represents the metallicity of the star [Fe/H]. See Section~\ref{sec:l_t_r} for details.}
  \label{fig:l_vs_tfeh}
\end{figure}

\begin{figure}[!hrt]		
  \centering 
  \includegraphics[width=0.9\textwidth] 
  {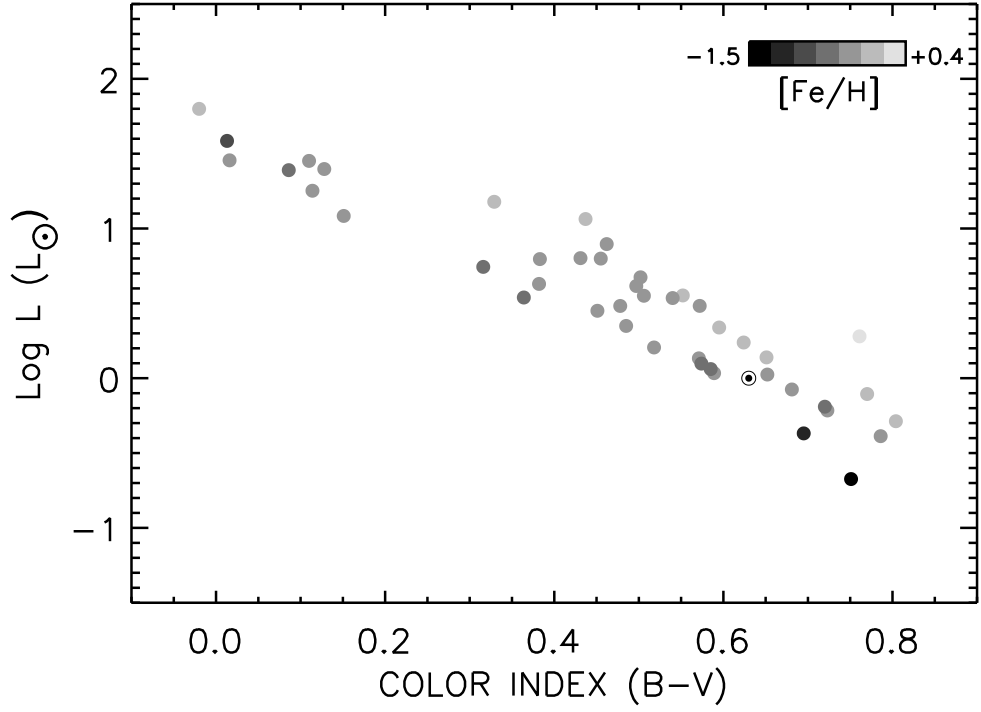}
  \caption[ ] {The luminosity and ($B-V$) color-index plotted of the stars in this survey.  The shading of the symbol represents the metallicity of the star [Fe/H]. See Section~\ref{sec:l_t_r} for details.}
  \label{fig:l_vs_bmvfeh}
\end{figure}

Finally, using {\it Hipparcos} parallaxes from \citet{van07}, we transform the measured angular diameters of these stars into linear radii $R$.  Each of these derived parameters are presented in Table~\ref{tab:r_l_t}.  Graphical representations of these data sets are presented in Figures~\ref{fig:l_vs_tfeh} and \ref{fig:l_vs_bmvfeh}.

\placetable{tab:r_l_t}	
\placefigure{fig:l_vs_tfeh}
\placefigure{fig:l_vs_bmvfeh}


\section{Discussion}

\subsection{Empirical Temperature Relations}
\label{sec:empirical_relations}

With the results in Table~\ref{tab:r_l_t}, we may begin to define a foundation of empirically based color-temperature relations.  These relationships are extremely useful in extending our knowledge to a larger number of stars, at distances too far to accurately resolve their sizes.  For giants and super-giants, temperature scales accurate to the 2.5\% level are obtainable, and are currently limited by the distances to these objects, not the sensitivities of our interferometric observations \citep{van99, van09a}. Recently, and more similarly comparable to this work, \citet{van09} present relations for main sequence stars, ranging from ($V-K$) $\sim$ 0.0 - 4.0 (spectral types of $\sim$ A-M).  These scales are slightly better than the aforementioned evolved classes of stars, and are accurate to the $\sim$ 2\% level.  The authors note here that the limitations to these relations are in the angular diameter measurements themselves.

\subsubsection{($B-V$) - $T_{\rm EFF}$}
\label{sec:empirical_relations_bmv}

Here, we derive color-temperature relations based on the precise $T_{\rm EFF}$ measurements presented in Table~\ref{tab:r_l_t} the in the form of a $6^{th}$ order polynomial.  The solution for the ($B-V$) color - temperature relation is expressed as:



\begin{eqnarray}
\log T_{\rm EFF}	& =	&	3.9680 \pm 0.0025 - 0.2633 \pm 0.0876 (B-V) - 3.2195 \pm 1.3407 (B-V)^2 + \nonumber	\\
 					&	&	15.3548 \pm 6.8551 (B-V)^3 - 27.2901\pm 15.7373 (B-V)^4 + \nonumber \\
					&	&	19.9193 \pm 16.8465 (B-V)^5 - 4.5127 \pm  6.8539 (B-V)^6  \\
					&	&	{\rm for} \; 0.05 < (B-V) < 0.80 \; {\rm and} \; {\rm [Fe/H]} > -0.75	\nonumber					
\label{eq:poly6_bmv}
\end{eqnarray}

\begin{figure}[!hrt]		
  \centering 
  \includegraphics[width=0.8\textwidth] 
  {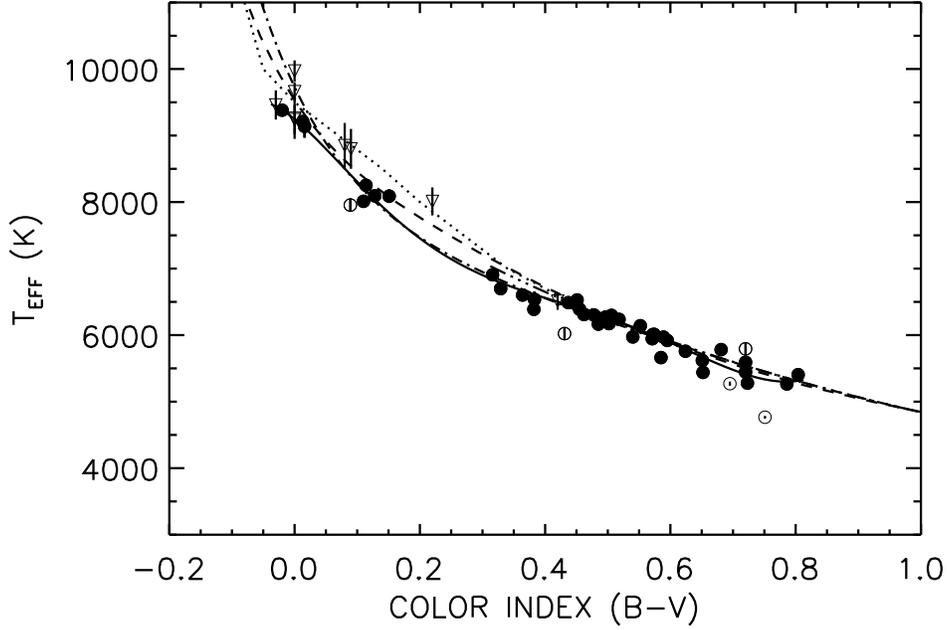}
  \caption{The data and solution for the temperature - ($B-V$) relation shown as circles and a solid line, respectively. The data omitted from the fit are plotted as open circles and can be identified (from left to right) as HD~210418, HD~162003, HD~6582, HD~182572, and HD~103095.  The inverted triangles show the data from \citet{cod76} and the empirical solution based on this data is plotted as a dotted line.  The 1-sigma errors in temperature are displayed, but are typically smaller than the data point.  The dash-dotted line is the calibration presented in \citet{gra92}, using a menagerie of data sources (see reference).  The solution for solar metallicity from \citet{cas10} is shown as a triple-dot dashed line.  The entirely model based solution from \citet{lej98} (dashed line) is also shown. For details, see Section~\ref{sec:empirical_relations_bmv} and Equation~\ref{eq:poly6_bmv}. }
  \label{fig:temp_VS_BmV}
\end{figure}

\noindent This solution, in the form of a $6^{th}$ order polynomial, defines the shape of the data inflection point at ($B-V$)~$\sim 0.3$ better than a lower order polynomial function, as well as a power law function (see Figure~\ref{fig:temp_VS_BmV}).  We are also cautious that the stars with low metallicity will affect the ($B-V$) - $T_{\rm EFF}$ transformation too severely to be useful for stars of solar-type abundances, and we refrain from using these in this analysis\footnote{There are two stars, HD~6582 and HD~103095, with metallicity [Fe/H]~$< -0.75$.}.  A preliminary fit to the data yields a median deviation in temperature of $68$~K, and a median deviation in temperature of 55~K is found for an identical solution if we omit three obvious outliers lying more than 5-$\sigma$ away, HD~210418, HD~182572 and and HD~162003 (offset of $-6.7, -5.7$ and $+6.5$~$\sigma$, respectively.  The statistical summary for the solution to the polynomial can be found in Table~\ref{tab:poly_coeff}\footnote{A fit to our data with a power relation yields an initial solution with a median deviation in temperature of 125~K, and 101~K removing the points with low metallicity.  A deviation of 92~K is found after removing outliers lying more than 3-sigma from the fit, almost double the value we find for the polynomial fit.}. 

 \placetable{tab:poly_coeff}	

In Figure~\ref{fig:temp_VS_BmV}, we show our data and the solution for the fit.  We also show the solution from several other sources, \citet{cod76, gra92} and \citet{lej98}. All the solutions shown here are approximately identical in the range of ($B-V$)~$> 0.45$.

Similar to our work, the results from \citet{cod76} are derived solely on empirical measurements.  In that milestone paper, \citet{cod76} measure the diameters of 32 stars using the Narrabri Stellar Intensity Interferometer (NSII), all being objects hotter than the Sun and most having evolved luminosity classes.  We show in Figure~\ref{fig:temp_VS_BmV} the 9 data points from \citet{cod76} that have a ($B-V$)~$>0$ ($\sim$A-type and later) that are luminosity class V or IV (8 A-type objects and 1 F-type object), as well as the fit from \citet{cod76}.  Comparing this fit to ours, we note that it is a few hundred Kelvin hotter than our own for $0.05 < $~($B-V$)~$< 0.3$, converging at the bluest range of ($B-V$)~$\sim 0$, as well as the reddest range ($B-V$)~$\sim 0.4$.  The offset is likely strongly connected to the fit's dependence on the sparse amount of data in this intermediate range. 

The function presented in \citet{gra92} also uses a $6^{th}$-order polynomial to fit ($B-V$) to $T_{\rm EFF}$ (their equation~15.14).  This is a fit to a compilation of data, including the NSII data from \citet{cod76}, but mostly data obtained via the infrared flux method (IRFM, see \citealt{bla77}), yielding a semi-empirical approach in obtaining the angular diameter, and consequentially the effective temperature, of a star.  There is an impressive correlation with our fit throughout the range of ($B-V$) colors.  A deviation only begins to appear at the bluest colors (($B-V$)~$\sim 0$).  Our sample stops here, but it can be readily seen in \citet{gra92} that there is another inflection point at this color index, producing a steeper curve at more negative color indices.  Because of this, we caution against the use of the relation presented here for colors bluer than ($B-V$)~$< 0.05$.  

As an update, the work presented in \citet{cas10} presents an excellent analysis and improvement to the IRFM temperature scales that have been proposed over the years.  In Figure~\ref{fig:temp_VS_BmV} we also show the \citet{cas10} solution for solar metallicity for comparison (valid in the ranges of $(B-V) > 0.3$).  This relation is distinguishable from ours only when $0.3 < (B-V) <0.4$, where at this point, the \citet{cas10} relation begins to predict $\sim 100$~K higher temperatures.  However, we suspect that the likely explanation for this difference is that their solution is in the form of a third order polynomial, which has been shown not to model the inflection point in this region properly.

For an additional comparison on the ($B-V$) to $T_{\rm EFF}$ relations, we also show the entirely model-based solution from \citet{lej98} (dashed line).  This solution is intermediate to ours (as well as the one in \citealt{gra92}) and the one from \citet{cod76} for the mid-A to mid-F-type stars.
 

The median error in the temperature measurement for these stars (45~K) is lower than the median deviation to the fit, suggesting the potential for improvement.  Iterating on the fit and removing outliers that lie further than 3-sigma gives an identical solution, but reduces the number of points used from 39 to 33. However, it does not improve the error of the fit by much.  The scatter in the points with a ($B-V$)$ > 0.6$ are clearly the cause of the ill-correlated relation, as illustrated in Figure~\ref{fig:temp_VS_BmV}. 

\placefigure{fig:temp_VS_BmV}	
\placefigure{fig:temp_VS_VmK}	

\subsubsection{($V-K$) - $T_{\rm EFF}$}
\label{sec:empirical_relations_vmk}

\begin{figure}[!hrt]		
  \centering 
  \includegraphics[width=0.8\textwidth] 
  {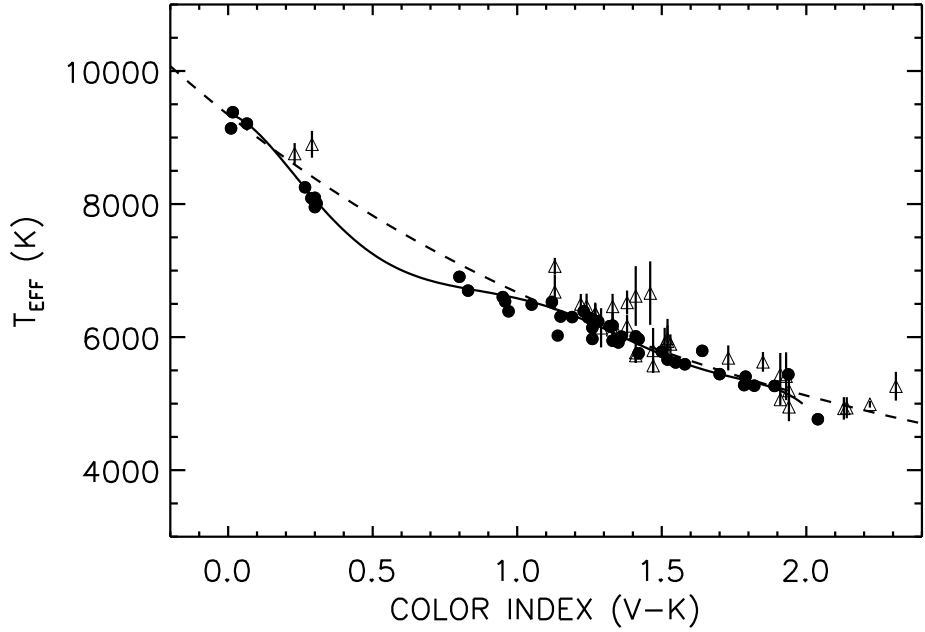}
  \caption{Data points plotted as filled circles depict the observations presented here, and open triangles are the measurements from \citet{van09}.  The 1-sigma errors in temperature are shown, but are typically smaller than the data point.  Our solution for the temperature-($V-K$) relation shown as a solid line (Equation~\ref{eq:poly6_vmk}, whereas the relation from \citet{van09} is shown as a dashed line. For details, see Section~\ref{sec:empirical_relations_vmk} and Equation~\ref{eq:poly6_vmk}.}
  \label{fig:temp_VS_VmK}
\end{figure}

We also present a fit of the temperature versus ($V-K$) color index (Equation ~\ref{eq:poly6_vmk} and Figure~\ref{fig:temp_VS_VmK}), the benefit here being that ($V-K$) colors are less sensitive to the stellar abundances than the ($B-V$) colors. A solution including all stars gives a median deviation in temperature of 67~K, with only one extreme outlier, HD~162003 ($+5.8$~$\sigma$)\footnote{We note that this star has a visual companion HD~162004, and thus could have bad photometry.  Note that HD~162003 was an outlier in the $B-V$ relation, with a negative offset from the fit derived.}.  Omitting this star, we arrive at our final solution, practically identical from the first, with a median temperature deviation of 64~K.  The statistical overview for this solution can be found in Table~\ref{tab:poly_coeff}, and Figure~\ref{fig:temp_VS_VmK} illustrates the fit.  We note that the solution is identical when clipping stars that lie more than $3.5$~$\sigma$ from the fit (instead of 5-$\sigma$) and has an improved median deviation in temperature of 56~K. As expected, the temperatures for the two previously mentioned metal poor stars, HD~6582 and HD~103095, agree exceptionally well with the temperature - ($V-K$) relation. 
 
 
 \begin{eqnarray}
 \log T_{\rm EFF} 	&	= 	&	3.9685 \pm 0.0034 + 0.0830 \pm  0.0570 (V-K) - 1.8948 \pm 0.2874 (V-K)^2 + \nonumber	\\
					&		&	4.0799 \pm 0.5438 (V-K)^3 - 3.7353 \pm 0.4739 (V-K)^4 + \nonumber	\\
					&		&	1.5651 \pm 0.1936 (V-K)^5 - 0.2472 \pm 0.0301 (V-K)^6	\\
 					&		&	{\rm for} \; 0.0 < (V-K) < 2.0 	\nonumber					
 \label{eq:poly6_vmk}
 \end{eqnarray}

 \subsubsection{Spectral Type - $T_{\rm EFF}$}
 \label{sec:empirical_relations_spty}

\begin{figure}[!hrt]		
  \centering 
  \includegraphics[width=0.8\textwidth] 
  {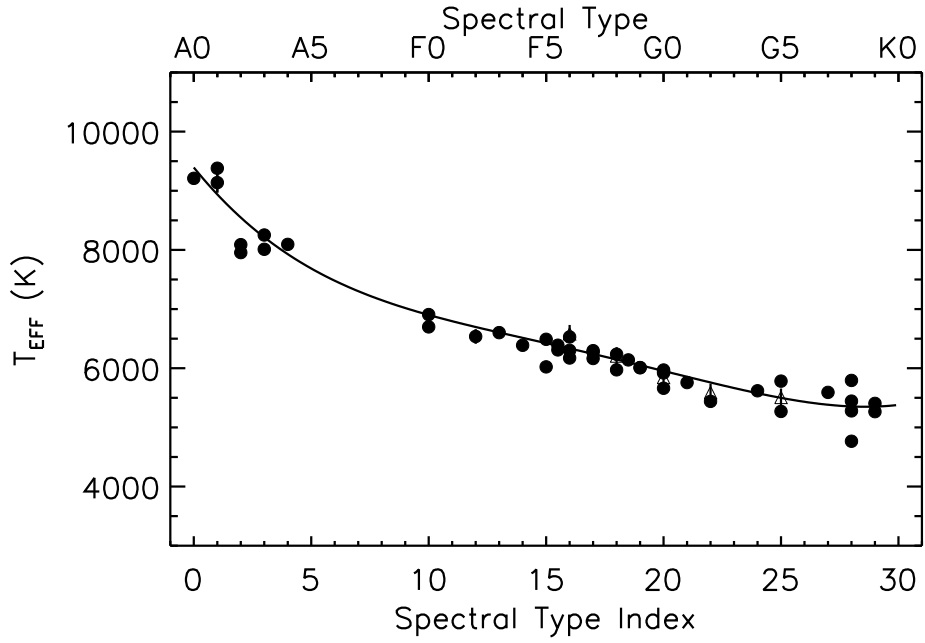}
  \caption{The data and solution for the temperature - spectral type relation shown as a circles and a solid line, respectively. The 1-sigma errors in temperature are shown, but are typically smaller than the data point.  The data from Table~7 in \citet{van09} are displayed as open triangles. For details, see Section~\ref{sec:empirical_relations_spty} and Equation~\ref{eq:temp_VS_SpTy}.}
  \label{fig:temp_VS_SpTy}
\end{figure}

\begin{figure}[!hrt]		
  \centering 
  \includegraphics[width=0.8\textwidth] 
  {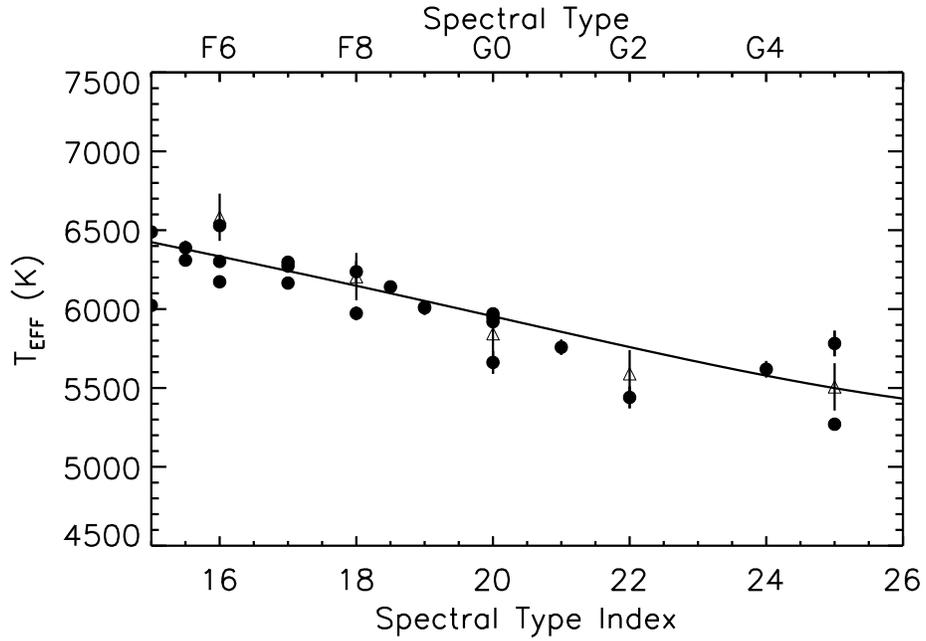}
  \caption{This is a zoomed-in version of Figure~\ref{fig:temp_VS_SpTy}, showing the region that overlaps with the solution presented in \citet{van09}.  Our data and solution for the temperature - spectral type relation are shown as circles and a solid line, respectively. The 1-sigma errors in temperature are shown, but are typically smaller than the data point. The data from Table~7 in \citet{van09} are displayed as open triangles.  For details, see Section~\ref{sec:empirical_relations_spty} and Equation~\ref{eq:temp_VS_SpTy}.}
  \label{fig:temp_VS_SpTy_closeup}
\end{figure}

We next derive a useful (albeit less accurate) relation between spectral type and temperature.  We do this by converting the spectral types for each star into a numerical value, following the scheme: A0, A1, A2 $\dots$ F0, F1, F2 $\dots$ G0, G1, G2, $\dots$, K0 $\rightarrow$ 0, 1, 2, $\dots$, 10, 11, 12, $\dots$, 20, 21, 22, $\dots$, 30.  Again, we omit the two metal-poor stars previously mentioned, and fit a forth order polynomial to arrive at the relation: 


\begin{eqnarray}
T_{\rm EFF}		&	=	&	9393.59 \pm 60.45 - 490.25 \pm 28.79 ST + 36.44 \pm 3.62 ST^2 - \nonumber 	\\
 				&		&	1.44 \pm 0.17 ST^3 + 0.0208 \pm 0.0027 ST^4	\\
				&		&	{\rm for} \; {\rm A0} < ST < {\rm K0} \; {\rm and} \; {\rm [Fe/H]} > -0.75	\nonumber					
\label{eq:temp_VS_SpTy}
\end{eqnarray}  

\noindent where in this equation, the variable $ST$ refers to the numerical value for the spectral type index. The fit for Equation~\ref{eq:temp_VS_SpTy} has a median absolute deviation of 90~K, and is plotted in Figure~\ref{fig:temp_VS_SpTy}. In the range from F6 to G5, we also show the fit from the data Table~7 from \citet{van09} (Figure~\ref{fig:temp_VS_SpTy_closeup} is a close-up view of this range), and our results are consistent with each other.

\placefigure{fig:temp_VS_SpTy}	
\placefigure{fig:temp_VS_SpTy_closeup}	


\subsection{Comparison to Indirectly Determined Temperatures}
\label{sec:comp_teff}

In the literature, there exist three surveys of nearby stars that include objects that overlap with the majority of stars in this sample: \citet{all99}, \citet{hol07} and \citet{tak07}.  The number of stars in common with each survey respectively are 37, 34 and 25.  We compare our results to these works in this section.

\citet[hereafter APL99]{all99} derive fundamental parameters for the stars in their survey by fitting model evolutionary tracks from \citet{ber94} to observed ($B-V$) photometry and absolute $V$ band magnitude $M_V$.  The Geneva-Copenhagen survey of the Solar neighbourhood II, done by \citet[hereafter GCS07]{hol07}, utilizes the Padova models \citep{gir00, sal00} to derive the stellar parameters based on Str\"{o}mgren $uvby$ calibrations to $T_{\rm EFF}$ and $M_V$. Lastly, \citet[hereafter Tak07]{tak07} use the Yonsei-Yale ($Y^2$) stellar isochrones \citep{yi2001, kim02, yi03, dem04} to fit their spectroscopically determined $T_{\rm EFF}$ along with the photometrically derived $L$ (from the absolute magnitude, and bolometric correction, see \citet{tak05} for details).  GCS07 demonstrate that these model isochrones (among others) show minimal differences when compared to each other, also seen in \citet{boy08}.  However, we choose to compare all three sources since the target overlap is not identical or complete with respect to our own, as well as the fact that different datasets and photometric indices were used for each group.

\begin{figure}[!hrt]		
  \centering 
  \includegraphics[width=0.8\textwidth] 
  {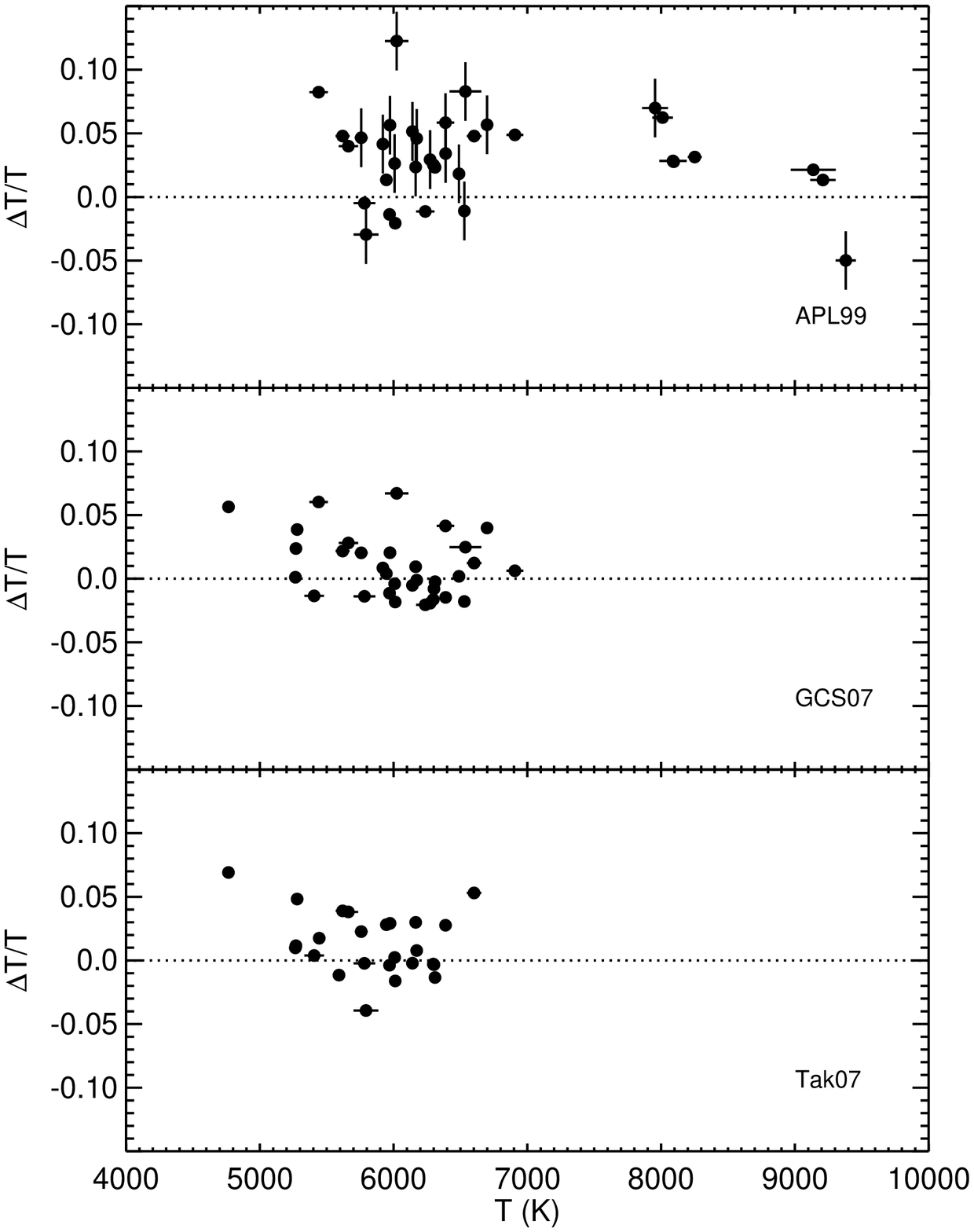}
  \caption{The data plotted show the dependence on the measured temperature versus the fractional differences between model temperatures determined by \citet{all99} (APL99), \citet{hol07} (GCS07), and \citet{tak07} (Tak07) and the empirical values determined in this project, along with 1-$\sigma$ errors for each.  The dotted line marks a zero deviation between each source. }
  \label{fig:CHARA_vs_All_temps}
\end{figure}

\begin{figure}[!hrt]		
  \centering 
  \includegraphics[width=0.8\textwidth] 
  {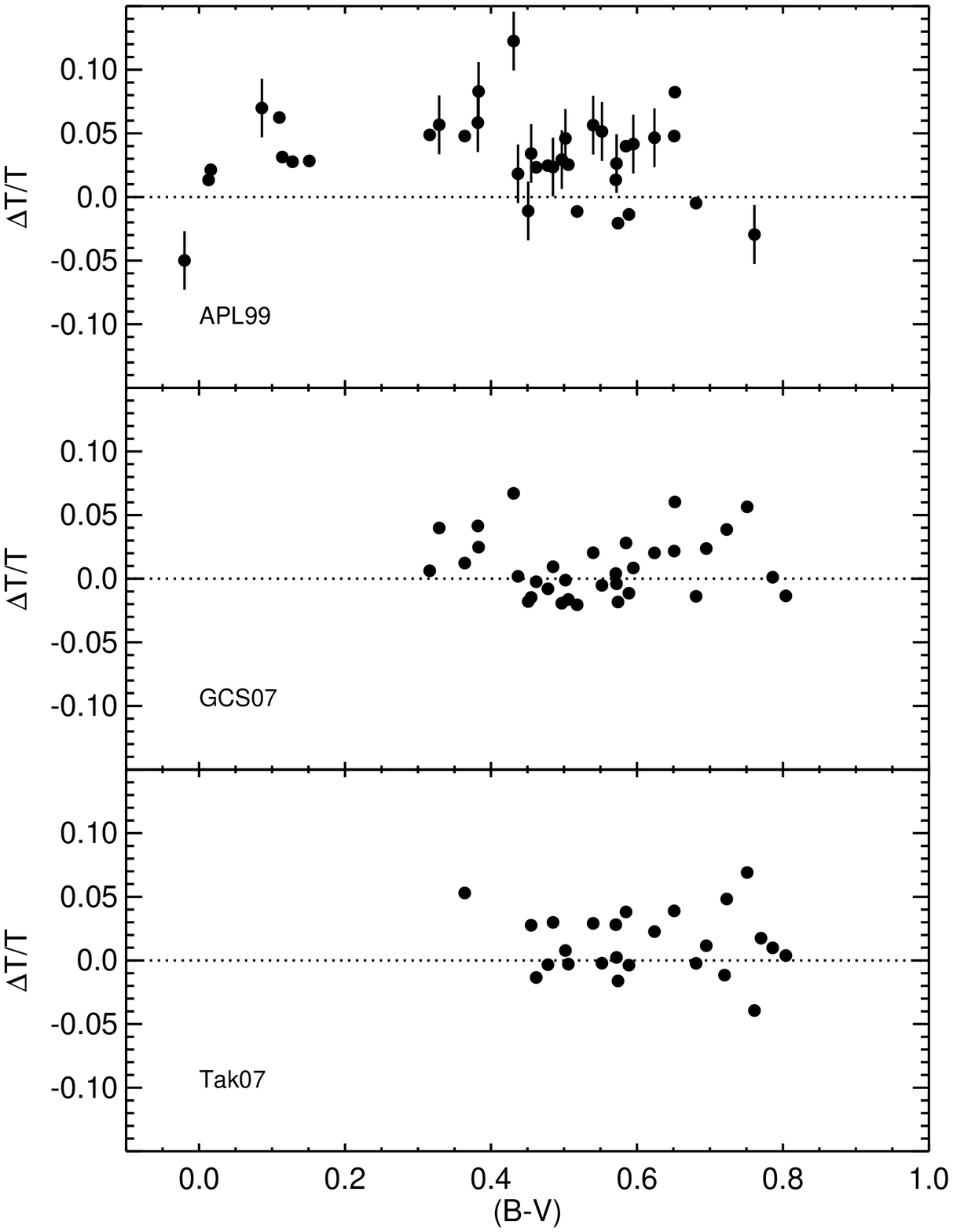}
  \caption{The data plotted show the dependence on the ($B-V$) color index versus the fractional differences between model temperatures determined by \citet{all99} (APL99), \citet{hol07} (GCS07), and \citet{tak07} (Tak07) and the empirical values determined in this project, along with 1-$\sigma$ errors for each.  The dotted line marks a zero deviation between each source.  }
  \label{fig:CHARA_vs_All_temps_bmv}
\end{figure}

\begin{figure}[!hrt]		
  \centering 
  \includegraphics[width=0.8\textwidth] 
  {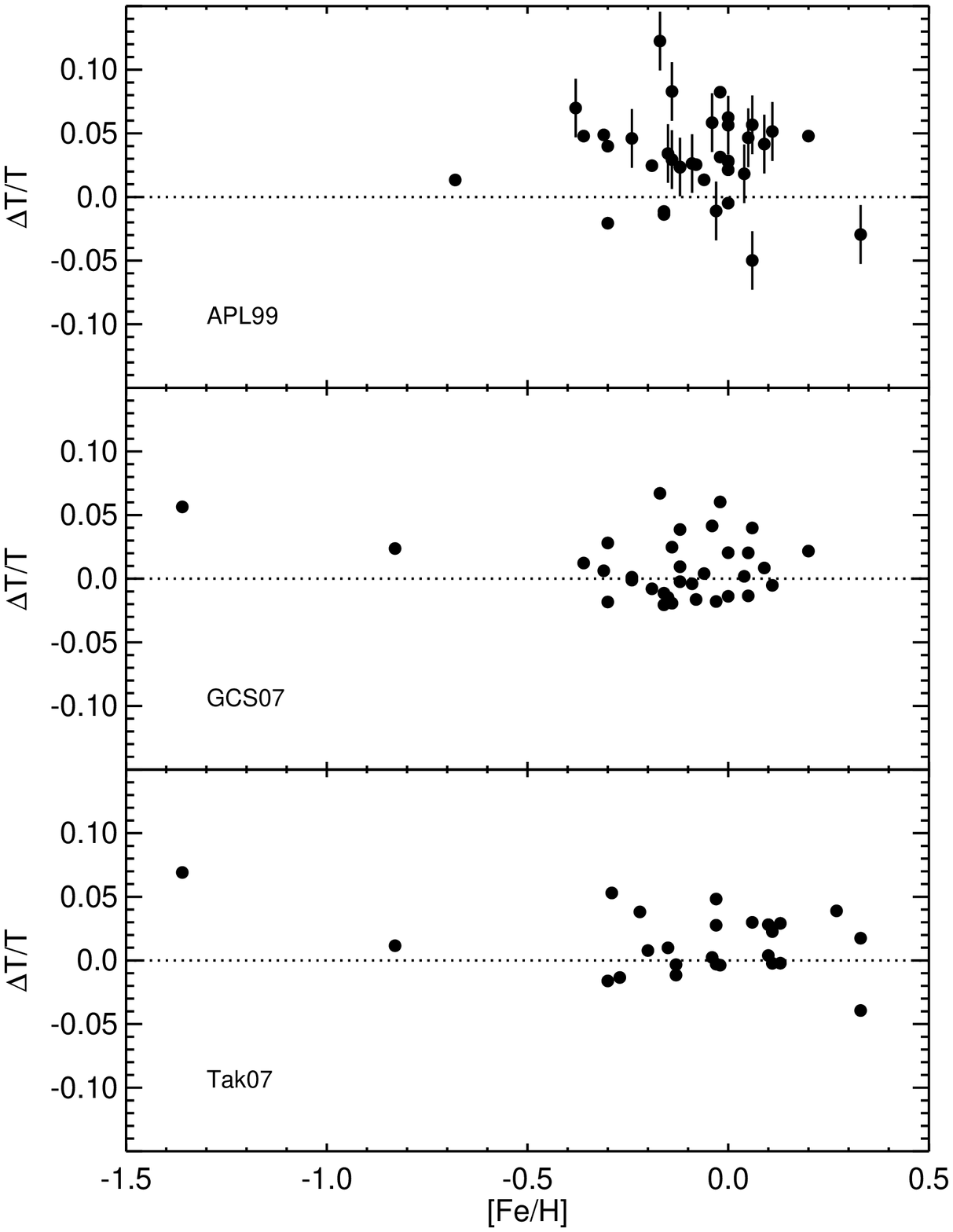}
  \caption{The data plotted show the dependence on the metallicity versus the fractional differences between model temperatures determined by \citet{all99} (APL99), \citet{hol07} (GCS07), and \citet{tak07} (Tak07) and the empirical values determined in this project, along with 1-$\sigma$ errors for each.  The dotted line marks a zero deviation between each source.  }
  \label{fig:CHARA_vs_All_temps_feh}
\end{figure}

In Figure~\ref{fig:CHARA_vs_All_temps}, we compare our temperatures to the temperatures of the stars in common in each reference. The most apparent discrepancies appear when comparing our results to those in APL99.  For most cases seen here, APL99 overestimates the effective temperature of the star through the entire range of effective temperatures by about 5\%, up to 15\%.  GCS07 and Tak07 are less drastic in comparison, but there is still a tendency of the models to overestimate the temperatures by a couple percent.  Figures~\ref{fig:CHARA_vs_All_temps_bmv} and \ref{fig:CHARA_vs_All_temps_feh} plot the fractional offset of the stellar temperature from each method versus ($B-V$) color index and metallicity. For the hottest/bluest of stars ($T_{\rm EFF} > 6500$, $B-V < 0.4$, a range mostly covered by APL99 only) a positive offset is seen for all but one measurement.  Likewise, temperatures of stars with the lowest metallicity are also overestimated in each reference compared to the temperatures presented here, but no trend can be identified with the sparse quantity of data available in this range.  

The stars with the largest offsets in the effective temperatures of GCS07 are HD~81937 (13\%), and HD~146233 (7\%). Interestingly enough, these stars also have high discrepancies between their SED diameter and the limb-darkened diameter measured with CHARA. However, stars such as HD~10780 and HD~109358 also have high deviation in the SED diameter versus the limb-darkened diameter measured by CHARA, but their agreement with the temperature from GCS07 is at the $\sim 1$\% level. It is interesting to note that the star HD~146233 (18 Sco), that was first identified by \citet{por97} to be a solar twin, is one of these stars with a large offset in effective temperature.

The agreement in temperature with Tak07 is better than both the APL99 and GCS07 surveys, but there is still a preference to higher temperatures than what we measure.  The largest outliers in temperature offsets compared to Tak07 are HD~128167 (6.5\%), HD~103095 (5.4\%), and HD~86728 (4.3\%).  Comparing these outliers to the GCS07 outliers, there are no two stars in each that show large deviations from the model versus CHARA temperature, with the exception of the very metal poor star, HD~103095. Again, it does not appear that a star's metallicity or color index is related to the deviation in temperatures of each source.


\placefigure{fig:CHARA_vs_All_temps}	
\placefigure{fig:CHARA_vs_All_temps_bmv}	
\placefigure{fig:CHARA_vs_All_temps_feh}	

\subsection{Isochrone Masses and Ages}
\label{sec:masses_and_ages}

\begin{figure}[!hrt]		
  \centering 
  \includegraphics[width=0.8\textwidth] 
  {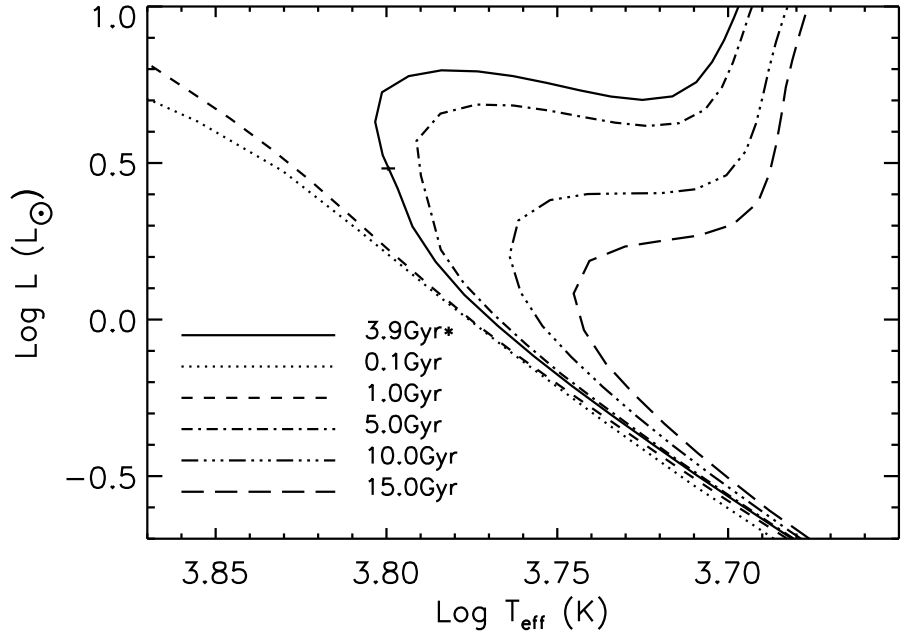}
  \caption{Stellar isochrones generated for HD~142860.  The asterisk depicts best age fit to our data and is plotted as a solid line. The luminosity and temperature and the 1-sigma errors for this star are also plotted.  These measurement errors show that the error in the x-direction (temperature) is the least constrained input for the isochrone fit.}
  \label{fig:142860_yy}
\end{figure}

We determine an estimate of the stellar mass and age by fitting our temperatures and luminosities presented in section~\ref{sec:l_t_r} to the Yonsei-Yale ($Y^{2}$) stellar isochrones \citep{yi2001, kim02, yi03, dem04}.  We are able to do this for most stars observed for this work, but unfortunately, useful results are not obtainable when fitting isochrones to stars with $L < 0.75 L_{\odot}$, because sensitivity to age in this region is minimal\footnote{Stars within this range are HD~6582, HD~10780, HD~101501, HD~103095, HD~131156, and HD~185144.}.  To run the model isochrones, input estimates are required for the abundance of iron [Fe/H] (Table~\ref{tab:sample}) and $\alpha$~elements [$\alpha$/Fe]\footnote{The [$\alpha$/Fe] for all stars are zero \citep{car96}. We assign [Fe/H] = 0 for HD~56537, HD~141795, and HD~213558 because no metallicity measurements are available in the literature for these stars.}.  For each star, isochrones were generated in increments of 0.1~Gyr, in the range of 0.1$-$15~Gyr.  We fit the model isochrones in the theoretical temperature-luminosity plane, where the solutions from the model are purely from the theory of stellar structure. This eliminates any dependence of the color table used in transforming the model isochrone temperature to observed photometric colors \citep{lej98}. Once a best-fit isochrone is established, an age along with the associated mass for this best fit isochrone is recorded for each star.  We show an example of this routine in Figure~\ref{fig:142860_yy}, and the results are listed in Table~\ref{tab:Y2_model_output}.  

We calculate the errors on the best fit isochrone mass and age using the 1-$\sigma$ temperature and luminosity maximum offset for the isochrone solution.  It is worth noting that the metallicity input for the model isochrones has an impact on the derived age (and in turn also on the derived mass). Lower metallicity isochrones shift down and to the left on these diagrams, so for a star with a true metallicity less than the input value, a higher isochrone age would be found. The opposite is true for stars with higher values of metallicity, where a younger age would result. For our isochrone fits, we adopt fixed values of metallicity measured from a uniform source (in this case \citealt{hol07}) that are used in the model input for computations. Thus, relative ages may be correct while absolute ages are highly uncertain\footnote{A characteristic error introduced by a $0.1$~dex uncertainty in the metallicity may introduce errors ranging from about $0.1 - 2.0$~Gyr and $0.06$~M$_{\odot}$ for the A to G type stars, respectively.}.

\placetable{tab:Y2_model_output}	
\placefigure{fig:142860_yy}	

\subsection{Gravity Masses}

With the linear radii known for all stars in the CHARA sample, we are able to determine the mass of a star using $\log g$ estimates published in APL99 and Tak07 using the relation

\begin{equation}
g_{\ast} = G M_{\ast} / R_{\ast}^{2}
\label{eq:gravity_mass}
\end{equation}  

\noindent where G is the gravitational constant, $M_{\ast}$ is the mass of the star, $R_{\ast}$ is the radius of the star, and $g_{\ast}$ is the surface gravity of the star. Hereafter, we will refer to the mass derived in this manner as $M_{\rm g-R}$.


\subsection{Comparative Analysis}
\label{sec:comp_ages}

\begin{figure}[!hrt]		
  \centering 
  \includegraphics[width=0.8\textwidth] 
  {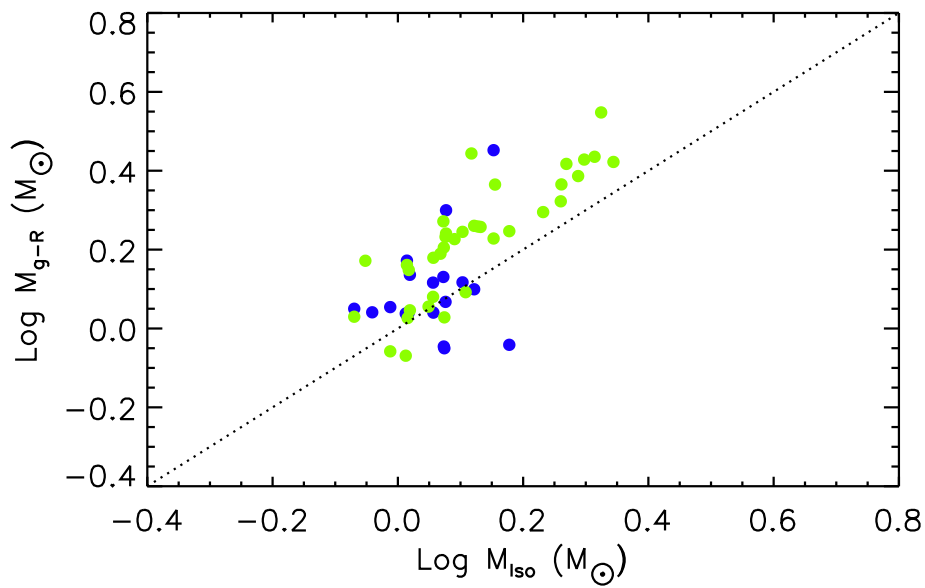}
  \caption{Masses derived from the $Y^2$ isochrones $M_{\rm Iso}$ compared to masses of stars calculated from the combination of $\log g$ estimates and our CHARA radii $M_{\rm g-R}$. Reference for $\log g$ data are for stars in common with the APL99 and Tak07 surveys are depicted green and blue, respectively. The dotted line shows a 1:1 relation. }
  \label{fig:YYmass_vs_CHARAmass}
\end{figure}

\begin{figure}[!hrt]		
  \centering 
  \includegraphics[width=0.8\textwidth] 
  {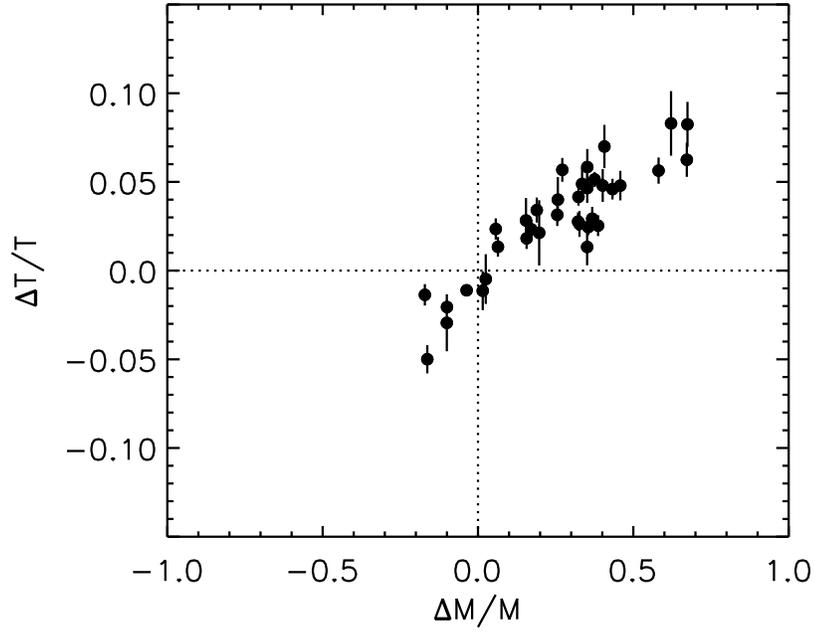}
  \caption{Fractional deviation of the masses derived from isochrone fitting and the ones derived when using APL99 $\log g$ in combination with CHARA radius measurements versus the fractional offset of our measured temperature versus the ones in APL99.  The dotted lines indicate zero deviation.}
  \label{fig:dToT_vs_dMoM}
\end{figure}

\begin{figure}[!hrt]		
  \centering 
  \includegraphics[width=0.8\textwidth] 
  {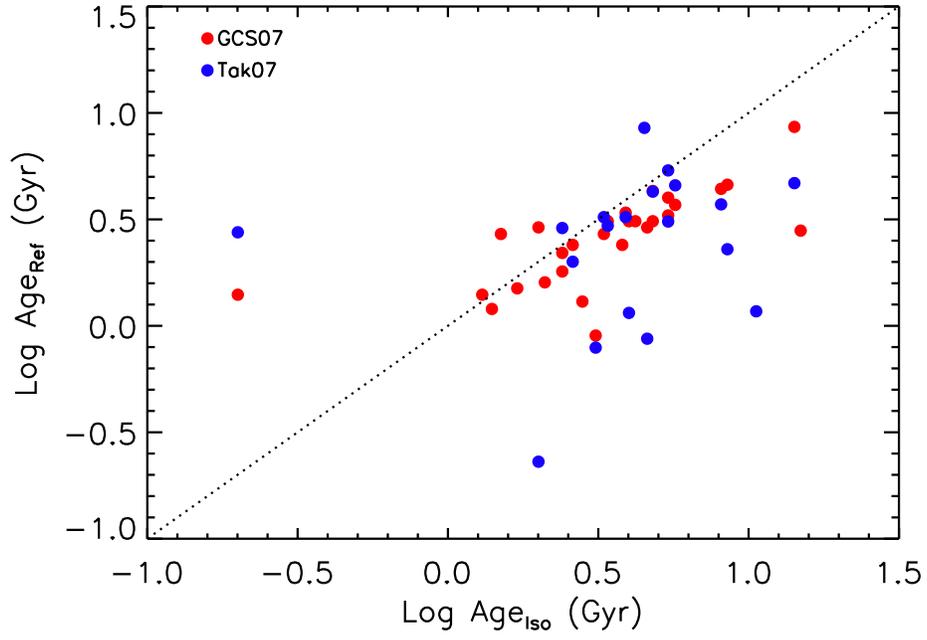}
  \caption{Ages derived from the $Y^2$ isochrones compared to ages of stars in common with GCS07 (red) and Tak07 (blue). The dotted line shows a 1:1 relation.}
  \label{fig:YYage_vs_Refage_log}
\end{figure}

Masses derived from isochrone fitting ($M_{\rm Iso}$) and those obtainable when the surface gravity and radius are known through Equation~\ref{eq:gravity_mass} ($M_{\rm g-R}$) are compared in Figure~\ref{fig:YYmass_vs_CHARAmass} for the stars with gravity measurements published by APL99 and Tak07.  There is significant scatter, however we find that for stars greater than $\sim 1.3 M_{\odot}$, the masses are over-predicted when using gravities from APL99. It is possible that the offset links back to the temperature offsets discussed in the previous section. For instance, if the model temperature used to fit the spectral lines to determine $\log g$ values for the stars is offset, this will in turn lead to spurious values of $\log g$.  Presumably these overestimated temperatures will lead to a slightly more massive star, because hotter stars on the main sequence are more massive than their cooler counterparts.  Sure enough, this trend can be seen in Figure~\ref{fig:dToT_vs_dMoM} when comparing the deviation in temperatures and masses for these stars. Ages derived are also affected in the sense that the stars will appear younger if their temperature or $\log g$ is artificially offset to higher values. In fact, a slight trend may be seen in Figure~\ref{fig:YYage_vs_Refage_log} to support this when comparing our ages to the ages in GCS07 and Tak07 (APL99 did not publish ages), where our ages are typically greater.  

The discussion in \citet{hol07} also gives several examples of how an offset in effective temperature will, in turn, offset the metallicity, and argues that these effects double up when determining the ages of the stars, thereby producing false age-metallicity relations.   

\placefigure{fig:dToT_vs_dMoM}	
\placefigure{fig:YYage_vs_Refage_log}	

For further comparison, we introduce the data set presented in \citet{and91}. \citet{and91} provides a compilation of data on all non-interacting eclipsing binaries (EB) known at the time - a total of 90 stars, most of which are on the main sequence. Section~4 in \citet{and91} argues that the motivation for compiling the EB data is to aid in the prediction of single-star properties where masses and radii are unobtainable by direct measurements for a large number of stars. We use these data for eclipsing binaries to compare with our results for single stars in this section.  

\begin{figure}[!hrt]		
  \centering 
  \includegraphics[width=0.8\textwidth] 
  {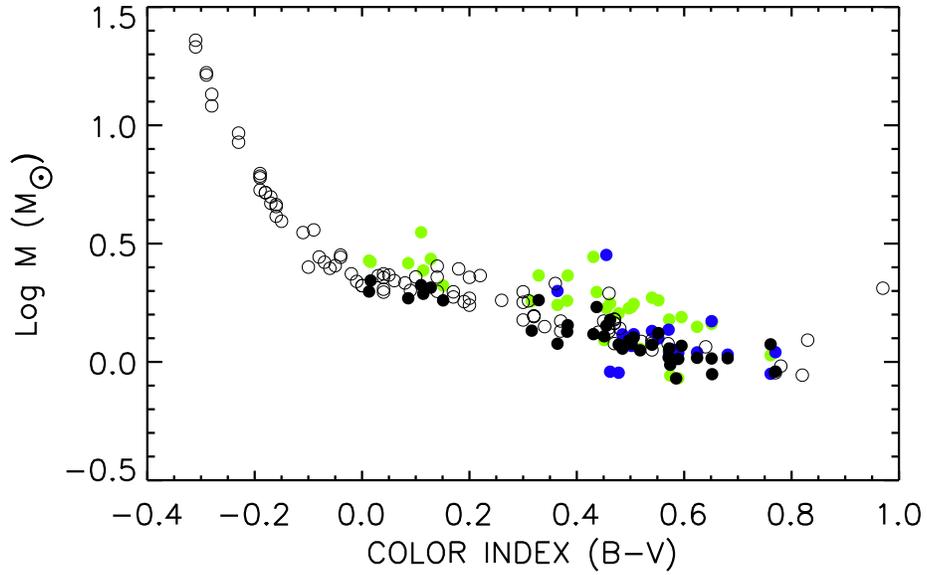}
  \caption{Mass versus color index for eclipsing binaries (black open circles) plotted with masses derived from our sample using the $Y^2$ isochrones (black filled circles). Also plotted are the calculated $M_{\rm g-R}$ for stars in common with the APL99 (green filled circles), and Tak07 (blue filled circles) surveys.}
  \label{fig:YYmass_vs_color_plusEB_plusGM}
\end{figure}

\begin{figure}[!hrt]		
  \centering 
  \includegraphics[width=0.8\textwidth] 
  {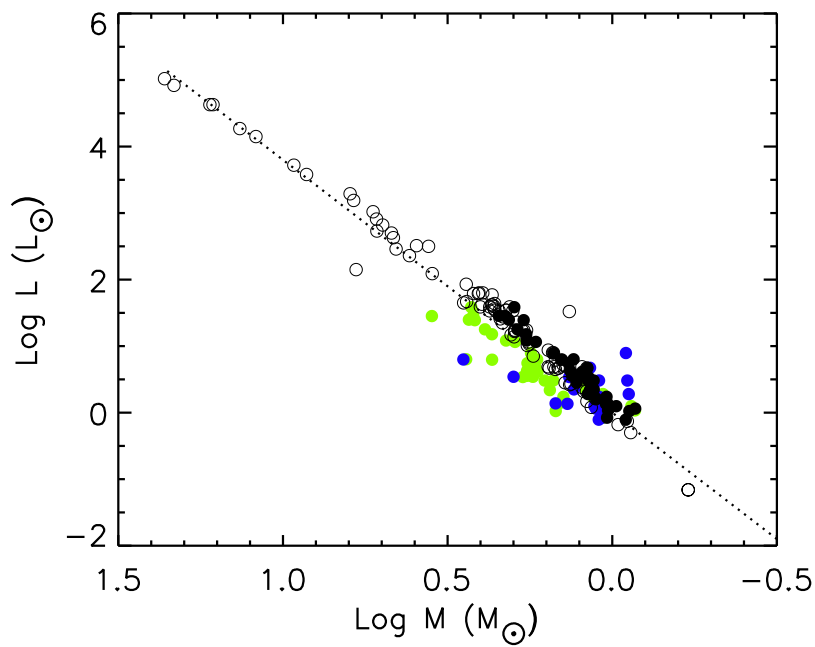}
  \caption{Mass versus luminosity for eclipsing binaries (black open circles) plotted with masses derived from our sample using the $Y^2$ isochrones (black filled circles). Also plotted are the calculated $M_{\rm g-R}$ for stars in common with the APL99 (green filled circles), and Tak07 (blue filled circles) surveys. The dotted line is the relation $L \propto M^{3.8}$.}
  \label{fig:YYmass_vs_Luminosity_plusEB_plusGM}
\end{figure}

In Figure~\ref{fig:YYmass_vs_color_plusEB_plusGM}, we show the relation between ($B-V$) color index and stellar mass.  The mass-luminosity relation is shown in Figure~\ref{fig:YYmass_vs_Luminosity_plusEB_plusGM}.  Eclipsing binary data from \citet{and91} are plotted as well as the masses for stars in this project derived from the $Y^2$ isochrones, and masses derived from the combination of the CHARA radii and $\log g$ estimates from each source (APL99 or Tak07).

The $Y^2$ masses we found are in excellent agreement with the sample of eclipsing binaries from \citet{and91} with respect to ($B-V$) color index and luminosity (Figure~\ref{fig:YYmass_vs_color_plusEB_plusGM} and Figure~\ref{fig:YYmass_vs_Luminosity_plusEB_plusGM}). However, comparing the $M_{\rm g-R}$ data points to the EB sample, we see that these masses are biased to larger masses, forcing them to appear under-luminous compared to the EB sample as well as the $Y^2$ data points.  The effect is again most apparent in the range where the APL99 gravities are used, where the systematics appear for stars with masses greater than $\sim 1.3 M_{\odot}$.

\placefigure{fig:YYmass_vs_color_plusEB_plusGM}	
\placefigure{fig:YYmass_vs_Luminosity_plusEB_plusGM}	

\section{Summary and Conclusion}

In this paper we present the angular diameters of nearby, main-sequence stars measured with the CHARA Array.  The survey includes a total of 44 stars, for which the angular diameter is measured to better than 4\%.  Robust bolometric flux measurements of the stars are presented, yielding absolute luminosities and effective temperatures measured to accuracies on the average of 1.8\% and 1\%, respectively. Lastly, we extract masses and ages of the stars by applying the Yonsei-Yale model isochrones to our data

We show that published values of such stellar parameters derived indirectly are not consistent with our results.  Our data compared to published values show that several sources overestimate the effective temperature by $1.5 - 4$\%.  The values for modelled stellar radius compensate this offset by being underestimated in order for the luminosity to come out right.  Generally speaking, this offset is most apparent in the metal-poor stars as well as the earlier type stars ($\gtrsim 1.3 M_{\odot}$) that we observed. We propose that this preference for models to derive hotter temperatures is the cause for further divergence in a star's surface gravity measurements, which consequently yields higher masses and younger ages.  Excellent agreement is seen when comparing our data to a large sample of eclipsing binaries.

This data sample is used to solve for empirically-based temperature relations for A-, F-, and G-type main-sequence stars, with precision close to the 1\% level on the color-temperature calibrations.  The results of the $B-V$ color - temperature relation are consistent with other relations for the stars redder than $B-V \sim 0.4$ (mid F- through G-type). However, for stars earlier than this (the A- and early F-types), our solution is a couple hundred Kelvin cooler than all but one of the relations we use for comparison.  This is likely due to the sparsity of good quality empirical data used for fitting in this region.  Our data is better correlated in the $V-K$ color - temperature relation than the interferometric data in \citet{van09} for the F- and G-type stars, due to the high precision of our measurements compared to theirs, but the fits to the data are indistinguishable in this range.  For the earlier-type stars, moderate disagreement is still present, again from lack of complete sampling in this region.  

Ongoing observations of more nearby main-sequence stars are under way by our group at the CHARA Array.  Color-temperature calibrations with various other color indices will be implemented in the forthcoming papers in the series.  Stellar metallicity was ignored in this work, however including it in temperature relations will almost certainly beat the errors down even more than what we have achieved with the current fit. In our future efforts, we aim to populate the data set enough to identify accurate color-temperature-metallicity relations, as well as extend the sample down to the lower end of the main-sequence.

\acknowledgments

TSB acknowledges support provided by NASA through Hubble Fellowship grant \#HST-HF-51252.01 awarded by the Space Telescope Science Institute, which is operated by the Association of Universities for Research in Astronomy, Inc., for NASA, under contract NAS 5-26555.  The CHARA Array is funded by the National Science Foundation through NSF grant AST-0908253 and by Georgia State University through the College of Arts and Sciences. This research has made use of the SIMBAD literature database, operated at CDS, Strasbourg, France, and of NASA's Astrophysics Data System. This publication makes use of data products from the Two Micron All Sky Survey, which is a joint project of the University of Massachusetts and the Infrared Processing and Analysis Center/California Institute of Technology, funded by the NASA and NSF.

\clearpage
\bibliographystyle{apj}            
\bibliography{apj-jour,paper}      

	\newpage

\begin{landscape}

\begin{deluxetable}{rrrrccccccccc}

\tabletypesize{\tiny}
\tablewidth{0pt}
\tablecaption{Target Sample of A, F, and G Dwarfs\label{tab:sample}}
\tablehead{
\colhead{\textbf{ }} &
\colhead{\textbf{ }} &
\colhead{\textbf{ }} &
\colhead{\textbf{Other}} &
\colhead{\textbf{RA}} &
\colhead{\textbf{DEC}} &
\colhead{\textbf{Spectral}} &
\colhead{\textbf{Spectral}} &
\colhead{\textbf{$V$}} &
\colhead{\textbf{$K$}} &
\colhead{\textbf{($B-V$)}}	& 
\colhead{\textbf{ }}	& 
\colhead{\textbf{$\pi \pm \sigma$ \tablenotemark{e}}} \\
\colhead{\textbf{HD}} &
\colhead{\textbf{HR}} &
\colhead{\textbf{HIP}} &
\colhead{\textbf{Name\tablenotemark{a}}} &
\colhead{\textbf{(hh mm ss.xx)}}	&
\colhead{\textbf{(dd mm ss)}}	&
\colhead{\textbf{Type\tablenotemark{b}}}	&
\colhead{\textbf{Type\tablenotemark{c}}}	&
\colhead{\textbf{(mag)}}	&
\colhead{\textbf{(mag)}}	&
\colhead{\textbf{(mag)}}	&
\colhead{\textbf{[Fe/H]\tablenotemark{d}}}	& 
\colhead{\textbf{(mas)}} 
}
\startdata
4614	&	219	&	3821	&	24 $\eta$~Cas~A	&	00 49 06.29	&	57 48 54.67	&	F9V	&	G0V	&	3.46	&	2.05	&	0.587	&	$-$0.30\phs	&	168.01$\pm$0.48\phn	\\
5015	&	244	&	4151	&	GJ 41	&	00 53 04.20	&	61 07 26.29	&	F8V	&	F8V	&	4.80	&	3.54	&	0.540	&	0.00	&	53.35$\pm$0.33	\\
6582	&	321	&	5336	&	34 $\mu$~Cas~A	&	01 08 16.39	&	54 55 13.22	&	G5Vb	&	G5Vp	&	5.17	&	3.36	&	0.704	&	$-$0.83\phs	&	132.40$\pm$0.60	\\
10780	&	511	&	8362	&	GJ 75	&	01 47 44.84	&	63 51 09.00	&	G9V	&	K0V	&	5.63	&	3.84	&	0.804	&	0.05	&	99.34$\pm$0.53	\\
16895	&	799	&	12777	&	13 $\theta$~Per~A	&	02 44 11.99	&	49 13 42.41	&	F7V	&	F7V	&	4.10	&	2.78	&	0.514	&	$-$0.12\phs	&	89.88$\pm$0.23	\\
19373	&	937	&	14632	&	$\iota$~Per	&	03 09 04.02	&	49 36 47.80	&	G0IV-V	&	G0V 	&	4.05	&	2.70	&	0.595	&	0.09	&	94.87$\pm$0.23	\\
20630	&	996	&	15457	&	$\kappa$~Cet	&	03 19 21.70	&	03 22 12.71	&	G5V 	&	G5Vvar	&	4.84	&	3.34	&	0.681	&	0.00	&	109.39$\pm$0.27\phn	\\
22484	&	1101	&	16852	&	10 Tau	&	03 36 52.38	&	00 24 05.98	&	F9IV-V	&	F9V	&	4.29	&	2.93	&	0.575	&	$-$0.09\phs	&	71.60$\pm$0.54	\\
30652	&	1543	&	22449	&	1 $\pi^{3}$~Ori	&	04 49 50.41	&	06 57 40.59	&	F6IV-V	&	F6V	&	3.19	&	2.07	&	0.484	&	$-$0.03\phs	&	123.94$\pm$0.17\phn	\\
34411	&	1729	&	24813	&	15 $\lambda$~Aur	&	05 19 08.47	&	40 05 56.59	&	G1V	&	G0V	&	4.69	&	3.24	&	0.630	&	0.05	&	79.18$\pm$0.28	\\
39587	&	2047	&	27913	&	54 $\chi^{1}$~Ori	&	05 54 22.98	&	20 16 34.23	&	G0IV-V	&	G0V	&	4.39	&	2.97	&	0.594	&	$-$0.16\phs	&	115.42$\pm$0.27\phn	\\
48737	&	2484	&	32362	&	31 $\xi$~Gem	&	06 45 17.37	&	12 53 44.13	&	F5IV-V	&	F5IV	&	3.35	&	2.30	&	0.443	&	0.01	&	55.55$\pm$0.19	\\
56537	&	2763	&	35350	&	54 $\lambda$ Gem	&	07 18 05.58	&	16 32 25.38	&	\nodata	&	A3V	&	3.58	&	3.27	&	0.106	&	\nodata	&	32.36$\pm$0.22	\\
58946	&	2852	&	36366	&	62 $\rho$ Gem	&	07 29 06.72	&	31 47 04.38	&	\nodata	&	F0V	&	4.16	&	3.32	&	0.320	&	$-$0.31\phs	&	55.41$\pm$0.25	\\
81937	&	3757	&	46733	&	23 h UMa	&	09 31 31.71	&	63 03 42.70	&	\nodata	&	F0IV	&	3.65	&	2.82	&	0.360	&	0.06	&	41.99$\pm$0.16	\\
82328	&	3775	&	46853	&	25 $\theta$~UMa	&	09 32 51.43	&	51 40 38.28	&	F5.5IV-V	&	F6IV	&	3.17	&	2.02	&	0.475	&	$-$0.12\phs	&	74.18$\pm$0.13	\\
82885	&	3815	&	47080	&	11 LMi	&	09 35 39.50	&	35 48 36.48	&	G8+V	&	G8IV-V	&	5.40	&	3.70	&	0.770	&	0.06	&	87.96$\pm$0.32	\\
86728	&	3951	&	49081	&	20 LMi	&	10 01 00.66	&	31 55 25.22	&	G4V	&	G3V	&	5.37	&	3.82	&	0.676	&	0.20	&	66.47$\pm$0.32	\\
90839	&	4112	&	51459	&	36 UMa	&	10 30 37.58	&	55 58 49.93	&	F8V	&	F8V	&	4.82	&	3.54	&	0.541	&	$-$0.16\phs	&	78.26$\pm$0.29	\\
95418	&	4295	&	53910	&	48 $\beta$ UMa	&	11 01 50.48	&	56 22 56.74	&	A1IV	&	A1V	&	2.34	&	2.35	&	0.033	&	0.06	&	40.89$\pm$0.16	\\
97603	&	4357	&	54872	&	68 $\delta$ Leo	&	11 14 06.50	&	20 31 25.38	&	A5IV(n)	&	A4V	&	2.56	&	2.26	&	0.128	&	0.00	&	55.82$\pm$0.25	\\
101501	&	4496	&	56997	&	61 UMa	&	11 41 03.02	&	34 12 05.89	&	G8V	&	G8V	&	5.31	&	3.53	&	0.723	&	$-$0.12\phs	&	104.03$\pm$0.26\phn	\\
102870	&	4540	&	57757	&	5 $\beta$ Vir	&	11 50 41.72	&	01 45 52.98	&	F8.5IV-V	&	F8V	&	3.59	&	2.33	&	0.518	&	0.11	&	91.50$\pm$0.22	\\
103095	&	4550	&	57939	&	CF UMa	&	11 52 58.77	&	37 43 07.24	&	K1V	&	G8Vp	&	6.42	&	4.38	&	0.754	&	$-$1.36\phs	&	109.98$\pm$0.41\phn	\\
109358	&	4785	&	61317	&	8 $\beta$ CVn	&	12 33 44.55	&	41 21 26.93	&	G0V	&	G0V	&	4.24	&	2.72	&	0.588	&	$-$0.30\phs	&	118.49$\pm$0.20\phn	\\
114710	&	4983	&	64394	&	43 $\beta$ Com	&	13 11 52.39	&	27 52 41.46	&	G0V	&	G0V	&	4.23	&	2.90	&	0.572	&	$-$0.06\phs	&	109.53$\pm$0.17\phn	\\
118098	&	5107	&	66249	&	79 $\zeta$ Vir	&	13 34 41.59	&	$-$00 35 44.95 \phs	&	A2Van	&	A3V	&	3.38	&	3.11	&	0.114	&	 $-$0.02\phs	&	44.01$\pm$0.19	\\
126660	&	5404	&	70497	&	23 $\theta$ Boo	&	14 25 11.80	&	51 51 02.68	&	F7V	&	F7V	&	4.04	&	2.78	&	0.497	&	$-$0.14\phs	&	68.83$\pm$0.14	\\
128167	&	5447	&	71284	&	28 $\sigma$ Boo	&	14 34 40.82	&	29 44 42.47	&	F4VkF2mF1	&	F3V	&	4.47	&	3.52	&	0.364	&	$-$0.36\phs	&	63.16$\pm$0.26	\\
131156	&	5544	&	72659	&	37 $\xi$ Boo	&	14 51 23.38	&	19 06 01.66	&	G7V	&	G8V	&	4.54	&	2.96	&	0.720	&	 $-$0.33\phs	&	149.03$\pm$0.48\phn	\\
141795	&	5892	&	77622	&	37 $\epsilon$ Ser	&	15 50 48.97	&	04 28 39.83	&	kA2hA5mA7V	&	A2m	&	3.71	&	3.43	&	0.147	&	\nodata	&	46.28$\pm$0.19	\\
142860	&	5933	&	78072	&	41 $\gamma$ Ser	&	15 56 27.18	&	15 39 41.82	&	F6V	&	F6V	&	3.85	&	2.66	&	0.478	&	$-$0.19\phs	&	88.85$\pm$0.18	\\
146233	&	6060	&	79672	&	18 Sco	&	16 15 37.27	&	$-$08 22 09.99 \phs	&	G2V	&	G1V	&	5.49	&	3.55	&	0.652	&	$-$0.02\phs	&	71.93$\pm$0.37	\\
162003	&	6636	&	86614	&	31 $\psi$ Dra	&	17 41 56.36	&	72 08 55.84	&	F5IV-V	&	F5IV-V	&	4.57	&	3.43	&	0.434	&	$-$0.17\phs	&	43.79$\pm$0.45	\\
164259	&	6710	&	88175	&	57 $\zeta$ Ser	&	18 00 29.01	&	$-$03 41 24.97 \phs	&	F2V	&	F3V	&	4.62	&	3.66	&	0.390	&	$-$0.14\phs	&	42.44$\pm$0.33	\\
173667	&	7061	&	92043	&	110 Her	&	18 45 39.73	&	20 32 46.71	&	F5.5IV-V	&	F6V	&	4.19	&	2.89	&	0.483	&	$-$0.15\phs	&	52.06$\pm$0.24	\\
177724	&	7235	&	93747	&	17 $\zeta$ Aql	&	19 05 24.61	&	13 51 48.52	&	A0IV-Vnn	&	A0Vn	&	2.99	&	2.92	&	0.014	&	 $-$0.68\phs	&	39.27$\pm$0.17	\\
182572	&	7373	&	95447	&	31 b Aql	&	19 24 58.20	&	11 56 39.90	&	\nodata	&	G8IV	&	5.17	&	3.53	&	0.761	&	0.33	&	65.89$\pm$0.26	\\
185144	&	7462	&	96100	&	61 $\sigma$ Dra	&	19 32 21.59	&	69 39 40.23	&	G9V	&	K0V	&	4.67	&	2.78	&	0.786	&	$-$0.24\phs	&	173.77$\pm$0.18\phn	\\
185395	&	7469	&	96441	&	13 $\theta$ Cyg	&	19 36 26.54	&	50 13 15.97	&	F3+V	&	F4V	&	4.49	&	3.52	&	0.395	&	$-$0.04\phs	&	54.55$\pm$0.15	\\
210418	&	8450	&	109427	&	26 $\theta$ Peg	&	22 10 11.99	&	06 11 52.31	&	\nodata	&	A2V	&	3.52	&	3.22	&	0.086	&	 $-$0.38\phs	&	35.34$\pm$0.85	\\
213558	&	8585	&	111169	&	7 $\alpha$ Lac	&	22 31 17.50	&	50 16 56.97	&	\nodata	&	A1V	&	3.76	&	3.75	&	0.031	&	\nodata	&	31.80$\pm$0.12	\\
215648	&	8665	&	112447	&	46 $\xi$ Peg	&	22 46 41.58	&	12 10 22.40	&	F6V	&	F7V	&	4.20	&	2.87	&	0.502	&	$-$0.24\phs	&	61.37$\pm$0.20	\\
222368	&	8969	&	116771	&	17 $\iota$ Psc	&	23 39 57.04	&	05 37 34.65	&	F7V	&	F7V	&	4.13	&	2.89	&	0.507	&	$-$0.08\phs	&	72.91$\pm$0.15	\\

\enddata

\tablenotetext{a}{Bayer-Flamsteed or GJ \citep{kos04}}
\tablenotetext{b}{\citet{gra01,gra03}}
\tablenotetext{c}{{\it SIMBAD} \citep{wen00}}
\tablenotetext{d}{\citet{hol07}, when available. For stars without metallicity estimates from \citet{hol07}, \\
	the [M/H] values from \citet{gra03, gra06} (HD~82885, HD~97603, HD~118098, HD~131156, HD~177724, HD 210418), \\
	and \citet{tak05} (HD~182572) are used.  \\
	Stars with no metallicity measurements are HD~56537, HD~141795, HD~213558.}	
\tablenotetext{e}{\citet{van07}}  

\end{deluxetable}

\end{landscape}
\newpage
\begin{landscape}
\begin{deluxetable}{rcccccccccccc}
\tabletypesize{\tiny}
\tablewidth{0pt}
\tablecaption{Calibrators Observed\label{tab:calibs}}
\tablehead{
\colhead{\textbf{Calibrator}} &
\colhead{\textbf{$V$}} &
\colhead{\textbf{$B-V$}} &
\colhead{\textbf{$U-B$}} &
\colhead{\textbf{$v$}} &
\colhead{\textbf{$b-y$}} &
\colhead{\textbf{$m1$}} &
\colhead{\textbf{$c1$}} &
\colhead{\textbf{$J$}} &
\colhead{\textbf{$H$}} &
\colhead{\textbf{$K$}} &
\colhead{\textbf{$\theta_{\rm SED} \pm \sigma$}}	& 
\colhead{\textbf{Target (s)}}	 \\
\colhead{\textbf{HD}} &
\colhead{\textbf{(mag)}}	&
\colhead{\textbf{(mag)}}	&
\colhead{\textbf{(mag)}}	&
\colhead{\textbf{(mag)}}	&
\colhead{\textbf{(mag)}}	&
\colhead{\textbf{(mag)}}	&
\colhead{\textbf{(mag)}}	&
\colhead{\textbf{(mag)}}	&
\colhead{\textbf{(mag)}}	&
\colhead{\textbf{(mag)}}	&
\colhead{\textbf{(mas)}}	&
\colhead{\textbf{HD}}	 
}
\startdata

71	&	6.990	&	1.187	&	\nodata	&	\nodata	&	\nodata	&	\nodata	&	\nodata	&	5.193	&	4.459	&	4.214	&	0.682 $\pm$ 0.024	&	4614	\\
6210	&	5.838	&	0.547	&	0.110	&	5.800	&	0.356	&	0.183	&	0.475	&	4.755	&	4.794	&	4.445	&	0.519  $\pm$  0.012	&	4614, 5015, 6582, 10780	\\
9407	&	6.530	&	0.684	&	0.236	&	\nodata	&	\nodata	&	\nodata	&	\nodata	&	5.296	&	4.941	&	4.888	&	0.430  $\pm$  0.017	&	4614	\\
20675	&	5.932	&	0.441	&	$-$0.015\phs	&	5.950	&	0.293	&	0.167	&	0.495	&	5.274	&	4.922	&	4.875	&	0.415  $\pm$  0.012	&	16895, 19373	\\
21790	&	4.727	&	$-$0.093\phs	&	$-$0.260\phs	&	4.738	&	$-$0.036\phs	&	0.107	&	0.833	&	5.282	&	4.960	&	4.886	&	0.308  $\pm$  0.009	&	20630, 22484	\\
22879	&	6.689	&	0.540	&	$-$0.086\phs	&	6.693	&	0.365	&	0.126	&	0.272	&	5.588	&	5.301	&	5.179	&	0.342  $\pm$  0.021	&	20630, 22484	\\
28355	&	5.025	&	0.218	&	0.120	&	5.023	&	0.115	&	0.225	&	0.909	&	4.793	&	4.656	&	4.534	&	0.425  $\pm$  0.030	&	30652	\\
30739	&	4.355	&	0.009	&	$-$0.016\phs	&	4.370	&	0.010	&	0.153	&	1.107	&	3.825	&	4.208	&	4.166	&	0.461  $\pm$  0.018	&	30652	\\
31295	&	4.644	&	0.085	&	0.095	&	4.661	&	0.044	&	0.178	&	1.007	&	4.846	&	4.517	&	4.416	&	0.439  $\pm$  0.043	&	30652	\\
34904	&	5.540	&	0.120	&	0.120	&	5.540	&	0.084	&	0.163	&	1.101	&	5.144	&	5.124	&	5.112	&	0.345  $\pm$  0.013	&	34411	\\
38558	&	5.527	&	0.274	&	0.249	&	5.400	&	0.196	&	0.148	&	1.176	&	4.944	&	4.746	&	4.483	&	0.422  $\pm$  0.008	&	39587	\\
42807	&	6.442	&	0.660	&	0.160	&	6.440	&	0.415	&	0.228	&	0.292	&	5.253	&	5.010	&	4.849	&	0.429  $\pm$  0.016	&	48737	\\
43042	&	5.201	&	0.434	&	$-$0.011\phs	&	5.207	&	0.291	&	0.166	&	0.443	&	4.008	&	3.827	&	4.129	&	0.591  $\pm$  0.030	&	39587	\\
43795	&	7.640	&	0.958	&	\nodata	&	7.645	&	0.593	&	0.293	&	0.483	&	5.940	&	5.522	&	5.409	&	0.376  $\pm$  0.008	&	48682	\\
50277	&	5.765	&	0.265	&	0.089	&	5.764	&	0.154	&	0.184	&	0.872	&	5.308	&	5.232	&	5.088	&	0.346  $\pm$  0.011 	&	48737	\\
50973	&	4.897	&	0.029	&	0.053	&	4.917	&	0.013	&	0.159	&	1.107	&	5.055	&	4.941	&	4.793	&	0.361 $\pm$ 0.026	&	48682	\\
58551	&	6.544	&	0.460	&	0.000	&	6.539	&	0.322	&	0.129	&	0.355	&	5.534	&	5.380	&	5.245	&	0.357  $\pm$  0.009	&	56537	\\
59037	&	5.011	&	0.112	&	0.119	&	5.084	&	0.063	&	0.201	&	1.015	&	4.818	&	4.793	&	4.744	&	0.389  $\pm$  0.018	&	58946	\\
65583	&	6.999	&	0.713	&	0.181	&	6.975	&	0.450	&	0.232	&	0.231	&	5.539	&	5.170	&	5.095	&	0.406  $\pm$  0.033	&	58946	\\
79439	&	4.832	&	0.186	&	0.087	&	0.113	&	0.196	&	0.892	&	2.833	&	4.481	&	4.353	&	4.291	&	0.482 $\pm$ 0.035	&	82328	\\
80290	&	6.160	&	0.420	&	$-$0.110\phs	&	6.132	&	0.300	&	0.139	&	0.392	&	5.219	&	5.067	&	4.972	&	0.385 $\pm$ 0.016	&	82328	\\
83951	&	6.140	&	0.360	&	0.000	&	6.000	&	0.244	&	0.162	&	0.594	&	5.355	&	5.246	&	5.169	&	0.360  $\pm$  0.006	&	82885, 86728	\\
87141	&	5.749	&	0.476	&	0.043	&	5.700	&	0.318	&	0.169	&	0.478	&	4.987	&	4.730	&	4.503	&	0.476  $\pm$  0.022	&	81937, 82328, 95418	\\
88986	&	6.460	&	0.600	&	0.160	&	6.440	&	0.397	&	0.209	&	0.363	&	5.247	&	4.946	&	4.884	&	0.432  $\pm$  0.013	&	82885, 86728	\\
89389	&	6.450	&	0.540	&	0.080	&	0.369	&	0.181	&	0.370	&	2.602	&	5.340	&	5.091	&	5.020	&	0.398  $\pm$  0.013	&	81937, 82328, 90839, 95418	\\
91480	&	5.159	&	0.335	&	$-$0.020\phs	&	5.140	&	0.228	&	0.159	&	0.574	&	4.922	&	4.688	&	4.334	&	0.518  $\pm$  0.014	&	81937, 90839, 95418	\\
99285	&	5.610	&	0.345	&	$-$0.017\phs	&	5.640	&	0.251	&	0.154	&	0.595	&	4.815	&	4.723	&	4.624	&	0.456  $\pm$  0.017	&	97603	\\
99984	&	5.964	&	0.493	&	$-$0.025\phs	&	5.800	&	0.340	&	0.148	&	0.429	&	4.900	&	4.657	&	4.591	&	0.483  $\pm$  0.020	&	103095	\\
102124	&	4.838	&	0.176	&	0.091	&	4.858	&	0.090	&	0.196	&	0.926	&	4.634	&	4.542	&	4.409	&	0.466  $\pm$  0.022	&	102870	\\
102634	&	6.145	&	0.520	&	0.069	&	6.153	&	0.329	&	0.176	&	0.439	&	5.212	&	5.081	&	4.921	&	0.404  $\pm$  0.010	&	102870	\\
103799	&	6.622	&	0.469	&	$-$0.026\phs	&	0.326	&	0.139	&	0.422	&	2.618	&	5.594	&	5.386	&	5.338	&	0.343  $\pm$  0.013	&	101501, 103095, 109358	\\
110897	&	5.956	&	0.548	&	$-$0.044\phs	&	5.958	&	0.374	&	0.147	&	0.284	&	5.173	&	4.667	&	4.465	&	0.492  $\pm$  0.022	&	109358	\\
114093	&	6.830	&	0.910	&	0.000	&	\nodata	&	\nodata	&	\nodata	&	\nodata	&	5.115	&	4.739	&	4.564	&	0.572  $\pm$  0.014	&	114710	\\
116831	&	5.956	&	0.187	&	0.125	&	5.973	&	0.095	&	0.206	&	0.972	&	5.668	&	5.577	&	5.531	&	0.278 $\pm$ 0.020	&	118098	\\
120066	&	6.329	&	0.621	&	0.151	&	6.329	&	0.399	&	0.188	&	0.397	&	5.212	&	4.997	&	4.851	&	0.428  $\pm$  0.013	&	118098	\\
128093	&	6.332	&	0.397	&	$-$0.022\phs	&	6.200	&	0.300	&	0.131	&	0.476	&	5.460	&	5.287	&	5.222	&	0.351  $\pm$  0.011	&	128167	\\
129153	&	5.915	&	0.218	&	0.045	&	0.131	&	0.206	&	0.813	&	2.816	&	5.447	&	5.402	&	5.365	&	0.309  $\pm$  0.010	&	131156	\\
132254	&	5.639	&	0.496	&	$-$0.003\phs	&	5.600	&	0.338	&	0.174	&	0.410	&	4.685	&	4.464	&	4.408	&	0.520  $\pm$  0.015	&	126660	\\
135101	&	6.689	&	0.680	&	0.260	&	6.685	&	0.433	&	0.220	&	0.368	&	5.403	&	5.090	&	5.030	&	0.409  $\pm$  0.014	&	131156	\\
139225	&	5.950	&	0.280	&	0.020	&	5.800	&	0.222	&	0.160	&	0.681	&	5.175	&	5.099	&	5.023	&	0.380  $\pm$  0.122	&	142860	\\
140775	&	5.571	&	0.035	&	0.059	&	5.568	&	0.024	&	0.150	&	1.107	&	5.466	&	5.463	&	5.428	&	0.275  $\pm$  0.013	&	141795	\\
145607	&	5.420	&	0.120	&	0.120	&	5.443	&	0.059	&	0.172	&	1.083	&	5.170	&	5.307	&	5.052	&	0.325  $\pm$  0.020	&	146233	\\
150177	&	6.390	&	0.490	&	$-$0.100\phs	&	6.333	&	0.334	&	0.119	&	0.395	&	5.353	&	5.064	&	4.977	&	0.391  $\pm$  0.019	&	146233	\\
154099	&	6.300	&	0.240	&	0.110	&	6.308	&	0.158	&	0.180	&	0.944	&	5.706	&	5.633	&	5.604	&	0.283  $\pm$  0.005	&	162003	\\
158352	&	5.418	&	0.227	&	0.094	&	5.420	&	0.148	&	0.183	&	0.923	&	4.813	&	4.883	&	4.805	&	0.407  $\pm$  0.013	&	164259	\\
162004	&	5.808	&	0.531	&	0.032	&	5.780	&	0.346	&	0.160	&	0.379	&	5.001	&	4.590	&	4.527	&	0.498  $\pm$  0.015	&	162003	\\
167564	&	6.350	&	0.200	&	0.150	&	6.354	&	0.123	&	0.158	&	1.148	&	5.891	&	5.791	&	5.750	&	0.259  $\pm$  0.004	&	165259	\\
174897	&	6.550	&	1.050	&	0.850	&	\nodata	&	\nodata	&	\nodata	&	\nodata	&	4.797	&	4.384	&	4.096	&	0.652  $\pm$  0.038	&	182572	\\
176303	&	5.239	&	0.530	&	0.066	&	5.267	&	0.356	&	0.168	&	0.452	&	4.324	&	4.039	&	3.930	&	0.659  $\pm$  0.016	&	173667, 177724, 182572	\\
180317	&	5.640	&	0.110	&	0.000	&	5.600	&	0.068	&	0.184	&	1.072	&	5.330	&	5.381	&	5.302	&	0.309  $\pm$  0.007	&	173667, 177724	\\
183534	&	5.750	&	0.000	&	$-$0.020\phs	&	5.750	&	$-$0.003\phs	&	0.157	&	1.023	&	5.632	&	5.690	&	5.674	&	0.241  $\pm$  0.012	&	185395	\\
191195	&	5.826	&	0.418	&	$-$0.030\phs	&	5.817	&	0.284	&	0.153	&	0.506	&	5.239	&	4.834	&	4.766	&	0.432  $\pm$  0.014	&	185395	\\
193664	&	5.919	&	0.585	&	0.058	&	5.922	&	0.382	&	0.180	&	0.323	&	4.879	&	4.690	&	4.451	&	0.494  $\pm$  0.019	&	185144	\\
204485	&	5.797	&	0.304	&	0.009	&	5.700	&	0.200	&	0.198	&	0.648	&	5.156	&	5.020	&	4.955	&	0.381  $\pm$  0.011	&	201091, 201092	\\
210715	&	5.393	&	0.154	&	0.069	&	5.400	&	0.076	&	0.200	&	0.967	&	5.013	&	5.016	&	4.959	&	0.366  $\pm$  0.015	&	213558	\\
211976	&	6.178	&	0.450	&	$-$0.052\phs	&	6.183	&	0.300	&	0.148	&	0.423	&	5.323	&	5.160	&	5.050	&	0.373  $\pm$  0.013	&	210418, 215648	\\
214923	&	3.406	&	$-$0.086\phs	&	$-$0.217\phs	&	3.406	&	$-$0.035\phs	&	0.113	&	0.868	&	3.538	&	3.527	&	3.566	&	0.611  $\pm$  0.029	&	215648	\\
216735	&	4.906	&	$-$0.002\phs	&	0.003	&	4.915	&	$-$0.006\phs	&	0.159	&	1.083	&	5.222	&	5.012	&	4.840	&	0.321  $\pm$  0.022	&	215648, 222368	\\
218470	&	5.680	&	0.417	&	$-$0.039\phs	&	5.687	&	0.290	&	0.146	&	0.486	&	4.819	&	4.670	&	4.649	&	0.462  $\pm$  0.014	&	213558	\\
222603	&	4.502	&	0.202	&	0.078	&	4.500	&	0.105	&	0.203	&	0.891	&	4.372	&	4.204	&	4.064	&	0.577  $\pm$  0.032	&	222368	\\
225003	&	5.704	&	0.329	&	$-$0.007\phs	&	5.699	&	0.209	&	0.155	&	0.645	&	5.077	&	5.008	&	4.910	&	0.386  $\pm$  0.017	&	222368	\\

\enddata
\tablecomments{Johnson $UBV$ \citep{mer97}, Stromgren $uvby$ \citep{hau98}, and 2MASS $JHK$ \citep{cut03} magnitudes for the calibrator stars. Refer to Section~\ref{sec:data_calibration} for details.}
\end{deluxetable}
\end{landscape}
\newpage
\begin{deluxetable}{rcccc}
\tabletypesize{\tiny}
\tablewidth{0pt}
\tablecaption{Observation Log\label{tab:observations}}
\tablehead{
\colhead{\textbf{Object}} &
\colhead{\textbf{UT}} &
\colhead{\textbf{ }} &
\colhead{\textbf{\# of}} &
\colhead{\textbf{Calibrator}}  \\
\colhead{\textbf{HD}} &
\colhead{\textbf{Date}} &
\colhead{\textbf{Baseline}} &
\colhead{\textbf{Brackets}} &
\colhead{\textbf{HD}}	
}

\startdata

4614	&	2007/06/29	&	W1/E1	&	2	&	6210	\\ 
	&	2007/06/30	&	W1/E1	&	5	&	6210	\\ 
	&	2007/07/01	&	W1/E1	&	3	&	6210	\\ 
	&	2007/07/18	&	S1/E1	&	3	&	6210	\\ 
	&	2007/07/19	&	S1/E1	&	3	&	6210	\\ 
	&	2007/11/16	&	S1/E1	&	4	&	6210	\\ 
	&	2008/10/02	&	W1/E1	&	4	&	6210, 9407	\\
	&	2009/11/21	&	S1/E1	&	3	&	71	\\ \\
5015	&	2007/10/10	&	W1/E1	&	10	&	6210	\\ 
	&	2007/11/03	&	W1/E1	&	7	&	6210	\\ 
	&	2007/11/17	&	S1/E1	&	8	&	6210	\\ \\
6582	&	2007/07/01	&	W1/E1	&	3	&	6210	\\ 
	&	2007/07/17	&	S1/E1	&	6	&	6210	\\ 
	&	2007/07/18	&	S1/E1	&	8	&	6210	\\ 
	&	2007/09/08	&	S1/E1	&	10	&	6210	\\ \\
10780	&	2007/06/29	&	W1/E1	&	2	&	6210	\\ 
	&	2007/07/19	&	S1/E1	&	10	&	6210	\\ 
	&	2007/10/10	&	W1/E1	&	10	&	6210	\\ \\
16895	&	2007/09/08	&	S1/E1	&	7	&	20675	\\ 
	&	2007/11/03	&	W1/E1	&	8	&	20675	\\ 
	&	2007/12/24	&	S1/E1	&	6	&	20675	\\ \\
19373	&	2007/01/25	&	S1/E1	&	8	&	20675	\\ 
	&	2007/08/28	&	W1/S1	&	2	&	20675	\\ 
	&	2007/09/08	&	S1/E1	&	10	&	20675	\\ 
	&	2007/11/04	&	W1/E1	&	6	&	20675	\\ \\
20630	&	2007/09/09	&	S1/E1	&	9	&	21790	\\ 
	&	2008/10/01	&	S1/E1	&	4	&	22879	\\ 
	&	2008/11/17	&	S1/E1	&	5	&	22879	\\ 
	&	2008/11/18	&	S1/E1	&	5	&	21790, 22879	\\ \\
22484	&	2006/12/05	&	S1/E1	&	2	&	21790	\\ 
	&	2006/12/07	&	S1/E1	&	3	&	21790	\\ 
	&	2007/09/09	&	S1/E1	&	8	&	21790	\\ 
	&	2008/10/01	&	S1/E1	&	6	&	22879	\\ 
	&	2008/10/02	&	W1/E1	&	4	&	22879	\\ \\
30652	&	2007/11/05	&	S1/E1	&	16	&	30739	\\ 
	&	2008/10/01	&	S1/E1	&	10	&	28355, 31295	\\ 
	&	2008/10/02	&	W1/E1	&	3	&	31295	\\ \\
34411	&	2007/01/26	&	S1/E1	&	5	&	34904	\\ 
	&	2007/11/03	&	W1/E1	&	8	&	34904	\\ 
	&	2007/11/15	&	S1/E1	&	4	&	34904	\\ 
	&	2007/11/17	&	S1/E1	&	7	&	34904	\\ \\
39587	&	2006/12/07	&	S1/E1	&	3	&	38558	\\ 
	&	2007/03/06	&	S1/E1	&	8	&	38558	\\ 
	&	2008/11/18	&	S1/E1	&	11	&	38558, 43042	\\ \\
48737	&	2006/12/07	&	S1/E1	&	4	&	50277	\\ 
	&	2008/11/17	&	S1/E1	&	12	&	42807, 50277	\\ 
	&	2008/11/18	&	S1/E1	&	11	&	42807, 50277	\\ \\
56537	&	2007/02/21	&	S1/E1	&	1	&	58551	\\ 
	&	2007/02/25	&	S1/E1	&	7	&	58551	\\ 
	&	2007/03/11	&	S1/E1	&	6	&	58551	\\ 
	&	2007/11/04	&	S1/E1	&	5	&	58551	\\ 
	&	2007/12/23	&	S1/E1	&	5	&	58551	\\ \\
58946	&	2007/01/25	&	S1/E1	&	6	&	65583	\\ 
	&	2007/11/16	&	S1/E1	&	7	&	59037	\\ 
	&	2007/11/17	&	S1/E1	&	7	&	59037	\\ \\
81937	&	2007/11/29	&	S2/E2	&	9	&	91480	\\
	&	2009/11/20	&	S1/E1	&	12	&	87141, 89389, 91480	\\
	&	2009/11/21	&	S1/E1	&	4	&	87141	\\
	&	2009/11/22	&	S1/E1	&	2	&	91480	\\ \\
82328	&	2007/11/02	&	W2/E2	&	9	&	87141	\\
	&	2009/11/20	&	S1/E1	&	7	&	79439, 80290, 89389	\\
	&	2009/11/22	&	S1/E1	&	3	&	79439	\\ \\
82885	&	2007/02/03	&	S1/E1	&	2	&	83951	\\ 
	&	2007/11/03	&	W1/E1	&	7	&	83951	\\ 
	&	2007/11/07	&	S1/E1	&	9	&	83951	\\ 
	&	2007/12/24	&	S1/E1	&	5	&	83951	\\
	&	2009/11/21	&	S1/E1	&	3	&	88986	\\ \\
86728	&	2007/11/15	&	S1/E1	&	10	&	83951	\\ 
	&	2007/11/16	&	S1/E1	&	2	&	83951	\\ 
	&	2007/12/24	&	S1/E1	&	6	&	83951	\\ 
	&	2008/11/16	&	S1/E1	&	10	&	83951, 88986	\\
	&	2009/11/21	&	S1/E1	&	4	&	88986	\\ \\
90839	&	2007/11/16	&	S1/E1	&	10	&	89389	\\ 
	&	2008/04/17	&	W1/S1	&	5	&	89389, 91480	\\ \\
95418	&	2007/04/04	&	S1/E1	&	7	&	91480	\\ 
	&	2007/11/07	&	S1/E1	&	6	&	91480	\\ 
	&	2008/04/17	&	W1/S1	&	5	&	89389, 91480	\\ 
	&	2009/11/21	&	S1/E1	&	3	&	87141	\\ 
	&	2009/11/22	&	S1/E1	&	4	&	91480	\\ \\
97603	&	2007/02/21	&	S1/E1	&	10	&	99285	\\ 
	&	2007/03/10	&	S1/E1	&	2	&	99285	\\ 
	&	2007/03/11	&	S1/E1	&	5	&	99285	\\ \\
101501	&	2007/11/15	&	S1/E1	&	7	&	103799	\\ 
	&	2007/12/24	&	S1/E1	&	3	&	103799	\\ \\
102870	&	2007/03/09	&	S1/E1	&	6	&	102124	\\ 
	&	2007/12/23	&	S1/E1	&	4	&	102124	\\ 
	&	2008/04/19	&	W1/S1	&	8	&	102124	\\ 
	&	2008/04/22	&	S1/E1	&	9	&	102124	\\
	&	2008/04/23	&	S1/E1	&	7	&	102634	\\ \\
103095	&	2007/11/16	&	S1/E1	&	7	&	103799	\\ 
	&	2007/12/24	&	S1/E1	&	10	&	103799	\\ \\
109358	&	2007/05/26	&	S1/E2	&	3	&	110897	\\ 
	&	2008/04/18	&	W1/S1	&	5	&	103799, 110897	\\ \\
114710	&	2008/04/21	&	W1/S1	&	10	&	114093	\\ 
	&	2008/06/27	&	S1/E1	&	6	&	114093	\\ \\
118098	&	2007/03/10	&	S1/E1	&	6	&	120066	\\ 
	&	2007/03/30	&	S1/E1	&	5	&	120066	\\ 
	&	2007/12/23	&	S1/E1	&	2	&	120066	\\
	&	2010/04/10	&	S1/E1	&	4	&	116831, 120066	\\ \\
126660	&	2007/05/24	&	W1/S1	&	5	&	132254	\\ 
	&	2007/07/16	&	S1/E1	&	6	&	132254	\\ 
	&	2008/07/25	&	S1/E1	&	4	&	132254	\\ \\
128167	&	2008/06/28	&	S1/E1	&	5	&	128093	\\ 
	&	2008/07/06	&	S1/E1	&	12	&	128093	\\ 
	&	2008/07/24	&	S1/E2	&	10	&	128093	\\ \\
131156	&	2007/03/12	&	S1/E1	&	5	&	135101	\\ 
	&	2008/04/18	&	W1/S1	&	5	&	135101, 129153	\\ 
	&	2008/04/19	&	W1/S1	&	6	&	135101	\\ 
	&	2008/06/27	&	S1/E1	&	9	&	135101, 129153	\\ \\
141795	&	2008/07/22	&	S1/E1	&	8	&	140775	\\ \\
142860	&	2007/07/20	&	S1/E1	&	3	&	139225	\\ 
	&	2007/07/21	&	S1/E1	&	6	&	139225	\\ 
	&	2008/04/21	&	W1/S1	&	10	&	139225	\\ \\
146233	&	2008/04/19	&	W1/S1	&	11	&	145607, 150177	\\ 
	&	2008/04/21	&	W1/S1	&	6	&	145607, 150177	\\ 
	&	2008/04/22	&	S1/E1	&	9	&	145607, 150177	\\ 
	&	2008/04/23	&	S1/E1	&	6	&	145607, 150177	\\ 
	&	2008/05/16	&	W1/E2	&	4	&	150177	\\ \\
162003	&	2007/07/17	&	S1/E1	&	8	&	154099	\\ 
	&	2007/07/18	&	S1/E1	&	2	&	162004	\\ 
	&	2007/10/10	&	W1/E1	&	6	&	162004	\\ 
	&	2007/11/17	&	S1/E1	&	4	&	162004	\\ 
	&	2008/06/26	&	S1/E1	&	5	&	162004	\\ \\
164259	&	2008/04/22	&	S1/E1	&	6	&	167564, 158352	\\ 
	&	2008/04/23	&	S1/E1	&	3	&	158352	\\ 
	&	2008/06/20	&	W1/S1	&	3	&	158352	\\ 
	&	2008/06/28	&	S1/E1	&	5	&	158352	\\ 
	&	2008/07/27	&	W1/S1	&	6	&	158352	\\ \\
173667	&	2007/07/20	&	S1/E1	&	3	&	180317	\\ 
	&	2007/07/21	&	S1/E1	&	9	&	176303	\\ 
	&	2008/04/21	&	W1/S1	&	3	&	176303	\\ 
	&	2008/06/28	&	S1/E1	&	8	&	176303	\\ 
	&	2008/07/07	&	W1/S1	&	1	&	176303	\\ 
	&	2008/07/21	&	W1/S1	&	1	&	176303	\\ 
	&	2008/07/22	&	S1/E1	&	6	&	176303	\\ 
	&	2008/07/23	&	W1/E1	&	6	&	176303	\\ \\
177724	&	2008/06/28	&	S1/E1	&	10	&	176303	\\ 
	&	2008/07/07	&	W1/S1	&	5	&	176303	\\ 
	&	2008/07/21	&	W1/S1	&	4	&	176303	\\ 
	&	2008/07/22	&	S1/E1	&	6	&	176303	\\ 
	&	2008/07/23	&	W1/E1	&	6	&	176303	\\ 
	&	2008/10/01	&	S1/E1	&	4	&	176303	\\ \\
182572	&	2007/07/21	&	S1/E1	&	6	&	174897	\\ 
	&	2007/09/09	&	S1/E1	&	10	&	174897	\\ 
	&	2008/07/22	&	S1/E1	&	5	&	174897	\\ 
	&	2008/07/24	&	S1/E2	&	5	&	174897	\\ 
	&	2008/09/30	&	S1/E1	&	7	&	176303	\\ \\
185144	&	2007/05/24	&	W1/S1	&	3	&	193664	\\ 
	&	2007/05/25	&	W1/S1	&	4	&	193664	\\ 
	&	2007/06/28	&	W1/E1	&	1	&	193664	\\ 
	&	2007/06/29	&	W1/E1	&	4	&	193664	\\ 
	&	2007/06/30	&	W1/E1	&	1	&	193664	\\ 
	&	2007/07/01	&	W1/E1	&	2	&	193664	\\ \\
185395	&	2007/05/26	&	S1/E2	&	3	&	183534	\\ 
	&	2007/07/19	&	S1/E1	&	11	&	191195	\\ 
	&	2007/11/02	&	W1/E2	&	5	&	191195	\\ 
	&	2008/07/25	&	S1/E1	&	8	&	191195	\\ \\
201091	&	2007/06/30	&	W1/E1	&	3	&	204485	\\ \\
201092	&	2007/06/30	&	W1/E1	&	3	&	204485	\\ \\
210418	&	2008/06/28	&	S1/E1	&	6	&	211976	\\ 
	&	2008/07/22	&	S1/E1	&	9	&	211976	\\ 
	&	2008/07/24	&	S1/E2	&	4	&	211976	\\ 
	&	2008/10/01	&	S1/E1	&	3	&	211976	\\ \\
213558	&	2007/09/08	&	S1/E1	&	7	&	218470	\\ 
	&	2007/10/10	&	W1/E1	&	10	&	210715	\\ 
	&	2007/12/24	&	S1/E1	&	6	&	218470	\\ 
	&	2008/07/21	&	S1/E1	&	5	&	218470	\\ \\
215648	&	2007/07/16	&	S1/E1	&	4	&	211976	\\ 
	&	2007/07/21	&	S1/E1	&	14	&	214923	\\ 
	&	2008/07/24	&	S1/E2	&	5	&	214923	\\ 
	&	2008/09/30	&	S1/E1	&	4	&	211976	\\ 
	&	2008/10/01	&	S1/E1	&	8	&	211976, 216735	\\ \\
222368	&	2006/12/07	&	S1/E1	&	4	&	222603	\\ 
	&	2007/07/20	&	S1/E1	&	11	&	222603	\\ 
	&	2007/09/09	&	S1/E1	&	5	&	222603	\\ 
	&	2008/09/30	&	S1/E1	&	10	&	222603, 225003	\\ 
	&	2008/10/01	&	S1/E1	&	8	&	216735	

\enddata

\tablecomments{Refer to Section~\ref{sec:data_calibration} for details.}

\end{deluxetable}
\newpage
\begin{deluxetable}{crr}
\tabletypesize{\tiny}
\tablewidth{0pt}
\tablecaption{CHARA Baseline Configurations\label{tab:CHARAbaselines}}
\tablehead{
\colhead{\textbf{Telescope}} &
\colhead{\textbf{B}} &
\colhead{\textbf{$\psi$}} 	 \\
\colhead{\textbf{Pair}} &
\colhead{\textbf{(m)}}	&
\colhead{\textbf{($^{\circ}$)}}
}
\startdata

W2/E2  & 156.28 & 63.3   \\
S2/E2  & 248.13 & 17.7   \\
W1/E2  & 251.34 & 77.6   \\
W1/S1  & 278.50 & 320.9  \\
S1/E2  & 278.77 & 14.5   \\
W1/E1  & 313.54 & 253.2  \\
S1/E1  & 330.67 & 22.1   \\

\enddata

\end{deluxetable}
\newpage
\begin{deluxetable}{rccccc}
\tabletypesize{\tiny}
\tablewidth{0pt}
\tablecaption{Angular Diameters\label{tab:diameters}}
\tablehead{
\colhead{\textbf{Star}} &
\colhead{\textbf{\# of}} &
\colhead{\textbf{Reduced}} &
\colhead{\textbf{$\theta_{\rm UD} \pm \sigma$}} &
\colhead{\textbf{ }} &
\colhead{\textbf{$\theta_{\rm LD} \pm \sigma$}}  \\
\colhead{\textbf{HD}} &
\colhead{\textbf{Obs.}} &
\colhead{\textbf{$\chi^2$}} &
\colhead{\textbf{(mas)}} &
\colhead{\textbf{$\mu_{K}$}}	&
\colhead{\textbf{(mas)}}	
}
\startdata

4614	&	27	&	1.43	&	$1.578\pm0.004$	&	0.280	&	$1.623\pm0.004$	\\
5015	&	22	&	0.45	&	$0.846\pm0.010$	&	0.270	&	$0.865\pm0.010$	\\
6582	&	26	&	0.96	&	$0.947\pm0.009$	&	0.320	&	$0.972\pm0.009$	\\
10780	&	22	&	0.78	&	$0.744\pm0.018$	&	0.310	&	$0.763\pm0.019$	\\
16895	&	21	&	1.02	&	$1.078\pm0.008$	&	0.270	&	$1.103\pm0.009$	\\
19373	&	22	&	0.90	&	$1.217\pm0.007$	&	0.270	&	$1.246\pm0.008$	\\
20630	&	21	&	1.15	&	$0.914\pm0.024$	&	0.290	&	$0.936\pm0.025$	\\
22484	&	23	&	1.05	&	$1.056\pm0.014$	&	0.280	&	$1.081\pm0.014$	\\
30652	&	34	&	0.60	&	$1.488\pm0.004$	&	0.260	&	$1.526\pm0.004$	\\
34411	&	18	&	1.07	&	$0.958\pm0.015$	&	0.290	&	$0.981\pm0.015$	\\
39587	&	17	&	0.33	&	$1.027\pm0.009$	&	0.280	&	$1.051\pm0.009$	\\
48737	&	24	&	1.49	&	$1.369\pm0.009$	&	0.180	&	$1.401\pm0.009$	\\
56537	&	20	&	0.52	&	$0.824\pm0.013$	&	0.250	&	$0.835\pm0.013$	\\
58946	&	15	&	0.51	&	$0.837\pm0.013$	&	0.230	&	$0.853\pm0.014$	\\
81937	&	18	&	0.34	&	$1.113\pm0.009$	&	0.260	&	$1.133\pm0.009$	\\
82328	&	19	&	0.54	&	$1.591\pm0.005$	&	0.320	&	$1.632\pm0.005$	\\
82885	&	25	&	0.45	&	$0.800\pm0.012$	&	0.290	&	$0.821\pm0.013$	\\
86728	&	28	&	0.62	&	$0.753\pm0.012$	&	0.260	&	$0.771\pm0.012$	\\
90839	&	19	&	0.38	&	$0.778\pm0.014$	&	0.180	&	$0.794\pm0.014$	\\
95418	&	29	&	1.41	&	$1.133\pm0.014$	&	0.210	&	$1.149\pm0.014$	\\
97603	&	16	&	1.59	&	$1.304\pm0.008$	&	0.310	&	$1.328\pm0.009$	\\
101501	&	10	&	0.22	&	$0.887\pm0.009$	&	0.270	&	$0.910\pm0.009$	\\
102870	&	32	&	0.54	&	$1.396\pm0.006$	&	0.320	&	$1.431\pm0.006$	\\
103095	&	16	&	0.08	&	$0.679\pm0.005$	&	0.280	&	$0.696\pm0.005$	\\
109358	&	12	&	1.97	&	$1.209\pm0.030$	&	0.280	&	$1.238\pm0.030$	\\
114710	&	16	&	0.40	&	$1.100\pm0.011$	&	0.180	&	$1.127\pm0.011$	\\
118098	&	15	&	0.17	&	$0.840\pm0.009$	&	0.260	&	$0.852\pm0.009$	\\
126660	&	15	&	0.43	&	$1.086\pm0.007$	&	0.250	&	$1.109\pm0.007$	\\
128167	&	26	&	0.40	&	$0.824\pm0.013$	&	0.320	&	$0.841\pm0.013$	\\
131156	&	30	&	1.92	&	$1.163\pm0.014$	&	0.210	&	$1.196\pm0.014$	\\
141795	&	8	&	0.10	&	$0.756\pm0.017$	&	0.260	&	$0.768\pm0.017$	\\
142860	&	19	&	0.11	&	$1.191\pm0.005$	&	0.290	&	$1.217\pm0.005$	\\
146233	&	25	&	0.46	&	$0.763\pm0.017$	&	0.250	&	$0.780\pm0.017$	\\
162003	&	25	&	2.23	&	$0.930\pm0.025$	&	0.240	&	$0.949\pm0.026$	\\
164259	&	19	&	0.56	&	$0.761\pm0.027$	&	0.250	&	$0.775\pm0.027$	\\
173667	&	42	&	1.06	&	$0.979\pm0.006$	&	0.170	&	$1.000\pm0.006$	\\
177724	&	31	&	1.05	&	$0.883\pm0.016$	&	0.320	&	$0.895\pm0.017$	\\
182572	&	33	&	1.91	&	$0.823\pm0.025$	&	0.320	&	$0.845\pm0.025$	\\
185144	&	15	&	1.00	&	$1.219\pm0.011$	&	0.240	&	$1.254\pm0.012$	\\
185395	&	25	&	0.86	&	$0.845\pm0.015$	&	0.200	&	$0.861\pm0.015$	\\
210418	&	20	&	0.32	&	$0.849\pm0.017$	&	0.180	&	$0.862\pm0.018$	\\
213558	&	27	&	0.84	&	$0.625\pm0.021$	&	0.260	&	$0.634\pm0.022$	\\
215648	&	34	&	1.06	&	$1.068\pm0.008$	&	0.260	&	$1.091\pm0.008$	\\
222368	&	36	&	0.90	&	$1.059\pm0.009$	&	0.260	&	$1.082\pm0.009$	

\enddata 

\tablecomments{Refer to Section~\ref{sec:diameters} for details.}

\end{deluxetable}

\newpage
\begin{deluxetable}{rccccc}
\tabletypesize{\tiny}
\tablewidth{0pt}
\tablecaption{CHARA Versus PTI Angular Diameters\label{tab:pti_chara}}
\tablehead{
\colhead{\textbf{ }} &
\colhead{\textbf{CHARA}} &
\colhead{\textbf{error}} &
\colhead{\textbf{PTI\tablenotemark{a}}} &
\colhead{\textbf{error}} &
\colhead{\textbf{ }} \\
\colhead{\textbf{HD}} &
\colhead{\textbf{$\theta_{\rm LD} \pm \sigma$}}  &
\colhead{\textbf{(\%)}} &
\colhead{\textbf{$\theta_{\rm LD} \pm \sigma$}}  &
\colhead{\textbf{(\%)}} &
\colhead{\textbf{$\Delta \theta_{\rm LD} / \sigma_{\rm C}$\tablenotemark{b}}} 
}
\startdata
16895	&	1.103	$\pm$	0.009	&	0.8	&	1.086	$\pm$	0.056	&	5.2	&	0.3	\\
19373	&	1.246	$\pm$	0.008	&	0.6	&	1.331	$\pm$	0.050	&	3.8	&	$-$1.7\phs	\\
20630	&	0.936	$\pm$	0.025	&	2.7	&	0.895	$\pm$	0.070	&	7.8	&	0.6	\\
22484	&	1.081	$\pm$	0.014	&	1.3	&	0.911	$\pm$	0.123	&	13.5\phn	&	1.4	\\
30652	&	1.526	$\pm$	0.004	&	0.3	&	1.409	$\pm$	0.048	&	3.4	&	2.4	\\
39587	&	1.051	$\pm$	0.009	&	0.9	&	1.124	$\pm$	0.056	&	5.0	&	$-$1.3\phs	\\
97603	&	1.328	$\pm$	0.009	&	0.7	&	1.198	$\pm$	0.053	&	4.4	&	2.4	\\
109358	&	1.238	$\pm$	0.030	&	2.4	&	1.138	$\pm$	0.055	&	4.8	&	1.6	\\
114710	&	1.127	$\pm$	0.011	&	1.0	&	1.071	$\pm$	0.057	&	5.3	&	1.0	\\
126660	&	1.109	$\pm$	0.007	&	0.6	&	1.130	$\pm$	0.055	&	4.9	&	$-$0.4\phs	\\
142860	&	1.217	$\pm$	0.005	&	0.4	&	1.161	$\pm$	0.054	&	4.7	&	1.0	\\
185144	&	1.254	$\pm$	0.012	&	1.0	&	1.092	$\pm$	0.057	&	5.2	&	2.8	\\
215648	&	1.091	$\pm$	0.008	&	0.7	&	1.022	$\pm$	0.059	&	5.8	&	1.2	\\
222368	&	1.082	$\pm$	0.009	&	0.8	&	1.062	$\pm$	0.057	&	5.4	&	0.3	

\enddata 

\tablenotetext{a}{From \citet{van09}.}
\tablenotetext{b}{Here, $\Delta \theta_{\rm LD}$ is the difference between PTI and CHARA limb darkened angular diameters, and $\sigma_{\rm C}$ is the combined error, $\sigma_{\rm C} = (\sigma_{\rm CHARA}^2 + \sigma_{\rm PTI}^2)^{0.5}$.}  

\tablecomments{Refer to Section~\ref{sec:CHARA_vs_PTI} for details.}

\end{deluxetable}

\newpage
\begin{deluxetable}{rcccrc}
\tabletypesize{\tiny}
\tablewidth{0pt}
\tablecaption{CHARA Versus PTI Calibrators\label{tab:pti_chara_calibs}}
\tablehead{
\colhead{\textbf{Calibrator}} &
\colhead{\textbf{CHARA}} &
\colhead{\textbf{PTI\tablenotemark{a}}} &
\colhead{\textbf{Calibrator SED}} &
\colhead{\textbf{Object}} &
\colhead{\textbf{Object Measured}} \\
\colhead{\textbf{HD}} &
\colhead{\textbf{$\theta_{\rm SED}$ (mas)}}  &
\colhead{\textbf{$\theta_{\rm SED}$ (mas)}} &
\colhead{\textbf{$\theta_{\rm CHARA} / \theta_{\rm PTI}$}}  &
\colhead{\textbf{HD}} &
\colhead{\textbf{$\theta_{\rm CHARA} / \theta_{\rm PTI}$\tablenotemark{b}}} 
}
\startdata

20675	&	0.415$\pm$0.012	&	0.424$\pm$0.020	&	0.98$\pm$0.05	&	16895	&	1.02$\pm$0.05	\\
20675	&	0.415$\pm$0.012	&	0.424$\pm$0.020	&	0.98$\pm$0.05	&	19373	&	0.94$\pm$0.04	\\
22879	&	0.342$\pm$0.021	&	0.369$\pm$0.009	&	0.93$\pm$0.06	&	20630	&	1.05$\pm$0.09	\\
22879	&	0.342$\pm$0.021	&	0.369$\pm$0.009	&	0.93$\pm$0.06	&	22484	&	1.19$\pm$0.16	\\
28355	&	0.425$\pm$0.030	&	0.401$\pm$0.012	&	1.06$\pm$0.08	&	30652	&	1.08$\pm$0.04	\\
30739	&	0.461$\pm$0.018	&	0.544$\pm$0.025	&	0.85$\pm$0.05	&	30652	&	1.08$\pm$0.04	\\
31295	&	0.439$\pm$0.043	&	0.470$\pm$0.022	&	0.93$\pm$0.10	&	30652	&	1.08$\pm$0.04	\\
38558	&	0.422$\pm$0.008	&	0.442$\pm$0.033	&	0.95$\pm$0.07	&	39587	&	0.94$\pm$0.05	\\
43042	&	0.591$\pm$0.030	&	0.655$\pm$0.017	&	0.90$\pm$0.05	&	39587	&	0.94$\pm$0.05	\\
99285	&	0.456$\pm$0.017	&	0.454$\pm$0.026	&	1.00$\pm$0.07	&	97603	&	1.11$\pm$0.05	\\
110897	&	0.492$\pm$0.022	&	0.504$\pm$0.009	&	0.98$\pm$0.05	&	109358	&	1.09$\pm$0.06	\\
132254	&	0.520$\pm$0.015	&	0.542$\pm$0.013	&	0.96$\pm$0.04	&	126660	&	0.98$\pm$0.05	\\
193664	&	0.494$\pm$0.019	&	0.552$\pm$0.011	&	0.89$\pm$0.04	&	185144	&	1.15$\pm$0.06	\\
211976	&	0.373$\pm$0.013	&	0.377$\pm$0.009	&	0.99$\pm$0.04	&	215648	&	1.07$\pm$0.06	\\
214923	&	0.611$\pm$0.029	&	0.552$\pm$0.094	&	1.11$\pm$0.20	&	215648	&	1.07$\pm$0.06	\\
216735	&	0.321$\pm$0.022	&	0.330$\pm$0.020	&	0.97$\pm$0.09	&	215648	&	1.07$\pm$0.06	\\
216735	&	0.321$\pm$0.022	&	0.330$\pm$0.020	&	0.97$\pm$0.09	&	222368	&	1.02$\pm$0.06	\\
222603	&	0.577$\pm$0.032	&	0.533$\pm$0.014	&	1.08$\pm$0.07	&	222368	&	1.02$\pm$0.06	

\enddata 
\tablenotetext{a}{From the PTICC \citep{van08}.}
\tablenotetext{b}{The limb-darkened diameter presented here versus the value in \citet{van09}.}
\tablecomments{Refer to Section~\ref{sec:CHARA_vs_PTI} for details.}

\end{deluxetable}

\newpage
\begin{deluxetable}{lccccc}
\tabletypesize{\tiny}
\tablewidth{0pt}
\tablecaption{Calibrators Observed\label{tab:Fbol_phot}}
\tablehead{
\colhead{\textbf{Object}} &
\colhead{\textbf{System /}} &
\colhead{\textbf{Bandpass}} &
\colhead{\textbf{Value}} &
\colhead{\textbf{Error}} &
\colhead{\textbf{ }}	 \\
\colhead{\textbf{HD}} &
\colhead{\textbf{Wavelength}}	&
\colhead{\textbf{Bandwidth}}	&
\colhead{\textbf{(mag)}}	&
\colhead{\textbf{(mag)}}	&
\colhead{\textbf{Reference}}	 
}
\startdata


HD 4614 & 1240  & 248 & 2.41 & 0.05 & \citet{1988A_AS...74..127S}\\

HD 4614 & 1250  & 250 & 2.36 & 0.05 & \citet{1965CoLPL...3...73J}\\

HD 4614 & 1250  & 250 & 2.36 & 0.05 & \citet{1966CoLPL...4...99J}\\

HD 4614 & 1250  & 250 & 2.35 & 0.05 & \citet{1968ApJ...152..465J}\\

HD 4614 & 1250  & 250 & 2.35 & 0.05 & \citet{1975A_AS...22....1V}\\

HD 4614 & 1250  & 250 & 2.41 & 0.05 & \citet{1990A_A...232..396B}\\

HD 4614 & 1270  & 254 & 2.35 & 0.05 & \citet{1980A_A....87..139B}\\

HD 4614 & 1620  & 324 & 2.02 & 0.05 & \citet{1980A_A....87..139B}\\
\enddata
\tablecomments{Refer to Section~\ref{sec:data_calibration} for details.}
\end{deluxetable} 
\newpage
\begin{deluxetable}{rccccc}
\tabletypesize{\tiny}
\tablewidth{0pt}
\tablecaption{Bolometric Fluxes\label{tab:sample_fbols}}
\tablehead{
\colhead{\textbf{Star}} &
\colhead{\textbf{Template}} &
\colhead{\textbf{\#}} &
\colhead{\textbf{Reduced}} &
\colhead{\textbf{$F_{\rm BOL} \pm \sigma$}} &
\colhead{\textbf{$A_{V} \pm \sigma$}}	\\
\colhead{\textbf{HD}} &
\colhead{\textbf{Sp.Ty.}}	&
\colhead{\textbf{PHOT}}	&
\colhead{\textbf{$\chi^{2}$}}	&
\colhead{\textbf{($10^{-8}$ erg~s$^{-1}$~cm$^{-2}$)}}	&
\colhead{\textbf{(mag)}} 
}
\startdata

4614	&	G0V	&	80	&	1.04	&	113.90$\pm$1.71\phn\phn	&	0.001$\pm$0.014	\\
5015	&	F8V	&	95	&	0.74	&	31.50$\pm$0.56\phn	&	0.005$\pm$0.016	\\
6582	&	G5V	&	165	&	2.63	&	24.48$\pm$0.39\phn	&	0.000$\pm$0.014	\\
10780	&	K0V	&	86	&	0.75	&	16.43$\pm$0.33\phn	&	0.000$\pm$0.018	\\
16895	&	F6V	&	94	&	0.70	&	58.21$\pm$1.04\phn	&	0.000$\pm$0.016	\\
19373	&	G0V	&	89	&	0.41	&	63.29$\pm$0.93\phn	&	0.016$\pm$0.013	\\
20630	&	G5V	&	205	&	0.59	&	32.46$\pm$0.55\phn	&	0.000$\pm$0.015	\\
22484	&	G0V	&	171	&	0.33	&	51.23$\pm$0.69\phn	&	0.000$\pm$0.012	\\
30652	&	F6V	&	251	&	0.72	&	139.80$\pm$1.50\phn\phn	&	0.000$\pm$0.009	\\
34411	&	G1V	&	151	&	0.49	&	35.01$\pm$0.45\phn	&	0.000$\pm$0.011	\\
39587	&	G0V	&	139	&	0.28	&	46.44$\pm$0.78\phn	&	0.011$\pm$0.015	\\
48737	&	F5III	&	85	&	0.60	&	115.20$\pm$2.37\phn\phn	&	0.000$\pm$0.018	\\
56537	&	A5V	&	136	&	1.11	&	95.40$\pm$2.18\phn	&	0.000$\pm$0.019	\\
58946	&	F0V	&	156	&	0.41	&	54.87$\pm$0.88\phn	&	0.020$\pm$0.014	\\
81937	&	F02IV	&	69	&	1.69	&	85.77$\pm$1.88\phn	&	0.035$\pm$0.019	\\
82328	&	F6.5IV	&	96	&	0.32	&	139.70$\pm$2.81\phn\phn	&	0.000$\pm$0.018	\\
82885	&	G8V	&	189	&	1.32	&	19.57$\pm$0.18\phn	&	0.032$\pm$0.008	\\
86728	&	G5V	&	150	&	0.26	&	19.63$\pm$0.39\phn	&	0.000$\pm$0.017	\\
90839	&	F8V	&	83	&	0.43	&	31.69$\pm$0.82\phn	&	0.000$\pm$0.023	\\
95418	&	A0IV	&	106	&	0.82	&	339.90$\pm$7.05\phn\phn	&	0.038$\pm$0.017	\\
97603	&	A4V	&	105	&	1.55	&	250.90$\pm$3.87\phn\phn	&	0.053$\pm$0.013	\\
101501	&	G6.5V	&	138	&	0.41	&	21.27$\pm$0.30\phn	&	0.011$\pm$0.012	\\
102870	&	G0V	&	222	&	0.35	&	96.43$\pm$1.41\phn	&	0.000$\pm$0.013	\\
103095	&	G9V	&	244	&	3.85	&	8.27$\pm$0.08	&	0.000$\pm$0.009	\\
109358	&	G0V	&	137	&	0.57	&	52.11$\pm$0.84\phn	&	0.000$\pm$0.014	\\
114710	&	G0V	&	216	&	0.43	&	52.49$\pm$0.56\phn	&	0.000$\pm$0.010	\\
118098	&	A5V	&	104	&	2.29	&	111.80$\pm$1.57\phn\phn	&	0.000$\pm$0.013	\\
126660	&	F7V	&	98	&	0.50	&	63.08$\pm$1.47\phn	&	0.000$\pm$0.020	\\
128167	&	F2V	&	182	&	1.88	&	44.52$\pm$0.54\phn	&	0.062$\pm$0.010	\\
131156	&	G8V	&	90	&	1.45	&	46.18$\pm$1.05\phn	&	0.043$\pm$0.019	\\
141795	&	A5V	&	101	&	1.13	&	83.87$\pm$2.05\phn	&	0.020$\pm$0.020	\\
142860	&	F6V	&	156	&	0.31	&	77.38$\pm$1.30\phn	&	0.014$\pm$0.014	\\
146233	&	G3.5V	&	116	&	0.62	&	17.65$\pm$0.46\phn	&	0.000$\pm$0.022	\\
162003	&	F6V	&	96	&	0.44	&	39.22$\pm$0.95\phn	&	0.000$\pm$0.021	\\
164259	&	F3.5V	&	134	&	0.43	&	36.34$\pm$0.74\phn	&	0.000$\pm$0.018	\\
173667	&	F6V	&	84	&	0.33	&	55.02$\pm$1.19\phn	&	0.002$\pm$0.019	\\
177724	&	A0V	&	163	&	1.16	&	191.50$\pm$3.12\phn\phn	&	0.002$\pm$0.014	\\
182572	&	G8IV	&	95	&	1.86	&	26.66$\pm$0.63\phn	&	0.000$\pm$0.021	\\
185144	&	K0V	&	240	&	0.72	&	39.93$\pm$0.56\phn	&	0.000$\pm$0.012	\\
185395	&	F2V	&	143	&	0.25	&	40.91$\pm$0.86\phn	&	0.015$\pm$0.018	\\
210418	&	A2IV	&	83	&	1.21	&	98.86$\pm$2.46\phn	&	0.000$\pm$0.022	\\
213558	&	A0V	&	109	&	0.94	&	93.04$\pm$2.21\phn	&	0.015$\pm$0.020	\\
215648	&	F8V	&	154	&	0.66	&	57.33$\pm$1.05\phn	&	0.000$\pm$0.016	\\
222368	&	F8V	&	279	&	0.40	&	60.95$\pm$0.99\phn	&	0.000$\pm$0.014	\\

\enddata

\tablecomments{For details, see Section~\ref{sec:fbols}.}

\end{deluxetable}
\newpage
\begin{deluxetable}{lccc}
\tabletypesize{\tiny}
\tablewidth{0pt}
\tablecaption{Radii, Luminosities, and Temperatures\label{tab:r_l_t}}
\tablehead{
\colhead{\textbf{Star}} &
\colhead{\textbf{$R$}} &
\colhead{\textbf{$L$}} &
\colhead{\textbf{$T_{\rm EFF}$}}	\\
\colhead{\textbf{HD}} &
\colhead{\textbf{($R_{\odot}$)}}	&
\colhead{\textbf{($L_{\odot}$)}}	&
\colhead{\textbf{(K)}} 
}
\startdata
4614	&	$1.039\pm0.004$	&	$1.252\pm0.019$	&	$6003\pm24$\phn\phn	\\
5015	&	$1.743\pm0.023$	&	$3.432\pm0.061$	&	$5963\pm44$\phn\phn	\\
6582	&	$0.790\pm0.009$	&	$0.428\pm0.007$	&	$5264\pm32$\phn\phn	\\
10780	&	$0.825\pm0.021$	&	$0.516\pm0.010$	&	$5396\pm72$\phn\phn	\\
16895	&	$1.319\pm0.011$	&	$2.235\pm0.040$	&	$6157\pm37$\phn\phn	\\
19373	&	$1.412\pm0.009$	&	$2.181\pm0.032$	&	$5915\pm29$\phn\phn	\\
20630	&	$0.919\pm0.025$	&	$0.841\pm0.014$	&	$5776\pm81$\phn\phn	\\
22484	&	$1.622\pm0.024$	&	$3.042\pm0.042$	&	$5997\pm44$\phn\phn	\\
30652	&	$1.323\pm0.004$	&	$2.822\pm0.030$	&	$6516\pm19$\phn\phn	\\
34411	&	$1.331\pm0.021$	&	$1.732\pm0.022$	&	$5749\pm48$\phn\phn	\\
39587	&	$0.979\pm0.009$	&	$1.081\pm0.018$	&	$5961\pm36$\phn\phn	\\
48737	&	$2.710\pm0.021$	&	$11.574\pm0.238$\phn	&	$6480\pm39$\phn\phn	\\
56537	&	$2.777\pm0.047$	&	$28.306\pm0.648$\phn	&	$8007\pm77$\phn\phn	\\
58946	&	$1.655\pm0.028$	&	$5.542\pm0.089$	&	$6899\pm63$\phn\phn	\\
81937	&	$2.902\pm0.026$	&	$15.086\pm0.330$\phn	&	$6693\pm45$\phn\phn	\\
82328	&	$2.365\pm0.008$	&	$7.871\pm0.158$	&	$6300\pm33$\phn\phn	\\
82885	&	$1.003\pm0.016$	&	$0.784\pm0.007$	&	$5434\pm45$\phn\phn	\\
86728	&	$1.247\pm0.021$	&	$1.378\pm0.027$	&	$5612\pm52$\phn\phn	\\
90839	&	$1.091\pm0.020$	&	$1.605\pm0.042$	&	$6233\pm68$\phn\phn	\\
95418	&	$3.021\pm0.038$	&	$63.015\pm1.307$\phn	&	$9377\pm75$\phn\phn	\\
97603	&	$2.557\pm0.020$	&	$24.973\pm0.385$\phn	&	$8085\pm42$\phn\phn	\\
101501	&	$0.940\pm0.010$	&	$0.609\pm0.009$	&	$5270\pm32$\phn\phn	\\
102870	&	$1.681\pm0.008$	&	$3.572\pm0.052$	&	$6132\pm26$\phn\phn	\\
103095	&	$0.681\pm0.006$	&	$0.212\pm0.002$	&	$4759\pm20$\phn\phn	\\
109358	&	$1.123\pm0.028$	&	$1.151\pm0.018$	&	$5653\pm72$\phn\phn	\\
114710	&	$1.106\pm0.011$	&	$1.357\pm0.014$	&	$5936\pm33$\phn\phn	\\
118098	&	$2.079\pm0.025$	&	$17.885\pm0.252$\phn	&	$8247\pm52$\phn\phn	\\
126660	&	$1.733\pm0.011$	&	$4.131\pm0.096$	&	$6265\pm41$\phn\phn	\\
128167	&	$1.431\pm0.023$	&	$3.461\pm0.042$	&	$6594\pm55$\phn\phn	\\
131156	&	$0.863\pm0.011$	&	$0.645\pm0.015$	&	$5580\pm46$\phn\phn	\\
141795	&	$1.783\pm0.040$	&	$12.134\pm0.296$\phn	&	$8084\pm102$\phn	\\
142860	&	$1.472\pm0.007$	&	$3.039\pm0.051$	&	$6294\pm29$\phn\phn	\\
146233	&	$1.166\pm0.026$	&	$1.058\pm0.028$	&	$5433\pm69$\phn\phn	\\
162003	&	$2.329\pm0.067$	&	$6.343\pm0.153$	&	$6014\pm90$\phn\phn	\\
164259	&	$1.961\pm0.071$	&	$6.251\pm0.127$	&	$6529\pm118$\phn	\\
173667	&	$2.064\pm0.017$	&	$6.296\pm0.136$	&	$6376\pm39$\phn\phn	\\
177724	&	$2.449\pm0.046$	&	$38.492\pm0.627$\phn	&	$9205\pm95$\phn\phn	\\
182572	&	$1.379\pm0.042$	&	$1.904\pm0.045$	&	$5787\pm92$\phn\phn	\\
185144	&	$0.776\pm0.008$	&	$0.410\pm0.006$	&	$5255\pm31$\phn\phn	\\
185395	&	$1.697\pm0.030$	&	$4.265\pm0.090$	&	$6381\pm65$\phn\phn	\\
210418	&	$2.623\pm0.083$	&	$24.549\pm0.610$\phn	&	$7951\pm97$\phn\phn	\\
213558	&	$2.143\pm0.074$	&	$28.552\pm0.678$\phn	&	$9131\pm167$\phn	\\
215648	&	$1.912\pm0.016$	&	$4.722\pm0.087$	&	$6167\pm36$\phn\phn	\\
222368	&	$1.595\pm0.014$	&	$3.555\pm0.058$	&	$6288\pm37$\phn\phn	
\enddata

\tablecomments{For details, see Section~\ref{sec:l_t_r}.} 

\end{deluxetable}
\newpage
\begin{deluxetable}{lccc}
\tablewidth{0pt}
\tablecaption{Solutions to $T_{\rm EFF}$ Relations \label{tab:poly_coeff}}
\tablehead{
\colhead{\textbf{ }} &
\colhead{\textbf{($B-V$)}} &
\colhead{\textbf{($V-K$)}} 	&
\colhead{\textbf{Spectral Type}} 	
}
\startdata
Equation in text	&	\ref{eq:poly6_bmv}	&	\ref{eq:poly6_vmk}		&	\ref{eq:temp_VS_SpTy}	\\
\hline 
\hline
$n$ \dotfill	&	39	&	44	&	41		\\
Range \dotfill	&	$0.05 - 0.80$\tablenotemark{a}	&	$0.0 - 2.0$	&	A0 - K0\tablenotemark{a}		\\
Reduced $\chi^2$ \dotfill	&	5.97	&	6.04	&	12.1		\\
Median d$T_{\rm EFF}$	\dotfill	&	55	&	64	&	90	\\
\enddata 
\tablenotetext{a}{Metallicity [Fe/H]~$> -0.75$}

\tablecomments{Refer to Section~\ref{sec:empirical_relations} for details.}

\end{deluxetable}

\newpage
\begin{deluxetable}{lcc}
\tabletypesize{\tiny}
\tablewidth{0pt}
\tablecaption{$Y^2$ Model Isochrone Results for Stars with $L > 0.75 L_{\rm \odot}$\label{tab:Y2_model_output}}
\tablehead{
\colhead{\textbf{Star}} &
\colhead{\textbf{Mass $\pm \sigma$}} &
\colhead{\textbf{Age $\pm \sigma$}}	 \\
\colhead{\textbf{HD}} &
\colhead{\textbf{($M_{\odot}$)}}	&
\colhead{\textbf{(Gyr)}} 
}

\startdata

4614&0.972$\pm$0.012&5.4$\pm$0.9\\

5015&1.182$\pm$0.011&5.4$\pm$0.3\\

16895&1.138$\pm$0.010&4.0$\pm$0.4\\

19373&1.169$\pm$0.013&4.8$\pm$0.5\\

20630&1.037$\pm$0.042&0.2$\pm$3.1\\

22484&1.139$\pm$0.016&5.7$\pm$0.4\\

30652&1.283$\pm$0.006&1.3$\pm$0.2\\

34411&1.041$\pm$0.015&8.1$\pm$0.8\\

39587&1.029$\pm$0.029&2.0$\pm$1.8\\

48737&1.706$\pm$0.012&1.7$\pm$0.1\\

56537&2.111$\pm$0.010&0.8$\pm$0.0\\

58946&1.355$\pm$0.013&2.1$\pm$0.2\\

81937&1.824$\pm$0.016&1.4$\pm$0.1\\

82328&1.506$\pm$0.095&2.4$\pm$0.7\\

82885&0.910$\pm$0.020&10.6$\pm$2.2\phn\\

86728&1.034$\pm$0.036&8.5$\pm$1.8\\

90839&1.119$\pm$0.035&1.5$\pm$1.4\\

97603&2.061$\pm$0.006&0.8$\pm$0.0\\

102870&1.324$\pm$0.005&3.3$\pm$0.1\\

109358&0.852$\pm$0.023&14.2$\pm$2.1\phn\\

114710&1.045$\pm$0.013&4.6$\pm$0.9\\

118098&1.940$\pm$0.006&0.7$\pm$0.0\\

126660&1.232$\pm$0.058&4.2$\pm$0.8\\

128167&1.194$\pm$0.013&3.1$\pm$0.4\\

141795&1.820$\pm$0.026&0.5$\pm$0.2\\

142860&1.184$\pm$0.012&3.9$\pm$0.3\\

146233&0.887$\pm$0.019&14.9$\pm$2.0\phn\\

162003&1.311$\pm$0.016&3.8$\pm$0.1\\

164259&1.429$\pm$0.013&2.4$\pm$0.2\\

173667&1.422$\pm$0.009&2.6$\pm$0.1\\

177724&1.984$\pm$0.006&0.8$\pm$0.0\\

182572&1.186$\pm$0.015&4.5$\pm$0.8\\

185395&1.342$\pm$0.011&2.8$\pm$0.2\\

210418&1.858$\pm$0.024&1.1$\pm$0.1\\

213558&2.209$\pm$0.037&0.4$\pm$0.1\\

215648&1.192$\pm$0.011&4.8$\pm$0.2\\

222368&1.268$\pm$0.009&3.4$\pm$0.2\\
\enddata

\tablecomments{Isochrones are not sensitive to stars with $L < 0.75$~L$_{\odot}$ which includes these stars in our survey: HD~6582, HD~10780, HD~101501, HD~103095, HD~131156 and HD~185144. Errors are calculated assuming the 1-$\sigma$ errors on temperature and luminosity.  Refer to Section~\ref{sec:masses_and_ages} for details.}

\end{deluxetable}

\newpage


 \clearpage 

 \clearpage 

 \clearpage 

 \clearpage 

 \clearpage 

 \clearpage 

\clearpage

\clearpage

\clearpage	
			
\clearpage
		

\clearpage

\clearpage

\clearpage

\clearpage

\clearpage

\clearpage

\clearpage

\clearpage

\clearpage

\clearpage

\clearpage

\clearpage

\clearpage

\clearpage

\clearpage

\clearpage

\clearpage

\end{document}